\documentclass[article,shortnames,nojss]{jss}

\usepackage{amsmath}
\usepackage{tikz}
\usepackage{subfig}
\usepackage{appendix}
\usepackage{rotating}
\usepackage{bigdelim}
\usepackage{multirow}
\usepackage{bigstrut} 
\usepackage{graphicx}
\usepackage[utf8]{inputenc}
\usepackage{amsfonts}
\usepackage{amssymb}
\usepackage[onehalfspacing]{setspace}
\usepackage[T1]{fontenc}
\usepackage{tabularx,ragged2e,booktabs}
\usepackage{dirtytalk}
\usepackage{natbib}
\usepackage{algorithmicx}
\usepackage[Algorithm,ruled]{algorithm}

\usepackage[noend]{algpseudocode}
\usepackage{pifont}
\usepackage{amsmath}

\newcommand{\bgamma}{\boldsymbol{\gamma}}
\newcommand{\bbeta}{\boldsymbol{\beta}}
\newcommand{\bepsilon}{\boldsymbol{\epsilon}}

\newcommand{\bone}{\boldsymbol{1}}

\newcommand{\bg}{\boldsymbol{g}}
\newcommand{\bd}{\boldsymbol{d}}
\newcommand{\bb}{\boldsymbol{b}}
\newcommand{\bff}{\boldsymbol{f}}

\newcommand{\bx}{\boldsymbol{x}}

\newcommand{\by}{\boldsymbol{y}}

\newcommand{\bc}{\boldsymbol{c}}
\newcommand{\bu}{\boldsymbol{u}}

\newcommand{\btheta}{\boldsymbol{\theta}}

\newcommand{\bzero}{\boldsymbol{0}}

\newlength{\RoundedBoxWidth}
\newsavebox{\GrayRoundedBox}
\newenvironment{GrayBox}[1][\dimexpr\textwidth-4.5ex]%
   {\setlength{\RoundedBoxWidth}{\dimexpr#1}
    \begin{lrbox}{\GrayRoundedBox}
       \begin{minipage}{\RoundedBoxWidth}}%
   {   \end{minipage}
    \end{lrbox}
    \begin{center}
    \begin{tikzpicture}%
       \draw node[draw=black,fill=black!10,rounded corners,%
             inner sep=2ex,text width=\RoundedBoxWidth]%
             {\usebox{\GrayRoundedBox}};
    \end{tikzpicture}
    \end{center}}

\author{Deniz Akdemir \\Michigan State University \\ Correspondence: deniz.akdemir.work@gmail.com}
\title{\pkg{STPGA}: Selection of training populations with a genetic algorithm}

\Plainauthor{Deniz Akdemir} 
\Plaintitle{\pkg{STPGA}: Selection of training populations (and other subset selection problems) with an accelerated genetic algorithm} 
\Shorttitle{\pkg{\pkg{STPGA}}: An R-package for selection of training populations with a genetic algorithm} 

\Abstract{
Optimal subset selection is an important task that has numerous algorithms designed for it and has many application areas. \pkg{STPGA} contains a special genetic algorithm supplemented with a tabu memory property (that keeps track of previously tried solutions and their fitness for a number of iterations), and with a regression of the fitness of the solutions on their coding that is used to form the ideal estimated solution (look ahead property) to search for solutions of generic optimal subset selection problems. I have initially developed the programs for the specific problem of selecting training populations for genomic prediction or association problems, therefore I give discussion of the theory behind optimal design of experiments to explain the default optimization criteria in \pkg{STPGA}, and illustrate the use of the programs in this endeavor.  Nevertheless, I have picked a few other areas of application: supervised and unsupervised variable selection based on kernel alignment, supervised variable selection with design criteria, influential observation identification for regression, solving mixed integer quadratic optimization problems, balancing gains and inbreeding in a breeding population. Some of these illustrations pertain new statistical approaches.
}

\Plainkeywords{Genomic Prediction, Genome-wide Association, Optimal Design, Optimal Experiments, High Dimensional Data, Sparse Models, Markers, Anchor Markers, Inbreeding, Genetic Gain, Kinship, Similarity, Subset Selection, Mixed Integer Quadratic Programming, R, Genomic Selection} 

\Address{
  Deniz Akdemir \\
  Department of Epidemiology and Biostatistics\\
  Michigan State University\\
  East Lansing MI USA\\
  E-mail: \email{deniz.akdemir.work@gmail.com}\\
}

\usepackage{Sweave}
\begin{document}


\section{Introduction}
\label{sec:Intro}

The paper introduces the \proglang{R} (R Core Team 2016) package \pkg{STPGA} that provides a genetic algorithm for subset selection. The package is available from the Comprehensive R Archive Network (CRAN) at \url{http://CRAN.R-project.org/package=STPGA}, and some of the underlying motivations, methodology and results were presented in \citep{akdemir2015optimization, isidro2015training, crossa2016genomic, akdemir2016efficient} and also some innovations that will be detailed in several subsequent articles.  This document details version $4.0$ which includes major upgrades and bug fixes compared to previous versions.

Numerous other algorithms have been proposed for the optimal subset selection problem, many of them are  heuristic exchange type algorithms \citep{fedorov1972theory,mitchell1974algorithm, nguyen1992review, rincent2012maximizing, isidro2015training}. In exchange type algorithms new solutions are obtained by adding one point and  removing another at a time (some exchange algorithms might allow exchange of more than one design point at once), these algorithms are greedy and are only proven to find the best subset for certain type of design criteria. In general, exchange algorithms are prone to getting stuck in local optimal solutions.  Branch and bound  (BB) \citep{furnival1974regressions} is a global exhaustive search method that has proven to be reasonably efficient on practical problems. BB searches the design region by iteratively dividing design region and searching each piece for an optimal solution. BB is often more efficient than straight enumeration because it can eliminate regions that provably do not contain an optimal solution. \cite{welch1982branch} uses a BB algorithm to find globally best $D$-optimal design for a given design criteria and a set of candidate points. Another method that has been applied to the subset selection problem is simulated annealing \citep{haines1987application}. Branch and bound and simulated annealing algorithms require appreciable computation time even for moderate sized problems. 

Genetic algorithms (GAs) are a class of evolutionary algorithms made popular by John Holland and his colleagues \citep{goldberg1988genetic, holland1992adaptation, holland1992genetic}, and have been applied to find exact or approximate solutions to optimization and search problems \citep{goldberg1988genetic, sivanandam2007introduction, akdemir2015optimization, akdemir2016efficient}. There are numerous packages that implement GAs or similar evolutionary algorithms. The packages \pkg{gafit} \citep{tendys2002gafit}, \pkg{galts} \citep{satman2013galts}, genalg \citep{willighagen2005genalg}, rgenoud \citep{mebane2015package},  DEoptim \citep{ardia2016package} and the \pkg{GA} \citep{scrucca2013ga} offer many options for using optimization routines based on evolutionary algorithms. The optimization algorithm that is used in \pkg{STPGA} (LA-GA-T algorithm) is a modified genetic algorithm with tabu search and look ahead property and it is specialized for solving subset selection problems.

\begin{figure}[h!]
\begin{center}
\includegraphics[width=10cm, angle=-90]{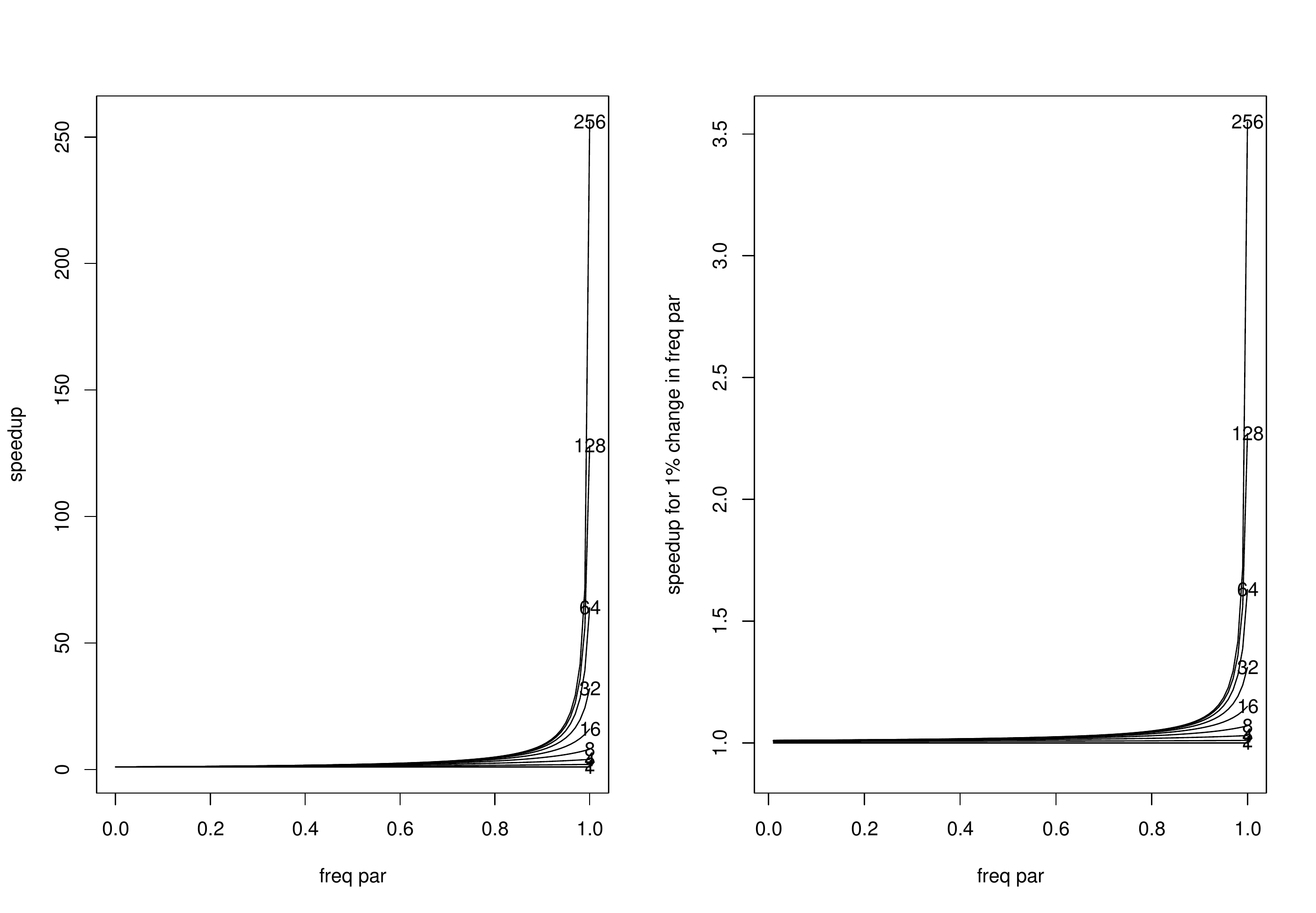}
\end{center}
\caption{(Left) Speedup for $\left \{1,2,4,8,16, 32, 64, 128, 256 \right \}$ processors for changing values of paralellization frequency, (Right) speedup for $1\%$ change in paralellization frequency.}\label{FigAmdahl}
\end{figure}

Today's trends in computation are towards computer architectures that integrate	many,	less	complex	processors, exploit	thread-level and	data-level	parallelism. This makes these computers perfect ground for implementation  of evolutionary algorithms for solving complex optimization problems since these algorithms can be easily to be run at parallel. To make my point more clear, lets remember Amdahl’s law \citep{amdahl1967validity} which puts a limit to the speed that can be gained by parallelizing a process: \[\text{speedup}=\frac{\text{serial processing time}}{\text{parallel processing time}}\leq \frac{1}{f_{par}/N_{p}+(1-f_{par})},\] where $f_{par}$ is the fraction of code which could be parallelized,$1-f_{par}$ is the serial fraction and $N_{p}$ is the number of processors. As it can be observed from the figures in Figure \ref{FigAmdahl}, obtained by applying this formula for varying values of $f_{par}$ between 0 and 1 and values of $N_p$ in $\left \{1,2,4,8,16, 32, 64, 128, 256 \right \},$ under the assumption that the number of processors doubles each year and the processing capabilities for each parallel node and everything else identical throughout, taking full advantage of parallelization requires methods that have paralelization frequency close to one. From the same figure we can read that $1\%$ change in parallelization frequency might cause up to 3.5 times speedup if there are 256 processors and a ideal parallelizable procedure might have speed more than to 50 times relative to a $80\%$ parallelizable procedure. Evolutionary algorithms, like GA, fall at the very right end of these figures. 

At the Michigan State University, at the beginning of year 2017, there were hundreds of nodes available for the researches through their high performance computer cluster (HPCC) system. It is not science fiction to claim that clusters with millions of nodes will be available to researchers with the technologies such as  cloud / grid computing. We are faced with the challenge of matching these these parallel resources with methods that can use them efficiently.

In my view, easy adaptation of GAs to parallel computation is a major advantage of GAs to other subset selection algorithms. GAs scale well with large and/or computationally expensive problems and can achieve reasonable running times by using parallel computing architectures with either shared or distributed memory systems, these systems are becoming increasingly available to the researchers. Scientific community prefer algorithms that run faster in serial. However, the direction the improvements in the computer technology seem be more in the parallelization rather than faster processors.

Some advantages of the GAs to other subset selection algorithms include the following:
\begin{itemize}
\item GA can be applied to many different problems and does not need to be reinvented for each new problem.
\item GA is a very flexible optimization algorithm, evolutionary mechanisms that are involved in GA can be modified at different stages of the algorithm; various selection strategies, various penalization of the objective function,  can be explored simultaneously or in a serial fashion. 
\item Adopting GAs in a parallel computing environment is easy. This might involve, for example, evaluation of the fitness function for the current GA population, or running many GAs at parallel to provide initial solutions generation of GAs.  
\end{itemize}

In addition the LA-GA-T algorithm in \pkg{STPGA} adds two more properties:
\begin{itemize}
\item Inferior solutions that were visited recently will not be visited. This property is akin to a memory in an intelligent system.
\item The state of current solutions are used to predict the best ideal solution. This gives the algorithm the look ahead property. This property is akin to inference in an intelligent system.
\end{itemize}

Nature solves problems through evolutionary processes, it works with communities of solutions that exploits the their communal information content to create new solutions, it is this information that persists, not the individual solutions. In addition, we have long standing theories explaining how and why such evolutionary processes work. There is also a vast amount of principled study of evolutionary mechanisms, both that are natural and artificial; the whole subject of evolutionary genetics; the methodology, theories and practices related to breeding; the theoretical and practical approaches of evolutionary algorithms and computation allows us humans to understand and manage these systems.    

In the next section, I briefly review the basic ideas behind simple GAs and the LA-GA-T that is used in STPGA. Then, I present the details of the interface to the \pkg{STPGA} package in Section 3, followed by several examples section and the conclusions. The examples section has been divided into two main parts: STPGA for selection of training populations, and STPGA in other subset selection problems. Both of these sections are rather long and detailed, especially the part that relates to optimal design that introduces some concepts and ideas of optimal design of experiments with a focus on predictive learning using regression models. Some of the design criteria discussed in this section are implemented in \pkg{STPGA}, a table listing of these criteria is provided. I also demonstrate how to write user defined criteria.

\section{Optimizer in STPGA}
\label{sec:opt}

The optimization algorithm that is used in \pkg{STPGA} is a modified genetic algorithm with tabu search and look ahead property. Genetic algorithms are stochastic search algorithms which are able to solve optimization problems using evolutionary strategies inspired by the basic principles of biological evolution. They use a population of candidate solutions that are represented as binary strings of $0$'s and $1$'s, this population evolving toward better solutions. At each iteration of the algorithm, a fitness function is used to evaluate and select the elite individuals and subsequently the next population is formed from the elites by genetically motivated operations such as crossover and mutation. The properties and prospects of genetic algorithms were first laid out in the cornerstone book of Holland \citep{holland1992adaptation}. 

GAs have an implicit parallelism property \citep{holland1992adaptation}. In addition, since GA uses a set of solutions at each iteration it couples well with the advanced computers (workstations with many processors and large memory) and computer systems (high performance computing clusters, cloud computing technologies) of today allowing it to be applied to very large scale optimization problems.  In my opinion, with the advent of new technologies like DNA computing \citep{liu2000dna,paun2005dna} that uses programmable molecular computing machines or quantum computers \citep{gruska1999quantum,leuenberger2001quantum} that operate on ''qubits'', parallelizable algorithms such as GA will have more and more important role in big scale optimization problems. The GA algorithm in STPGA is supplemented with two additional principles, tabu (memory) and inference through prediction based on a current population of solutions. I refer to it as the LA-GA-T (look ahead genetic algorithm with tabu) algorithm.

Tabu search is a search where most recently visited solutions are avoided by keeping a track of the previously tried solutions. This avoids many function evaluations and decreases the number of iterations till convergence, it is especially useful for generating new solutions around local optima.

The LA-GA-T algorithm in \pkg{STPGA} also uses the binary coding of the current population of solutions and their fitness to fit a linear ridge regression model from which the effects of individual digits in this binary code are estimated assuming that the contribution of an individual to the criterion value does not change much in relation to different subsets. The predicted ideal solution based on this model is constructed and included in the elite population of solutions. This gives the algorithm a look ahead property and improves the speed of convergence especially in the initial steps of the optimization. I should note here that the idea of regressing the fitnesses of solutions on their designs was inspired by the genomic selection methodology recently put into use in plant and animal breeding with the promise of increasing genetics gains from selection per unit of time. 

\begin{figure}[h!]
\begin{center}
\includegraphics[width=10cm, angle=-90]{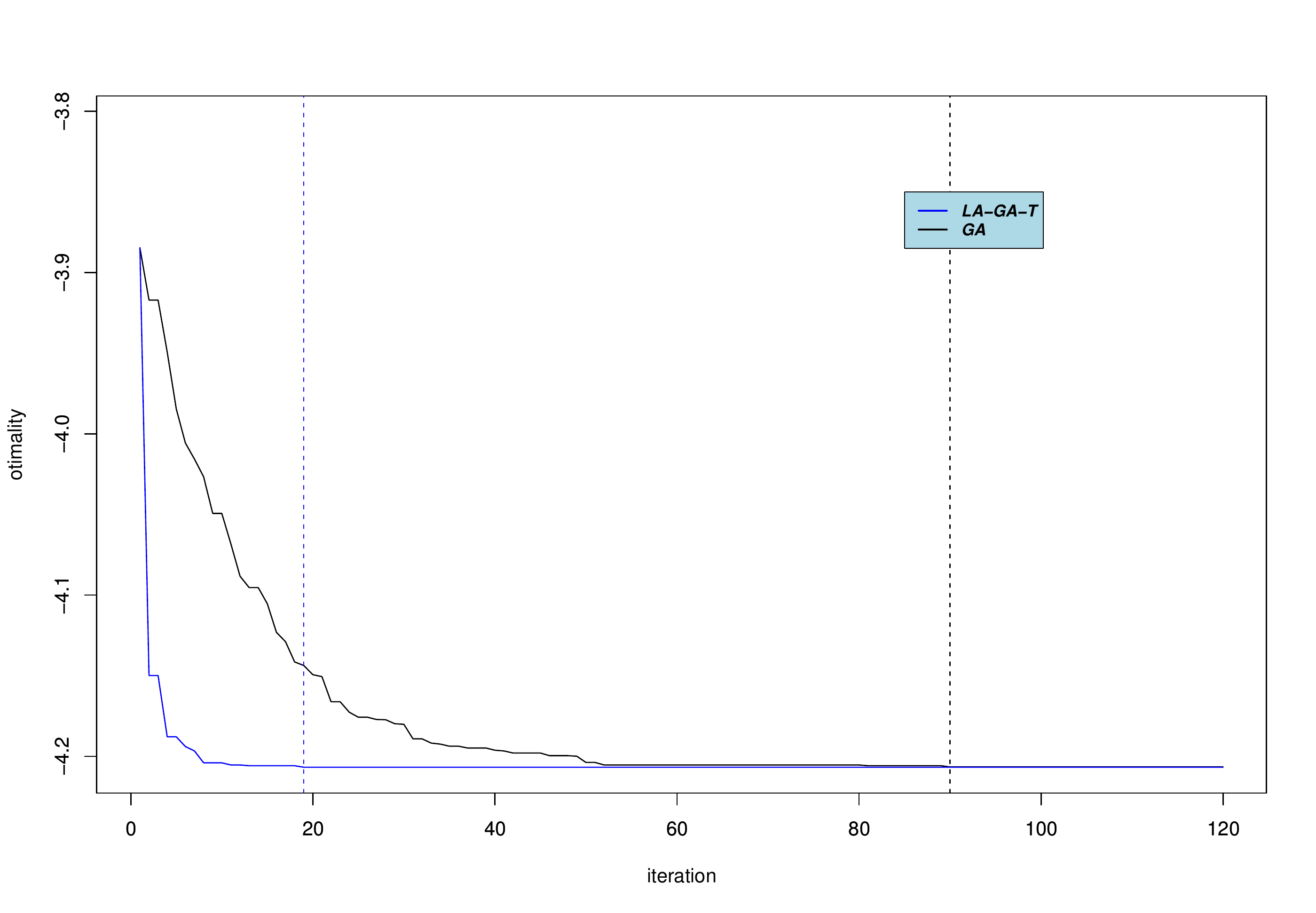}
\end{center}
\caption{The convergence of GA  and LA-GA-T for no test scenario.}\label{fig:COMPARESTPGAALGS1}
\end{figure}

\begin{figure}[h!]
\begin{center}
\includegraphics[width=10cm, angle=-90]{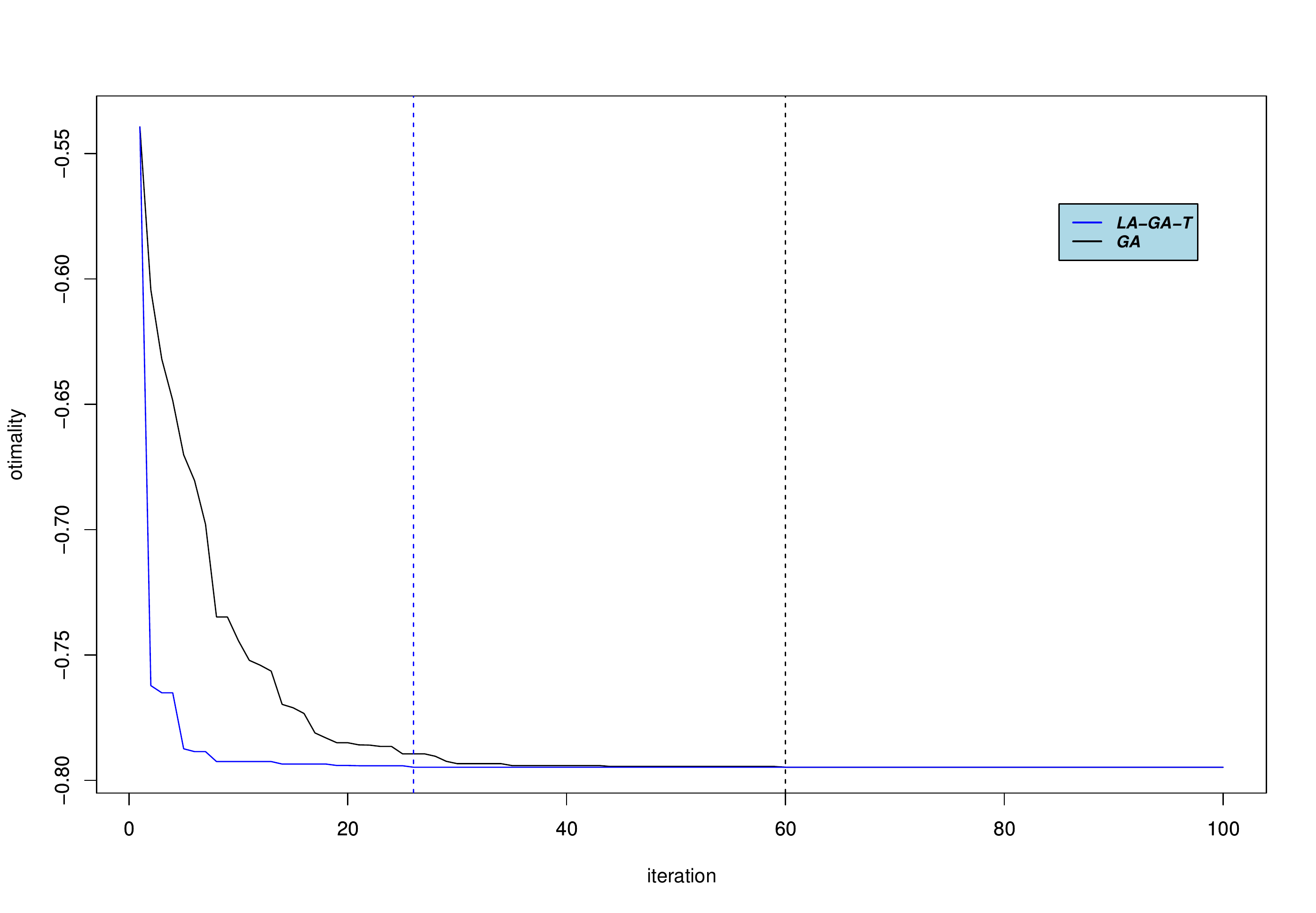}
\end{center}
\caption{The convergence of GA  and LA-GA-T when there is a target.}\label{fig:COMPARESTPGAALGS2}
\end{figure}

As can be seen from the Figures \ref{fig:COMPARESTPGAALGS1} and \ref{fig:COMPARESTPGAALGS2}, LA-GA-T converges in much fewer iterations compared to a simple GA. However, I have to note that the per iteration computation time for LA-GA-T algorithm is slightly higher compared to a simple GA. 

\begin{algorithm}
 \floatname{algorithm}{Procedure}
 \caption{Genetic Algorithm}
\label{GAalgorithm}
\begin{algorithmic}[1]
\State $t=0.$
\State initialization -Create an initial population of solutions of desired size, $S_t.$
\State Memory for tabu is empty, $MemTabu_t=NULL;$
\Repeat
\State $t=t+1,$
\State Evaluation -For each solution in $S_{t-1}$ calculate the criterion value,
\State Look ahead -Use the binary coding of $S_{t-1}$ and their fitness to fit a linear ridge regression model from which the effects of individual digits in this binary code are estimated. Put this solution in $S_t,$
\State Selection -Identify the best solutions by the ordering of criterion values, these are denoted by $E_t,$
\State Elitism -Let the best solution in $E_{t}$ be $s_{t}.$ Put $s_{t}$ in $S_t,$
\State Tabu -Update memory for tabu by letting  $MemTabu_t=S_{t-1}.$
\Repeat 
\State Crossover-Randomly pick two solutions in $E_t.$ Recombination of these two solutions are obtained by summing the frequency distributions of these solutions and sampling with new solutions using probabilities corresponding to this combined frequency distribution.
\State Mutation - With a given probability decrease the frequency of a mate that has positive frequency by some integer value less than the current frequency of that mate and increase the frequency of some other mate pair is by the same amount.
\If {the resulting solution is in $MemTabu_t$} 
\State  eliminate solution
\Else 
\State  Insert solution into $S_{t}.$
\EndIf
 \Until{$S_t$ has $N_{pop}$ solutions.}
\Until{Convergence is reached}
\State Evaluation -For each solution in $S_{t}$ calculate the criterion value,
\State Selection - Identify the best solutions by the ordering of criterion values, these are denoted by $E_t,$
\State Elitism -Let the best solution in $E_{t}$ be $s_{t}.$ Return $s_{t}.$ 
\end{algorithmic}
\end{algorithm}

The solutions obtained by any run of GA may be sub-optimal and different solutions can be obtained given different starting populations. Another layer of safety is obtained if the algorithm is started from multiple initial populations and an island model of evolution is used where separate populations are evolved independently for several steps and then the best solutions from these algorithms becomes the initial solutions to evolutionary algorithm. Since the functions in STPGA can start from user provided initial values, island models and other strategies can be combined when using the algorithm. I give an example code for doing this in the Appendix section.

\section{Software interface, computational considerations}
\label{sec:interface}

There are two main functions in \pkg{STPGA}, these are \pkg{GenAlgForSubsetSelection} and \pkg{GenAlgForSubsetSelectionNoTest}. The function \pkg{GenAlgForSubsetSelection} uses a simple genetic algorithm to identify a training set of a specified size from a larger set of candidates which minimizes an optimization criterion (for a known test set). The function \pkg{GenAlgForSubsetSelectionNoTest} is for identifying a training set of a specified size from a larger set of candidates which minimizes an optimization criterion, no test set is specified. These functions share a lot of common parameters, except \pkg{GenAlgForSubsetSelection} requires an additional input that specifies the target set of individuals and the data matrix should be supplemented to include the observed value of the variables for these target individuals. The inputs for these functions are described below:

\textbf{Inputs}
{
\begin{itemize}
  \item \textbf{P}{
depending on the criterion this is either a numeric data matrix or a symmetric similarity matrix. When it is a data matrix, the union of the identifiers of the candidate (and test) individuals should be put as row names (and column names in case of a similarity matrix). For methods using the relationships, this is the inverse of the relationship matrix with row and column names as the the identifiers of the candidate (and test) individuals.
}
  \item \textbf{Candidates}{ 
  vector of identifiers for the individuals in the candidate set.
}
  \item \textbf{Test}{
  vector of identifiers for the individuals in the test set.
}
  \item \textbf{ntoselect}{
 $n_{Train}:$ number of individuals to select in the training set.
}
  \item \textbf{npop}{
 genetic algorithm parameter, number of solutions at each iteration
}
  \item \textbf{nelite}{
 genetic algorithm parameter, number of solutions selected as elite parents which will  generate the next set of solutions. 
}
 \item \textbf{keepbest}{
 genetic algorithm parameter, TRUE or FALSE. If TRUE then the best solution is always kept in the next generation of solutions (elitism). 
}
 \item \textbf{tabu}{
 genetic algorithm parameter, TRUE or FALSE. If TRUE then the solutions that are saved in tabu memory will not be retried.}
 \item \textbf{tabumemsize}{
 genetic algorithm parameter, integer>0. Number of generations to hold in tabu memory. 
}

  \item \textbf{mutprob}{
 genetic algorithm parameter, probability of mutation for each generated solution.
}
 \item \textbf{mutintensity}{
mean of the Poisson variable that is used to decide the number of mutations for each cross. 
}
  \item \textbf{niterations}{
 genetic algorithm parameter, number of iterations.  
}
  \item \textbf{minitbefstop}{
 genetic algorithm parameter, number of iterations before stopping if no change is observed in criterion value.  
}

 \item \textbf{niterreg}{
 genetic algorithm parameter, number of iterations to use regressions}
  \item \textbf{lambda}{
 scalar shrinkage parameter  ($\lambda>0$).
}
\item \textbf{plotiters}{
 plot the convergence: TRUE or FALSE. Default is FALSE.
}
\item \textbf{plottype}{
 type of plot, default is 1. possible values 1,2,3.
}

\item \textbf{errorstat}{
 optimality criterion: One of the optimality criterion. Default is ''PEVMEAN''. It is possible to use user defined functions as shown in the examples.
}

\item \textbf{mc.cores}{
 number of cores to use.
}
\item \textbf{InitPop}{
a list of initial solutions
}
\item \textbf{tolconv}{
if the algorithm cannot improve the errorstat more than tolconv for the last minitbefstop iterations it will stop.
}
 \item \textbf{C}{
contrast matrix.
}
\item \textbf{Vg}{
covariance matrix between traits generated by the relationship K (only for multi-trait version of PEVMEANMM).
}
 \item \textbf{Ve}{
residual covariance matrix for the traits (only for multi-trait version of PEVMEANMM). 
}
\end{itemize}
}

All these inputs except \textbf{P}, \textbf{ntoselect} (also \textbf{Candidates} and \textbf{Test} for  \pkg{GenAlgForSubsetSelection}) have default values of NULL meaning that they are internally assigned to the default suggested settings. These settings are as follows: npop = 100, nelite = 5, keepbest = TRUE, tabu = FALSE, tabumemsize = 1, mutprob = .8, mutintensity = 1, niterations = 500, minitbefstop=100, niterreg = 5, lambda = 1e-6, plotiters = FALSE, plottype = 1, errorstat = ''PEVMEAN'', C = NULL, mc.cores = 1, InitPop = NULL, tolconv = 1e-7, Vg = NULL, Ve = NULL. In a specific application of \pkg{STPGA}, we recommend the users to change these options until they are satisfied with the final results.  Especially, when used with large data sets (many columns or rows), the parameters npop, niterations, minitbefstop should be increased.

Both functions return a named list of length $nelite+1.$ The first nelite elements of the list are optimized training samples of size $n_{train}$ and they are listed in increasing order of the optimization criterion. The last item on the list is a vector that stores the minimum values of the objective function at each iteration. The solution with best criterion value has name `Solution with rank 1`, the second `Solution with rank 2`, etc, ... The minimum values of the objective function through the iterations has name  `Best criterion values over iterations`.

The function \pkg{GenAlgForSubsetSelection} in the package uses this algorithm to identify a training set of a specified size from a larger set of candidates which minimizes an optimization criterion (for a known test set). The function ''GenAlgForSubsetSelectionNoTest'' tries to identify a training set of a specified size from a larger set of candidates which minimizes an optimization criterion, no test set is specified. 

The subset selection algorithms in ''GenAlgForSubsetSelectionNoTest'' and ''GenAlgForSubsetSelection'' have somewhat different inner workings. ''GenAlgForSubsetSelectionNoTest'' splits the individuals given in row names of the input matrix \textbf{P} into two parts: a set called Train of size ''ntoselect'' and its complement. The ''GenAlgForSubsetSelection'' starts with an input defining of split of the individuals given in row names of the input matrix \textbf{P} into three parts: a set called ''Candidates'', a set called ''Test'' and their complement, after this the algorithm splits the set ''Candidates'' into a set called ''Train'' and its complement. 

These two functions can be used with any user defined fitness functions and in the examples section, I will illustrate how these mechanisms can be used for general subset selection problems.

I have developed this package for my interest in solving certain design problems. Therefore, I have included  several of my favorite design criteria in \pkg{STPGA}. A list of the names, required input parameters and the corresponding formulas are summarized with the Table \ref{tab:STPGAOPTFUNCS} for reference. More explanation about the usage, examples and other details of these criterion can be found in the package help documentations. 
 
When using a design criterion that uses the design matrix of the target individuals along with the candidates, the ''GenAlgForSubsetSelection'' function uses the individuals listed in ''Test'' extract the design matrix of these individuals from the design matrix of all individuals \textbf{P}. If you use a similar criterion with the ''GenAlgForSubsetSelection'' function 'the design of the target set is implicitly assigned as the rows in \textbf{P} not in ''Train''.

Many modern statistical learning problems involve the analysis of high dimensional data. For example, in genomic prediction problems phenotypes are regressed on large numbers of genome-wide markers. \pkg{STPGA} was initially prepared for working with high dimensional data related to such whole-genome-regression  \citep{Meuwissen:2001, akdemir2015locally, daetwyler2013genomic} and association \citep{risch1996future, burton2007genome,rietveld2013gwas, tian2011genome} approaches that are becoming increasingly popular for the analysis and prediction of complex traits in plants  \citep[e.g.][]{Crossa:2010}, animals \citep[e.g.][]{VanRaden:2009, Hayes:2009} and humans \citep[e.g.][]{Yang:2010,makowsky2011beyond}. The design criteria in \pkg{STPGA} and their use were motivated by the practical problem of selecting the best genotypes for a phenotypic experiment so that the inferences made based on the data obtained by the experiment are optimally informative for genomic prediction and association problems \citep{isidro2015training, akdemir2015optimization, crossa2016genomic}.   These high dimensional design problems pose additional computational challenges and the selection and use of the design criteria has a big influence on computational requirements.  I recommended the use of dimension reduction techniques, whether they are supervised, unsupervised  or based on algebraic manipulation, such as the use of dimension reduction methods like principle components analysis or variable clustering, use of methods based on similarity or distance matrices, before running \pkg{STPGA}. Several design optimization criteria in \pkg{STPGA} are equivalent and will produce the same or similar designs. However, calculation of one might be easier than the other based on the relative number of rows or columns of the data matrix.   

STPGA software is written purely in \proglang{R}, however the computationally demanding criteria can be programmed by the user in \proglang{C}, \proglang{C++} or \proglang{Fortran}.  \pkg{STPGA} also benefits from multi-thread computing. The computational performance of the algorithm can be greatly improved if \proglang{R} is linked against a tuned \proglang{BLAS} implementation with multi-thread support, for example \proglang{OpenBLAS}, \proglang{ATLAS}, \proglang{Intel mkl}, etc.

\section{Illustrations}
\label{subsec:illustrations}

In this section, I am going to illustrate use of \pkg{STPGA}. The first group of examples are related to selection of training populations. The second part, includes examples of other subset selection problems.

\subsection{STPGA for selection of training populations}\label{sec:crit}

Experiments provide useful information to scientists as long as they are properly designed and analysed. The history of the theoretical work on the design problem goes way back. Fisher \citep{fisher1992arrangement, fisher1960design} gives a mathematical treatment of the determination of designs for some models. The first extended presentation of the ideas of optimum experimental designs appear in \citep{kiefer1959optimum} and \citep{yates1935complex}.  A brief history of statistical work on optimum experimental design is given by Wynn (wynn1984jack) and the subject continues to develop, recently at an increasing rate.  Box et al. \citep{box1978statistics}, Box and Draper \citep{draper1996overview, atkinson1992optimum, pukelsheim2006optimal} are a few of the authoritative texts on the subject. The alphabetical naming of designs is due to \citep{kiefer1985jack}. For a detailed discussion of standard criteria reference is made to these references.

In this paper we focus only on a the narrow design problem of selecting an optimal set of $n$ design points, $X_{Train}=\{\bx_{1},\ldots, \bx_n\},$ from a set of candidate design points $X_{C}=\{\bx_1,\ldots, \bx_N\}.$ The design defined by these $n$ points, can be viewed as a measure on the candidate set $X_{C}=\{\bx_1,\ldots, \bx_N\}.$ Let $\zeta$ be a probability measure on $X_{C}$ such that 
*  $\zeta(x_i) = 0$ if  $x_i\notin X_{Train}$, and
*  $\zeta(x_i)= 1/n$ if  $x_i\in X_{Train}.$

When dealing with problems of supervised learning where the resulting model of the experiment will be used to make inferences about a known set of individuals, we can distinguish between the candidate set and a target set $X_{Target}=\{\bx_{1},\ldots, \bx_{t}\}:$ $X_{Target}$ describes the focused design region for which predictions about the dependent variables based on the models trained on $X_{Train}$  are required.  Let's assume that $1 \leq n \leq N$ and $1 \leq t$ are fixed integers, $x_i$ are $p$-vectors and we denote the matrix form of $X_{Train}$ as $X,$ this is called the design matrix; and $X_{Target}$ as $X^*$, this is called the design matrix for the target space.

The first component of a design optimization problem is the objective function. For example the objective function might be chosen as theoretically or numerically obtained sampling variance of a prespecified estimator of a population quantity of interest. The second component of an optimization problem is the set of decision variables and the constraints on the values of these variables. Once the objective function and the set of constraints are known the next step is to use a method to look for solutions that optimize the objective function and also adhering to the constraints.

Parametric design criteria usually depend on a function of the information matrix for the model parameters that gives some indication about the sampling variance and covariance of the estimated parameters.  Let $I_{\btheta} (\zeta)$ denote the information matrix of the parameters $\btheta$ for a given design $\zeta.$ In order to be able to achieve a criteria that orders designs with respect to their information matrices, usually, a scalar function of the information matrix is used. These designs criteria have alphabetical names, the designs obtained by optimizing these criteria are referred to as A-, D, E-, G-, etc,... optimal designs. The list of design criteria that are implemented in \pkg{STPGA} are described by Table \ref{tab:STPGAOPTFUNCS} with references to the equations in this manuscript from which these were inspired.  

Many practical and theoretical problems in science treat relationships of the type $y = g(\bx, \btheta),$ where the response, $y,$ is thought of as a particular value of a real-valued model function or response function, $g,$ evaluated at the pair of arguments $(x, \btheta).$  The parameter value  $\btheta,$ unknown to the experimenter, is assumed to lie in a parameter domain $\Theta.$ This is called the regression of $y$ on $x.$ The \pkg{STPGA} is not confined to regression, but we use regression analysis to do most of the explaining and demonstrations. 

\subsubsection{Linear models}

The choice of the function $g$ is central to the regression model building process. One of the simplest regression models is the linear model. Let $X_{n\times p}$ be the design matrix, $\bbeta_{p\times 1}$ the vector of regression parameters, $\by_{n\times 1}$ the vector of observations, and $\bepsilon_{n\times 1}=(\epsilon_1,\epsilon_2, \ldots,\epsilon_n)'$ our error vector giving \[\by=X\bbeta+\bepsilon.\]

With $I_n$ as the $n \times n$ identity matrix, the model is represented by the expectation vector and covariance matrix of $\by,$ \[E(\by) = X\bbeta, cov(\by)=\sigma^2 I_n,\] and is termed the classical linear model with moment assumptions. We assume $\epsilon_i,$ $i = 1, 2,\ldots, n$ will be iid with mean zero and $cov(\bepsilon) = \sigma^2 I_n.$ Under the additional normality assumption we write $\by\sim N(\bzero, \sigma^2 I_n).$

We now concentrate on determining the optimal estimator for $\bc' \bbeta$ in the linear regression model.  If $X$ is not of full rank, it is not possible to estimate $\bbeta$ uniquely. However, $X\bbeta$ is uniquely estimable, and so is $\bc'X\bbeta$ for any conformable vector $\bc$ that is in the row space of $X.$ If estimability holds then the Gauss-Markov Theorem determines the optimal estimator for $\bc'\bbeta$ to be $\bc'(X'X)^{-}X'\by,$ where $A^{-}$ denotes any generalized inverse of $A$ that satisfies $A=AA^{-}A.$  The variance of this estimator depends only on the matrix $X'X,$  \[var_{\bbeta,\sigma^2}(\bc'(X'X)^{-}X'Y) = (\sigma^2)\bc'(X'X)^{-}X'X(X'X)^{-}\bc.\] Up to the common factor $\sigma^2/n,$ the optimal estimator has variance $\bc'(X'X)^{-}X'X(X'X)^{-}\bc.$ 

Assuming estimability, the optimal estimator for the linear function of the coefficients $\gamma = C\bbeta$ is also given by the Gauss-Markov Theorem: $\hat{\gamma} = C(X'X)^{-}X'Y.$ The covariance matrix of the estimator $\hat{\gamma}$ is  $C(X'X)^{-}X'X(X'X)^{-} C'.$ The covariance matrix becomes invertible provided the coefficient matrix $C$ has full row rank. 

A closely related task is that of prediction. Suppose we wish to predict additional responses $\by^*= X^*\bbeta + \bepsilon^*.$ If we take the random vector $\hat{\gamma}$ from above as a predictor for the random vector $\by^*,$  to obtain precise estimators for $X^*\bbeta,$ we would like to choose the design so as to  maximize the relevant information matrix (minimize the covariance matrix). For example, $G$-optimal designs are obtained by minimizing the maximum variance of the predicted values, i.e., the maximum entry in the diagonal of the matrix $X(X'X)^{-}X'.$ 

\subsubsection{Ridge Regression}

Note the following: If A is a symmetric matrix, then the limit \[
 \lim_{\lambda \to 0}(A+\lambda^2 I)^{-1}
\]
is a generalized inverse of A, and also \[\lim_{\lambda \to 0}(A+\lambda^2 I)^{-1}A\lim_{\lambda \to 0}(A+\lambda^2 I)^{-1}=\lim_{\lambda \to 0}(A+\lambda^2 I)^{-1}.\] This means, for small $\lambda>0,$   $\hat{\bgamma} \approx C(X'X+\lambda I)^{-1}X'Y.$ 

This estimator is called the ridge estimator, the coefficients have covariance matrix  approximately proportional to \begin{equation}\label{eq:ridge1}C(X'X +\lambda I)^{-1}C'.\end{equation} 

Furthermore, prediction error variance for estimating the $CX^{*}\bbeta$ with ridge regression is approximately proportional to  \begin{equation}\label{eq:ridge2}CX^{*}(X'X +\lambda I)^{-1}X'^{*}C'.\end{equation}  Ridge estimators have smaller variance than BLUE's  but they are biased since the estimators are ''shrunk'' towards zero. 

Ridge estimators are especially useful when $X'X$ is singular. In some cases, the ridge estimation is only applied to a subset of the explanatory variables in $X,$ for example it is customary to not shrink the mean term. 

Splitting the design matrix $X$ as $X=(X_F, X_R),$ where $X_F$ contains the effects modeled without ridge penalty and $X_R$ contains the terms modeled with ridge penalty, a design criterion concerning the estimation of shrunk coefficients can be written as \[C(X'M^{-1}X +\lambda^2 I)^{-1}C'\] with $M=I-X_F(X'_FX_F)^{-}X'_F.$

\subsubsection{RKHS}
Using the matrix identity $(P^{-1} + B'R^{-1}B)^{-1}B' R^{-1} = PB' (BPB' + R)^{-1},$ we can write  $(X'X+\lambda I )^{-1}X' = X'(XX' +\lambda I )^{-1}.$ The ridge regression solution for $\bgamma$ can then be written as follows:\[\hat{\bgamma}=C\hat{\beta}\approx C(X'X+\lambda I)^{-1}X'\by=CX'(XX'+\lambda I)^{-1}\by.\] The important message here is that we only need access partitions of the matrix \[K_{C,X}(\zeta)= \begin{bmatrix}
 XX' & XC' \\
CX' & CC'
\end{bmatrix}= \begin{bmatrix}
 K_{11} & K_{12} \\
K'_{12} & K_{22}
\end{bmatrix}\]  since $\hat{\gamma}=CX'(XX'+\lambda I)^{-1}\by=K_{21}(K_{11}+\lambda I)^{-1}\by.$ Using
\[(X'X+\lambda \mathbf{I})^{-1} =\frac{1}{\lambda}(\mathbf{I}-X'(XX'+\lambda \mathbf{I})^{-1}X\] we have \begin{equation*} 
\begin{split}
 C(XX+\lambda \mathbf{I})^{-1} C' & =C(\lambda(\frac{X'X}{\lambda}+\mathbf{I}))^{-1}C'\\
 & =\frac{1}{\lambda}C(\mathbf{I}-X'(XX'+\lambda \mathbf{I})^{-1}X) C'\\
  & = \frac{1}{\lambda}\left[CC'\right]-\frac{1}{\lambda}[CX'(XX' +\lambda \mathbf{I})^{-1} XC'] \\
  & \propto \mathbf{K}_{22} -\mathbf{K}_{21}(\mathbf{K}_{11}+m\lambda \mathbf{I})^{-1}\mathbf{K}'_{21}.
\end{split}
\end{equation*}
The variance covariance matrix $\hat{\bgamma}$ for proportional to \begin{equation}\label{eq:rkhs1}(\mathbf{K}_{22}-\mathbf{K}_{21}(\mathbf{K}_{11}+\lambda I)^{-1}\mathbf{K}_{12}).\end{equation} 

Reproducing Kernel Hilbert Spaces (RKHS) regression methods  replace the inner products by kernels, it is as if we are performing ridge regression on a transformed data $\phi(x),$ where $\phi$ is a feature map associated to the chosen kernel function and the associated kernel matrix. The resulting predictor is now nonlinear in $x$ and agrees with the predictor derived from the RKHS perspective (Schölkopf and Smola 2002). RKHS regression extends ridge regression allowing a wide variety of kernel matrices, not necessarily additive in the input variables, calculated using a variety of kernel functions. A kernel function, $k(.,.)$ maps a pair of input points $\bx$ and $\bx'$ into real numbers. It is by definition symmetric ($k(\bx,\bx')=k(\bx',\bx)$) and non-negative. Given the inputs for the $n$ individuals we can compute a kernel matrix $K$ whose entries are $K_{ij}=k(\bx_i,\bx_j).$ The common choices for kernel functions are the linear ($k(\bx; \by) = \bx'\by.$), polynomial ($k(\bx; \by) =(\bx'\by+ c)^d$ for $c$ and  $d$ $\in$ $R$), Gaussian kernel functions ($k(\bx; \by) = exp(-h(\bx'-\by)'(\bx'-\by))$ where $h>0.$), though many other options are available \citep{scholkopflearning}.  Reproducing Kernel Hilbert Spaces Regressions (RKHS) have been used for regression \citep[e.g., Smoothing Spline][]{Wahba:1990}, spatial smoothing \citep[e.g., Kriging][]{Cressie:1988} and classification problems \citep[e.g., Support Vector Machine, ][]{Vapnik:1998}. \citet{Gianola:2006}, proposed to use this approach for genomic prediction and, since then several follow-up articles with focus on the application of these methods to various genome-wide regression problems have also been published \citep{Gonzalez:2008}.

\subsubsection{Gaussian Linear Mixed Models}
The linear mixed model methodology was first developed within the context of animal genetics and breeding research by \cite{Henderson:1975, kempthorne1957introduction, henderson1959estimation}, many important statistical models can be expressed as mixed effects models and it is the most widely used model in prediction of quantitative traits, and genome-wide association studies.  In studies on linear mixed models it is usual to consider the estimation of the fixed effects $\bbeta$ and the variance components, and also the prediction of the random effects $\bu.$ For a given data vector $\by,$ the vector of random effects $\bu$ is a realization of random variables which are  observed and these effects must therefore necessarily be predicted from the data \citep{henderson1953estimation}. 

In the linear mixed-effects model, the observations are assumed to result from a hierarchical linear model: \[y = W\bbeta + Z \bu + \bepsilon;\] and $\bepsilon\sim N(0,R)$ is independent of $\bu \sim N(0;G).$  These assumptions imply $E (y|W;Z) = W\bbeta,$ $y \sim N(W\bbeta; ZGZ' + R) = N(W\bbeta; V ).$

The similarity of the mixed models and RKHS regression models has been stressed many times. However, mixed modeling approach provides a formalized approach since he inferences are based on a probabilistic model, and therefore, allows legitimate inferences about the parameters and predictions.

Henderson et al. show that maximizing the joint density of $\by$ and $\bu$ yields the MLEs of the parameters $\bbeta$ and EBLUPs (estimated BLUPs) $\hat{\bu}$ that solve: $W'R^{-1}W \hat{\bbeta} + W'R^{-1}Z\hat{\bu}= W'R^{-1}y$ and $Z'R^{-1}W \hat{\bbeta}+ Z'R^{-1}Z\hat{\bu} + G^{-1}\hat{\bu} = Z'R^{-1}y,$ this leads to the Henderson's mixed model equations:

Henderson's mixed-model equations can be used to estimate the standard errors of the fixed and random effects. For a given design, the inverse of the coefficient matrix is written as 
 \[\begin{bmatrix}
 W' R^{-1}W & W'R^{-1}Z \\
Z' R^{-1}W & Z'R^{-1}Z + G^{-1}
\end{bmatrix}^{-1}
=
\begin{bmatrix}
    H_{11} & H_{12}\\
    H'_{12} & H_{22}
\end{bmatrix}
 \] where $H_{11},$ $H_{12},$ and $H_{22}$ are, respectively, $p\times p,$ $p\times q,$ and $q\times q$ sub-matrices. Note that referring to the coefficient matrix is an abuse of notation since the parameters of the mixed effects model does not include the vector $\bu.$  Using this notation, the sampling covariance matrix for the BLUE (best linear unbiased estimator) of $\bbeta$ is given by $\sigma(\bbeta ) = H_{11}=(W'V^{-1}W)^{-}$ that the sampling covariance matrix of the prediction errors $(\hat{u}-u)$ is given by \begin{equation}\label{eq:mixed1} cov(\hat{\bu}-\bu ) = H_{22}=G- GZ'PZG\end{equation} for $P = V^{-1} - V^{-1}W(W'V^{-1}W)^-W'V^{-1}$ and that the sampling covariance of estimated effects and prediction errors is given by $\sigma(\bbeta, \hat{\bu}-\bu ) = H_{12}=-(W'V^{-1}W)^{-}W'V^{-1}ZG$
(We consider $\hat{\bu}-\bu$ rather than $\hat{\bu}$ as the latter includes variance from both the prediction error and the random effects $\bu$ themselves.). The standard errors of the fixed and random effects are obtained, respectively, as the square roots of the diagonal elements of $H_{11}$ and $H_{22.}$  In addition, using the above definitions, $cov(\bu |\by)=G- GZ'V^{-1}ZG=(Z'R^{-1}Z + G^{-1})^{-1}.$

Optimal design of experiments with mixed models involve determination of the design matrices $W$ and $Z;$ however, in many applications, estimates of only one of $\bbeta$ or $\bu$ is needed. For example, design criterion is obtained by considering the variance-covariance matrix of $C'(\hat{\bu}-\bu)$ given by $C'H_{22}C$ is named the prediction error variance. A more recent design criterion is the generalized coefficient of determination \citep{laloe1993precision,Laloe2003241,rincent2012maximizing} for the random terms $\bc'_i(\hat{\bu}-\bu),$ $i =1,\dots, l:$
\[\sum_{i=1}^l \frac{\bc'_iH_{22}\bc_i}{\bc'_iG\bc_i}\] for a set of contrasts $\bc_i.$

If the mixed model is simplified such that $\bepsilon \sim N(0,R=\sigma^2_{\bepsilon} I)$ and  $\bu \sim N(0;G=\sigma^2_{\bu} A),$ and the rows of $C$ have zeros corresponding to fixed effects, the formula for prediction error variance becomes:  \[C(Z'MZ+\lambda A^{-1})^{-1}C'\] and the corresponding formula for coefficient of determination becomes: \begin{equation}\label{eq:mixed2}\sum_{i=1}^l \frac{\bc'_i(A-\lambda(Z'MZ+\lambda A^{-1})^{-1})\bc_i}{\bc'_iA\bc_i}, \end{equation} where $\lambda=\sigma^2_{\bepsilon}/\sigma^2_{\bu}.$ Furthermore, when we assume  $\bu\sim N(\bzero, G=\sigma^2_{\bu}I)$, then the above formulas simplify further to  \[C(Z'MZ+\lambda I)^{-1}C'\] and \[\sum_{i=1}^l (1-\frac{\lambda\bc'_i (Z'MZ+\lambda I)^{-1}\bc_i}{\bc'_i\bc_i}).\] Here, $M=I-W(W'W)^{-}W'$ is  is a projection matrix orthogonal to the vector subspace spanned by the columns of $W,$ so that $MW=0.$

\subsubsection{Some generalizations and extensions of parametric design criteria}

A generalization of the $D$-, $A$-, $G$- optimal criteria is provided by the Keifer's $\phi_p$ criteria: given by \[\phi_p(\zeta)=(trace(I_{\btheta} (\zeta))^{-p})^{1/p},\] where $-1\leq p<\infty.$ $p=0, 1, \infty$ gives the $D$-, $A$-, $E$- optimal criteria correspondingly.

Another extension of $D$-optimality deals with minimizing\[\sum_{j=1}^h \alpha_j  log|A_jI(\zeta)^{-}A'_j|.\] The criterion permits designs for $h$ different models which may be fitted to the data, for the $j$th of which the information matrix is $I_j(\zeta).$ The matrix $A_j$ defines $S_j$ linear combinations of the $p_j$ parameters in model $j$ which are of experimental importance and the non-negative weights $\alpha_j$ express the relative importance of the different aspects of the design. Examples of compound $D$-optimum designs for linear models are given by \cite{atkinson1992optimum}. 

\cite{pritchard1978prospects} proposed a new criterion alternative to the traditional $D$-optimal design, which has a measure of the overall correlation among the parameters directly as objective function to be minimized i.e. the root square of the individual correlations between pair of parameters:  \[F={\left(\sum_{,i; i\neq j}\frac{corr^2_{ij}}{p^2-p} \right)^{1/2}}.\]

\subsubsection{Non-parametric design criteria}

The design criteria of the previous sections started from a parametric model. There are some optimal design approaches that does not make any parametric assumptions, leading to non-parametric design criteria. Most of these methods are based on a distance matrix.  

A design criteria that aims to achieve a high spread  among its support points within the design region, i.e., make the smallest distance between neighboring points in the design as large as possible is called the maximin-distance criterion. Let $D={d_{ij}}_{i=1,...,N}$ denote the distance matrix among the possible design points. Maximin distance criteria is finding the subset of $n$ points such that  \[\phi_1(\zeta)=min(d_{ij}), i\neq j\] to be maximized among these $n$ points. Another possibility is to pick the $n$ design points so that the the maximum distance from all the points in a target set of points $X^*$ is as small as possible. These designs are called space filling designs, some performance bench-marking for various space-filling designs can be found in \citep{pronzato2008optimal} and \citep{pronzato2012design}.

\textbf{Example 1:} In this example, we want to find the best D-optimal 13 point design for a second order regression model over a grid design region defined by two variables both with possible values in the set ${-2, -1, 0, 1, 2 }.$ Naming these variables as $x_1$ and $x_2$ and the generic response as $y,$ we can write this model as \[y=\beta_0+\beta_1 x_1+\beta_2 x_2+ \beta_{11} x_1^2+\beta_{22} x_2^2 + \beta_{12} x_1x_2+\epsilon.\] First, we crate the design matrix for the design space. 

\begin{GrayBox}
 \scriptsize
\textbf{Box 1: Creating the design matrix for grid, selecting ''best'' subset by enumeration and using STPGA}
\begin{Schunk}
\begin{Sinput}
> library(STPGA)
> set.seed(1234)
> X<-matrix(0,nrow=5^2,ncol=5)
> ij=0
> for (i in -2:2){
+   for (j in -2:2){
+     ij=ij+1
+     X[ij,]<-c(i,j, i^2,j^2, i*j)
+   }
+ }
> X<-cbind(1,X)
> rownames(X)<-paste("x",1:5^2, sep="")
> #lisofallsubsetsofsize13<-combn(rownames(X), 13)
> #dim(lisofallsubsetsofsize13)
> ########complete enumeration of
> ########(5^2 choose 13)=5200300 possibilities
> #I have done this once, you dont need to do it.
> #DOPTVALS<-apply(lisofallsubsetsofsize13, 2, 
> #        function(x){DOPT(Train=x, Test=NULL, P=X, lambda = 1e-09, C=NULL)})
> BESTSOL<-c("x1","x2","x3","x5","x6","x10",
+            "x11","x13","x15","x21","x22","x24","x25")
> #BESTSOL<-lisofallsubsetsofsize13[,which.min(DOPTVALS)]
> #best solution is not unique for this problem
> mindoptvals<--21.3096195830339709687
> #mindoptvals<-min(DOPTVALS)
> ListTrain1<-GenAlgForSubsetSelectionNoTest(P=X,ntoselect=13, InitPop=NULL,
+             npop=200, nelite=5, mutprob=.5, mutintensity = 1,
+             niterations=200,minitbefstop=50, tabu=FALSE,
+             tabumemsize = 0,plotiters=FALSE,
+             lambda=1e-9,errorstat="DOPT", mc.cores=4)
> length(intersect(ListTrain1$`Solution with rank 1`,BESTSOL))
\end{Sinput}
\begin{Soutput}
[1] 12
\end{Soutput}
\begin{Sinput}
> mindoptvals==min(ListTrain1$`Best criterion values over iterarions`)
\end{Sinput}
\begin{Soutput}
[1] TRUE
\end{Soutput}
\end{Schunk}
\end{GrayBox}

\begin{GrayBox}
 \scriptsize
\textbf{Box 2: Plotting the results on the grid}
\begin{Schunk}
\begin{Sinput}
> par(mfrow=c(1,2))
> labelling1<-rownames(X)%in%ListTrain1$`Solution with rank 1`+1
> plot(X[,2], X[,3], col=labelling1,
+      pch=2*labelling1,cex=2*(labelling1-1),
+      xlab="", ylab="", main="STPGA solution",
+      cex.main=.7,xaxt='n',yaxt='n')
> text(x=X[,2]-.1, y = X[,3]-.1, labels = rownames(X), cex=.5)
> for (i in -2:2){
+   abline(v=i, lty=2)
+   abline(h=i,lty=2)
+ }
> labelling2<-rownames(X)%in%BESTSOL+1
> plot(X[,2], X[,3], col=labelling2,
+      pch=2*labelling2,cex=2*(labelling2-1),
+      xlab="", ylab="", main="Best solution", 
+      cex.main=.7,xaxt='n',yaxt='n')
> text(x=X[,2]-.1, y = X[,3]-.1, labels = rownames(X), cex=.5)
> for (i in -2:2){
+   abline(v=i, lty=2)
+   abline(h=i,lty=2)
+ }
> par(mfrow=c(1,1))
\end{Sinput}
\end{Schunk}
\includegraphics{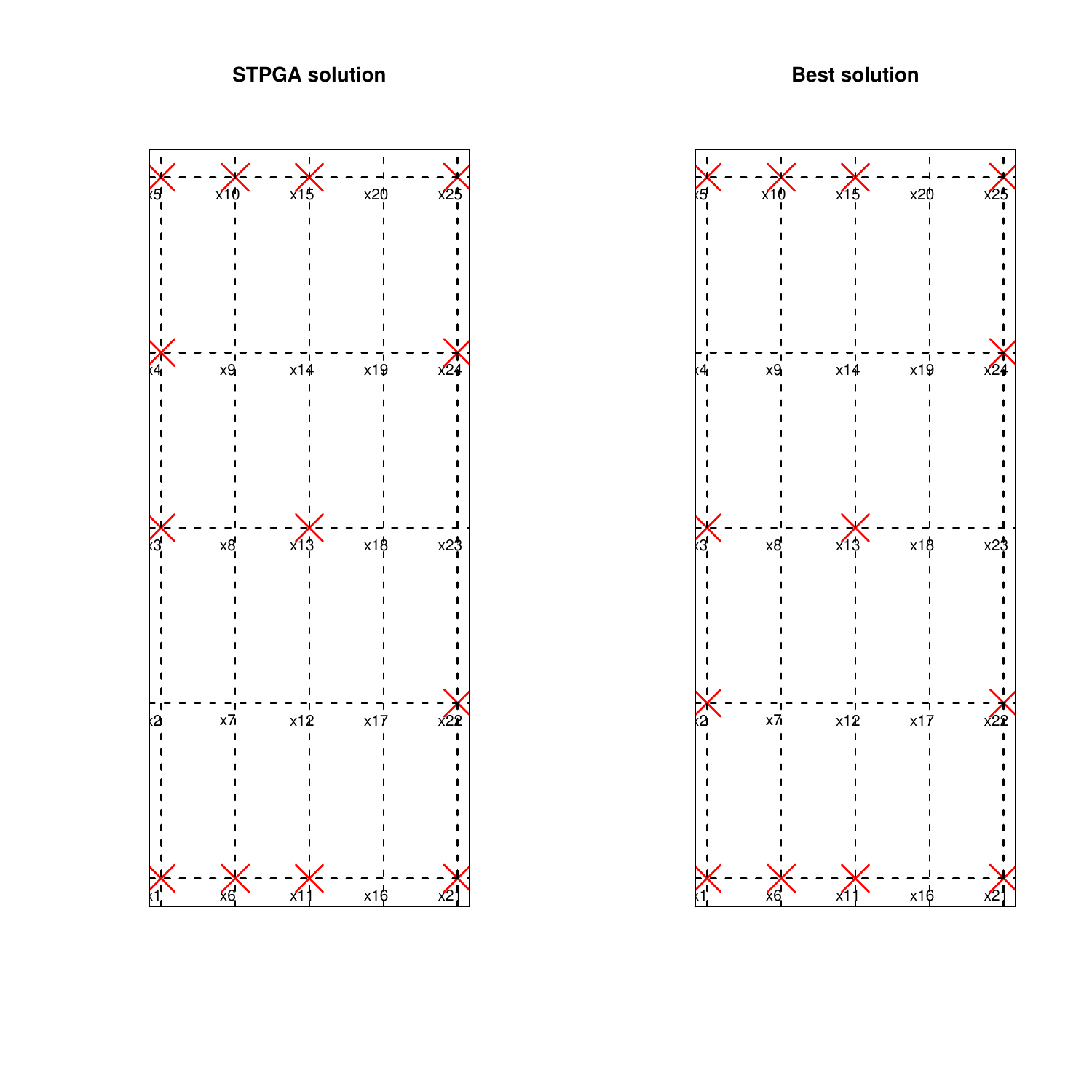}
\end{GrayBox}

\textbf{Example 2:} In statistical genetics, an important task involves building predictive models of the genotype-phenotype relationship to be able to make genomic predictions and also to attribute a proportion of the total phenotypic variance to locations on the genome (genomewide association studies (GWAS)).  If the genotypic information for the candidates (and the target) are available, phenotypic experiments can be executed for a subset of these individuals that is optimal according to a design criteria that only uses the available genotypic information.

\pkg{STPGA} package comes with a genomic data set called \textbf{WheatData} involving phenotypes and markers for 200 elite wheat lines selected at random from a larger genetic pool. Data was downloaded from the website \url{triticeaetoolbox.org}. The 4670 markers available for these 200 genotypes were preprocessed for missingness and minor allele frequencies, coded numerically as 0, 1, and 2; the relationship genomic relationship  matrix was calculated according to the formula in Van Raden \citep{VanRaden:2008}: $\frac{M_cM'_c}{k}$ where $k=\sum_{j=1}^{m}2p_j(1-p_j)$ is twice the sum of heterozygosities  of the markers and $M_c$ the allele counts matrix M centered by the mean frequencies of alleles. The genotypic values for plant heights were predicted using a mixed effects model that is fitted to a multi-environmental trial involving these genotypes and the corresponding phenotypic observations.

As long as the model assumptions are correct for the data and a suitable criterion is employed, the prediction accuracies of models built on optimal sets are expected to be better than average prediction accuracies of models based on a random set of the same size. In addition, if a target set is specified, further gains might be achieved using this knowledge. GWAS results based on a genetic information based optimal design is expected to improve the association results compared to models built on phenotypic experiments performed on a random set of the same size in a similar way. Obtaining genotypic information is becoming cheaper by the day, however the high costs and challenges related to phenotypic experiments persist. To see this is the case consider what might be involved in a longitudinal study on human subjects.

Optimal design of phenotypic experiments based on prior genotypic information can lead to high information value at low costs.  To illustrate these points, lets use the wheat data set first in prediction and association settings. We begin by loading the data and doing some preprocessing necessary to run the experiment:

\begin{GrayBox}
 \scriptsize
\textbf{Box 3: Loading and preprocessing the wheat data set included in STPGA}
\begin{Schunk}
\begin{Sinput}
> data(WheatData)
> svdWheat<-svd(Wheat.K, nu=50, nv=50)
> PC50WHeat<-Wheat.K%*%svdWheat$v
> rownames(PC50WHeat)<-rownames(Wheat.K)
> DistWheat<-dist(PC50WHeat)
> TreeWheat<-cutree(hclust(DistWheat), k=4)
\end{Sinput}
\end{Schunk}
\end{GrayBox}

''TreeWheat'' partitions the data into four sets, lets observe this grouping using a plot of first two principal components.

\begin{GrayBox}
 \scriptsize
\textbf{Box 4: Plotting the observations in wheat data using first two principal components}
\begin{Schunk}
\begin{Sinput}
> plot(PC50WHeat[,1],PC50WHeat[,2], col=TreeWheat,
+      pch=as.character(TreeWheat), xlab="pc1", ylab="pc2")
\end{Sinput}
\end{Schunk}
\includegraphics{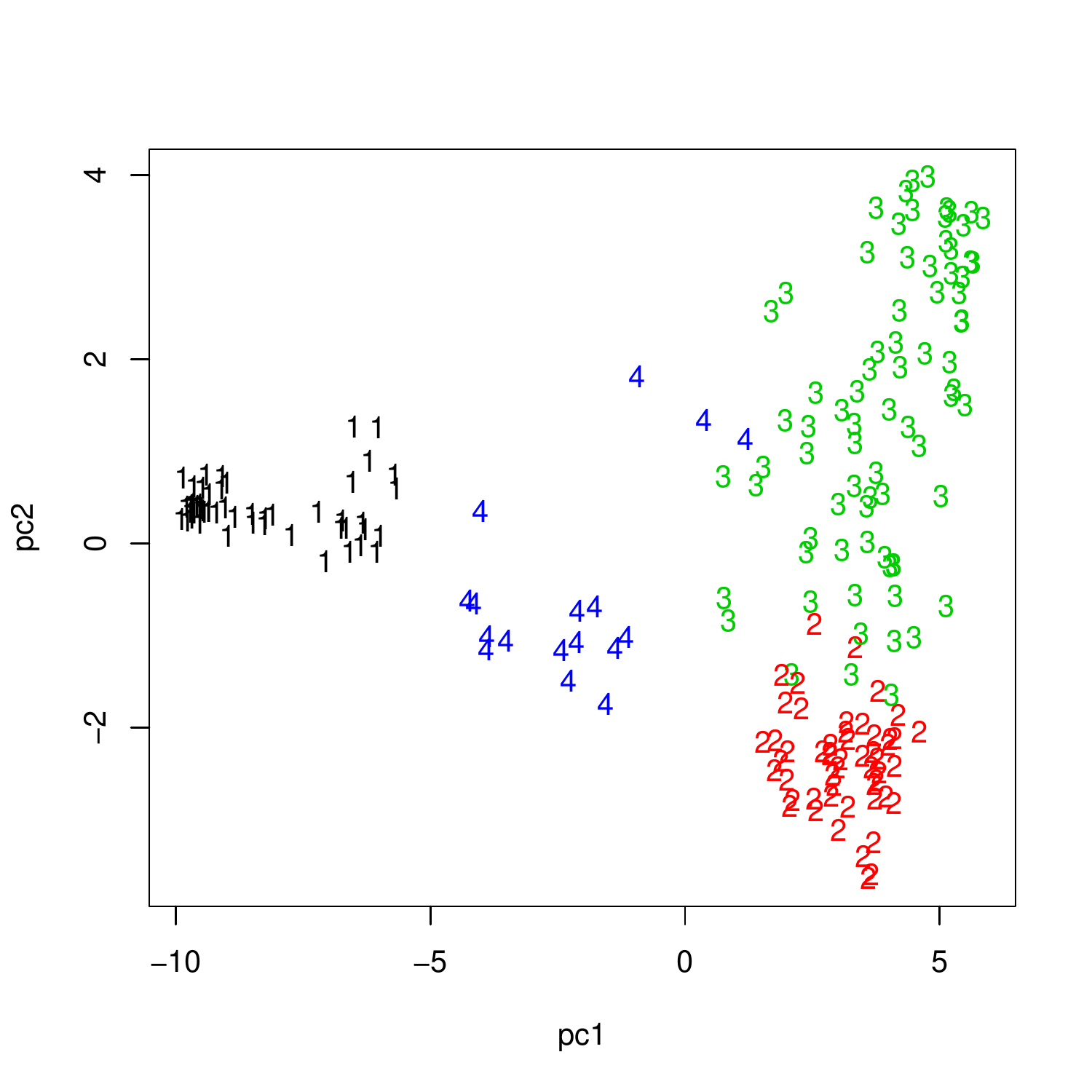}
\end{GrayBox}

We will consider a scenario where the final aim is to accurately predict the genotypic values in group 2 and we want to establish this by conducting a phenotypic experiment on 50 genotypes selected from the remaining groups. 

\begin{GrayBox}
 \scriptsize
\textbf{Box 5: Splitting the data into Candidates and Test}
\begin{Schunk}
\begin{Sinput}
> Test<-rownames(PC50WHeat)[TreeWheat==2]
> length(Test)
\end{Sinput}
\begin{Soutput}
[1] 53
\end{Soutput}
\begin{Sinput}
> Candidates<-setdiff(rownames(PC50WHeat), Test)
> Ztrainfull<-as.matrix(model.matrix(~-1+factor(Candidates,
+                     levels=rownames(Wheat.M))))
> deptrainfull<-Wheat.Y[Wheat.Y$id%in%Candidates,]
> dim(deptrainfull)
\end{Sinput}
\begin{Soutput}
[1] 147   2
\end{Soutput}
\end{Schunk}
\end{GrayBox}

Once the data is ready, it is easy to call STPGA with the default options. Note that there are two options, we can use the information about the target genotypes or we could ignore this. I will do both:

\begin{GrayBox}
 \scriptsize
\textbf{Box 6: Optimization of training populations with STPGA}
\begin{Schunk}
\begin{Sinput}
> Train1<-GenAlgForSubsetSelection(P=PC50WHeat,Candidates=Candidates, 
+         Test=Test, ntoselect=50, mc.cores=4)
> Train2<-GenAlgForSubsetSelectionNoTest(P=PC50WHeat,
+         ntoselect=50, mc.cores=4)
\end{Sinput}
\end{Schunk}
\end{GrayBox}

However, it is important to be able to specify GA parameters:

\begin{GrayBox}
 \scriptsize
\textbf{Box 7:  Optimization of training populations with STPGA, specifying algorithm parameters}
\begin{Schunk}
\begin{Sinput}
> Train3<-GenAlgForSubsetSelection(P=PC50WHeat,Candidates=Candidates, 
+         Test=Test,ntoselect=50, 
+         InitPop=NULL,npop=200, 
+         nelite=10, mutprob=.5, mutintensity = 1,niterations=200,
+         minitbefstop=50, tabumemsize = 1,plotiters=FALSE,
+         lambda=1e-9,errorstat="PEVMEAN", mc.cores=4)
> Train4<-GenAlgForSubsetSelectionNoTest(
+         P=PC50WHeat[rownames(PC50WHeat)%in%Candidates,],ntoselect=50, 
+         InitPop=NULL,npop=200, 
+         nelite=10, mutprob=.5, mutintensity = 1,niterations=200,
+         minitbefstop=50, tabumemsize = 1,plotiters=FALSE,
+         lambda=1e-9,errorstat="PEVMEAN", mc.cores=4)
\end{Sinput}
\end{Schunk}
\end{GrayBox}

We finally want to compare the prediction accuracy for predicting the target data compared to the average accuracy that would be obtained using a sample size of same size. I will only use ''Train3'' and ''Train4'' below, we will also need the the package \pkg{EMMREML} \citep{akdemir2015EMMREML} for fitting the mixed effects model:

\begin{GrayBox}
 \scriptsize
\textbf{Box 8: Building models based on optimal samples, getting predictions}
\begin{Schunk}
\begin{Sinput}
> require("EMMREML")
> deptest<-Wheat.Y[Wheat.Y$id%in%Test,]
> Ztest<-model.matrix(~-1+deptest$id)
> ##predictions by optimized sample
> deptrainopt3<-Wheat.Y[(Wheat.Y$id%in%Train3$`Solution with rank 1`),]
> Ztrain3<-model.matrix(~-1+deptrainopt3$id)
> modelopt3<-emmreml(y=deptrainopt3$plant.height,
+                    X=matrix(1, nrow=nrow(deptrainopt3), ncol=1),
+                    Z=Ztrain3, K=Wheat.K)
> predictopt3<-Ztest%*%modelopt3$uhat
> #####
> deptrainopt4<-Wheat.Y[(Wheat.Y$id%in%Train4$`Solution with rank 1`),]
> Ztrain4<-model.matrix(~-1+deptrainopt4$id)
> modelopt4<-emmreml(y=deptrainopt4$plant.height,
+                    X=matrix(1, nrow=nrow(deptrainopt4), ncol=1),
+                    Z=Ztrain4, K=Wheat.K)
> predictopt4<-Ztest%*%modelopt4$uhat
\end{Sinput}
\end{Schunk}
\end{GrayBox}

We will repeat estimation with random sample $300$ times to obtain mean performance:

\begin{GrayBox}
 \scriptsize
\textbf{Box 9: Estimating the accuracy of a random sample of the same size}
\begin{Schunk}
\begin{Sinput}
> corvecrs<-c()
> for (rep in 1:300){
+   rs<-sample(Candidates, 50)
+   
+   deptrainrs<-Wheat.Y[(Wheat.Y$id%in%rs),]
+   
+   Ztrainrs<-model.matrix(~-1+deptrainrs$id)
+ 
+   modelrs<-emmreml(y=deptrainrs$plant.height,
+                    X=matrix(1, nrow=nrow(deptrainrs), ncol=1), 
+                    Z=Ztrainrs, K=Wheat.K)
+   predictrs<-Ztest%*%modelrs$uhat
+   corvecrs<-c(corvecrs,cor(predictrs, deptest$plant.height))
+ }
\end{Sinput}
\end{Schunk}
\end{GrayBox}

Here are the results:
\begin{GrayBox}
 \scriptsize
\textbf{Box 10: Comparisons of accuracies}
\begin{Schunk}
\begin{Sinput}
> ##average accuracy random sample
> mean(corvecrs)
\end{Sinput}
\begin{Soutput}
[1] 0.303162
\end{Soutput}
\begin{Sinput}
> ##accuracy of Train3$`Solution with rank 1`
> cor(predictopt3, deptest$plant.height)
\end{Sinput}
\begin{Soutput}
          [,1]
[1,] 0.3146401
\end{Soutput}
\begin{Sinput}
> ##accuracy of Train3$`Solution with rank 1`
> cor(predictopt4, deptest$plant.height)
\end{Sinput}
\begin{Soutput}
          [,1]
[1,] 0.3936563
\end{Soutput}
\end{Schunk}
\end{GrayBox}

These results are as expected: Optimally designed phenotypic experiments are more informative, they result in higher accuracies compared to a random sample of the same size. If the researcher also has access to the genotypic information for the individuals in the target set, then this information when properly used might lead to gains in per unit information that will come from a phenotypic experiment.

I also expect that the association (GWAS) results from an optimized sample to be better than a random sample. I can not verify this with a simple example. However, here is a comparison of the marker effects estimated from a full set, compared to a random sample and an optimized sample of the same size.

\begin{GrayBox}
 \scriptsize
\textbf{Box 11: Using STPGA in training population selection for GWA studies}
\begin{Schunk}
\begin{Sinput}
> modelrrblupfull<-emmreml(y=deptrainfull$plant.height,
+                    X=matrix(1, nrow=nrow(deptrainfull), ncol=1),
+                   Z=Ztrainfull%*%Wheat.M, K=diag(ncol(Wheat.M)))
> Trainopt<-GenAlgForSubsetSelectionNoTest(
+         P=PC50WHeat[rownames(PC50WHeat)%in%Candidates,],ntoselect=50, 
+         InitPop=NULL,npop=200, 
+         nelite=10, mutprob=.5, mutintensity = 1,niterations=100,
+         minitbefstop=50, tabumemsize = 1,plotiters=FALSE,
+         lambda=1e-9,errorstat="DOPT", mc.cores=4)
> deptrainopt<-Wheat.Y[(Wheat.Y$id%in%Trainopt$`Solution with rank 1`),]
> Ztrainopt<-model.matrix(~-1+deptrainopt$id)
> modelrrblupopt<-emmreml(y=deptrainopt$plant.height,
+                    X=matrix(1, nrow=nrow(deptrainopt), ncol=1), 
+                   Z=Ztrainopt%*%Wheat.M, K=diag(ncol(Wheat.M)))
> modelrrbluprs<-emmreml(y=deptrainopt4$plant.height,
+                    X=matrix(1, nrow=nrow(deptrainopt4), ncol=1), 
+                   Z=Ztrainrs%*%Wheat.M, K=diag(ncol(Wheat.M)))
> orderfull<-order(abs(modelrrblupfull$uhat), decreasing=T)
> orderopt<-order(abs(modelrrblupopt$uhat), decreasing=T)
> orderrs<-order(abs(modelrrbluprs$uhat), decreasing=T)
> mean(abs(orderrs-orderfull))
\end{Sinput}
\begin{Soutput}
[1] 1580.187
\end{Soutput}
\begin{Sinput}
> mean(abs(orderopt-orderfull))
\end{Sinput}
\begin{Soutput}
[1] 1567.686
\end{Soutput}
\end{Schunk}
\end{GrayBox}

As noted before, the subset selection optimization problem is a combinatorial one. We need to see if the algorithm got close to convergence, we can do this by plotting the criterion values over the iterations, these values are stored in the output of STPGA with the name `Best criterion values over iterations`.

\begin{GrayBox}
 \scriptsize
\textbf{Box 12: Plotting the progress of the optimization}
\begin{Schunk}
\begin{Sinput}
> plot(Train3$`Best criterion values over iterarions`,
+      xlab="iteration", ylab="criterion value")
\end{Sinput}
\end{Schunk}
\includegraphics{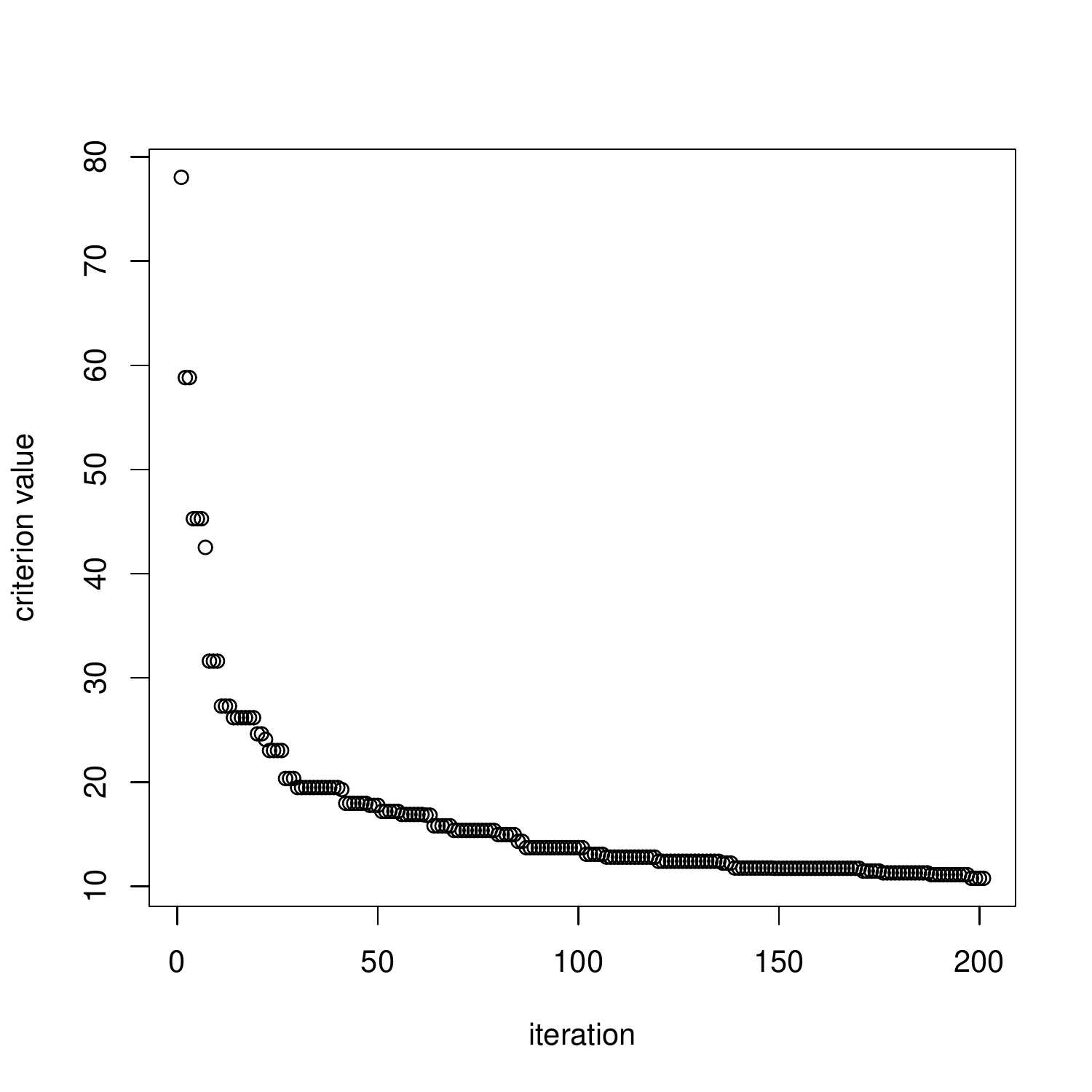}
\end{GrayBox}

The users are recommended to check the convergence of the algorithm by checking the plot, the last iterations should show little or no improvement. If the algorithm has still room to improve the solutions from the last run can be used as new starting points and the iteration can be restarted, perhaps with different settings for the GA parameters. For difficult problems, a good strategy is to run the algorithm several times and select the best solution among these. It is also possible to implement the genetic algorithm in an island model, I have included the code for  a simple island model in \textbf{Appendix}.

\subsection{STPGA in other subset selection problems}
Optimal subset selection algorithm in STPGA can be used with user supplied optimization criterion, and therefore, it has a wide area of application. I am going to try to give a few examples: supervised unsupervised variable selection, Minimize inbreeding while maximizing gain, mixed integer programming, influential observation selection. These examples can easily be extended. 

The following example illustrates how the users can define their own criteria and use it with STPGA for a variable selection problem. It involves aligning kernels to select variables and as far as I know this wasn't done before. 

\textbf{Selecting most representative marker set (markers that represent most of the variability in a given marker data)}

A genetic relationship matrix measures the amount individuals in a certain group are genetically similar. Genetic relationship matrices can be constructed using pedigrees, or using genome-wide markers. In this example, we will try to find a fixed size subset of available genome-wide markers that results in a genetic relationship matrix that is as close to the genetic relationship matrix as possible. These selected markers can be called the \emph{genetic anchor markers}, since they explain most of the properties of the genome-wide relationship matrix. The main question is if there is a subset of markers that can explain a big part of all of the variation captured by all the markers (or even the genome sequence), since this relates to many important genetic concepts like effective population size, effective number of independent chromosome segments, population structure and its effects on predictability within and between sub-populations.  Note that the selection of the anchor markers does not involve any phenotypic observations, therefore this is an unsupervised marker selection approach, similar to some recent approaches expressed in sparse principal components analysis \citep{witten2009penalized} or sparse partial least squares \citep{chun2010sparse}. However, the interpretation of the factors extracted by sparse PCA and sparse PLS might be difficult since these are linear combinations of the original variables. 

The following is a simple function for obtaining a genetic relationship matrix given the matrix of markers (n x m) (n:number of genotypes), (m:number of markers) coded as 0, 1, 2 representing the number of minor alleles. A function to calculate the relationship genomic relationship  matrix according to the formula in Van Raden \citep{VanRaden:2008} is given below:

\begin{GrayBox}
 \scriptsize
\textbf{Box 13: A function to calculate Van Raden's relationship matrix from minor allele frequency scores (markers coded as 0,1 and 2)}
\begin{Schunk}
\begin{Sinput}
> A.mat<-function(M){
+ pks<-colMeans(M)/2
+ W<-scale(M, center=TRUE, scale=FALSE)
+ c<-2*sum(pks*(1-pks))
+ Amat<-tcrossprod(W)/c
+ rownames(Amat)<-colnames(Amat)<-rownames(M)
+ return(Amat)
+ }
\end{Sinput}
\end{Schunk}
\end{GrayBox}

I will only use the lines in the second cluster, the whole data would take too much time to process. 
\begin{GrayBox}
 \scriptsize
\textbf{Box 14: Obtaining a subset of wheat data set for a quick analysis}
\begin{Schunk}
\begin{Sinput}
> #the problem can be soved for all the genotypes to make the 
> #problem computationally easier, we will pick only the 
> #genotypes in forth cluster
> library(Matrix)
> Wheat.M4<-Wheat.M[rownames(Wheat.M)%in%names(TreeWheat[TreeWheat==2]),]
> Wheat.M4<-Matrix(Wheat.M4+1)
> #relationship using all markers
> ##A.mat requires the markers are coded as 0, 1, 2
> Afull<-A.mat(M=Wheat.M4)
\end{Sinput}
\end{Schunk}
\end{GrayBox}

We can see how this optimally selected markers compare with the randomly selected marker sets of the same size.
\begin{GrayBox}
 \scriptsize
\textbf{Box 15: Optimally selected anchor markers versus randomly selected markers}
\begin{Schunk}
\begin{Sinput}
> n<-nrow(Wheat.M4)
> diffvecrs<-c()
> for (i in 1:100){
+ rssmallM<-Wheat.M4[,sample(1:ncol(Wheat.M4),50)]
+ Ars<-A.mat(M=rssmallM)
+ diffvecrs<-c(diffvecrs,mean((c(Afull[lower.tri(Afull, diag=TRUE)])
+                              -c(Ars[lower.tri(Ars, diag=TRUE)]))^2))
+ }
> #User defined criterion
> STPGAUSERDEFFUNC<-function(Train,Test=NULL, P, lambda=1e-6, C=NULL){
+  trsmallM<-t(P[rownames(P)%in%Train,])
+  Atr<-A.mat(M=trsmallM)
+  return(mean((c(Afull[lower.tri(Afull, diag=TRUE)])
+               -c(Atr[lower.tri(Atr, diag=TRUE)]))^2))
+ }
> GAOUT<-GenAlgForSubsetSelectionNoTest(P=t(Wheat.M4),
+         ntoselect=50,npop=300, 
+         nelite=10, mutprob=.5, mutintensity = 1,
+         niterations=400, minitbefstop=50,tabu=FALSE,
+         tabumemsize = 1,plotiters=F,lambda=1e-6,
+         errorstat="STPGAUSERDEFFUNC",mc.cores=4)
> min(GAOUT$`Best criterion values over iterarions`);mean(diffvecrs)
\end{Sinput}
\begin{Soutput}
[1] 0.03310715
\end{Soutput}
\begin{Soutput}
[1] 0.1044086
\end{Soutput}
\begin{Sinput}
> optsmallM<-Wheat.M4[, colnames(Wheat.M4)%in%GAOUT$`Solution with rank 1`]
> optA<-A.mat(optsmallM)
\end{Sinput}
\end{Schunk}
\end{GrayBox}

\begin{GrayBox}
 \scriptsize
\textbf{Box 16: Plotting the results for optimally selected ''anchor markers'' versus randomly selected markers}
\begin{Schunk}
\begin{Sinput}
> layout(matrix(c(1,2,3,4,5,6),2,3, byrow=TRUE), widths=c(2,2,2),
+        heights=c(2,2), respect=TRUE)
> par(mar=c(3,2,2,1))
> # turn off the axes
> image(Ars, axes=FALSE, main="Random Markers")
> image(optA, axes=FALSE, main="Optimal Markers")
> image(Afull, axes=FALSE, main="All Markers")
> image((Ars-Afull)^2, axes=FALSE, 
+       main="Squared errors for random")
> image((optA-Afull)^2, axes=FALSE, 
+       main="Squared errors for optimal")
> par(mfrow=c(1,1))
\end{Sinput}
\end{Schunk}
\includegraphics{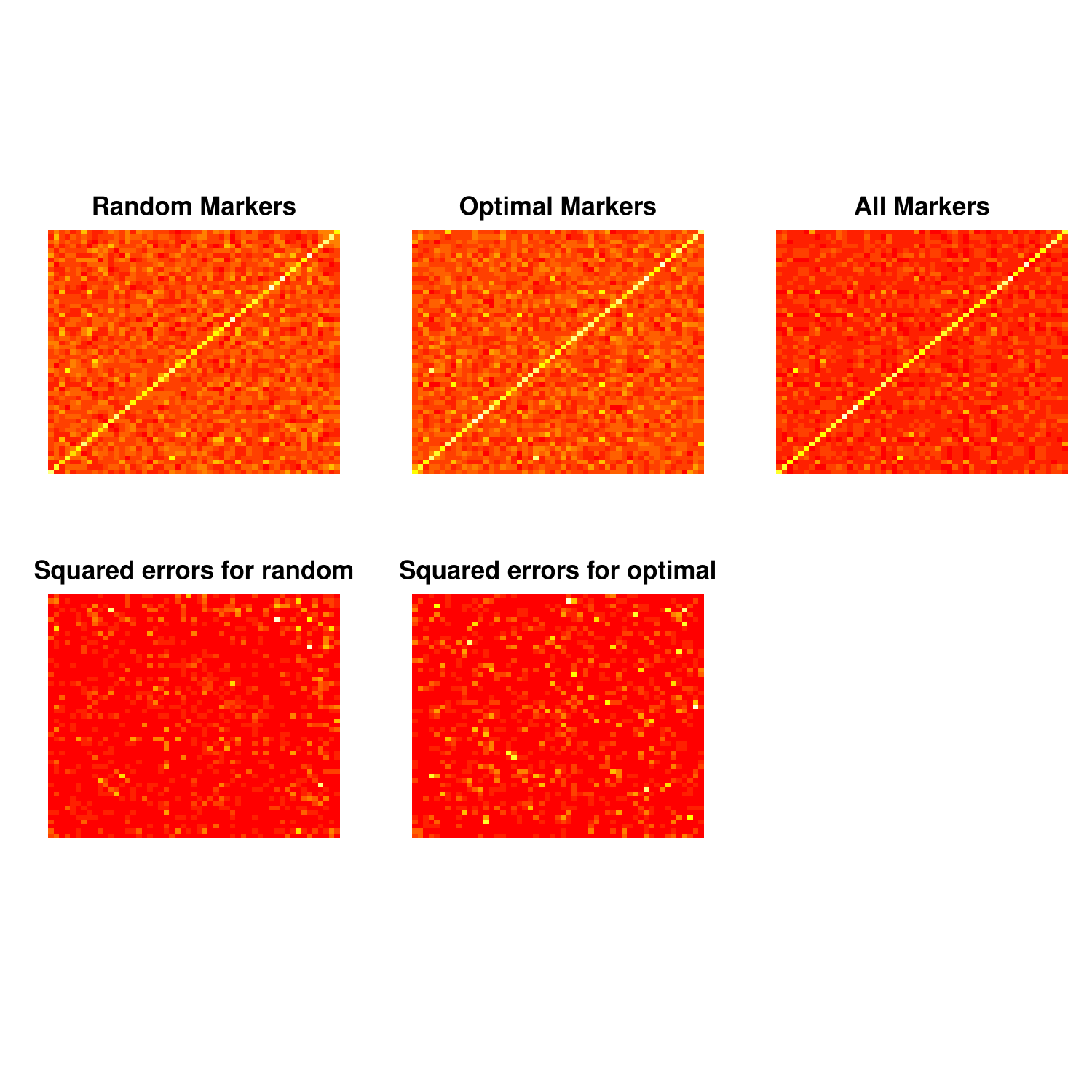}
\end{GrayBox}

According to these results, the optimally selected markers sets result in a genetic relationship matrix that is closer to the full marker genetic relationship matrix than the  relationship matrices calculated from random sets of markers.  I also want to see how the variance captured by these relationship matrices and the accuracies of the models built using these relationship matrices compare.
\begin{GrayBox}
 \scriptsize
\textbf{Box 17: Comparing accuracy of predictions for optimally selected anchor markers, randomly selected markers and all markers}
\begin{Schunk}
\begin{Sinput}
> linenames<-rownames(Afull)
> Test<-sample(linenames, 20)
> Train<-setdiff(linenames, Test)
> Wheat.Y4<-Wheat.Y[Wheat.Y$id%in%rownames(Afull),]
> Wheat.Y4$id<-factor(as.character(Wheat.Y4$id), levels=linenames)
> Wheat.Y4Train<-Wheat.Y4[Wheat.Y4$id%in%Train,]
> Wheat.Y4Test<-Wheat.Y4[Wheat.Y4$id%in%Train,]
> Ztrain<-model.matrix(~-1+Wheat.Y4Train$id)
> Ztest<-model.matrix(~-1+Wheat.Y4Test$id)
> library(EMMREML)
> modelfull<-emmreml(y=Wheat.Y4Train$plant.height,
+                    X=matrix(rep(1,nrow(Ztrain)), ncol=1),
+                    Z=Ztrain, K=Afull+1e-9*diag(nrow(Afull)))
> modelfull$Vu 
\end{Sinput}
\begin{Soutput}
[1] 63.14514
\end{Soutput}
\begin{Sinput}
> rsVus<-c()
> for( i in 1:100){
+ rssmallM<-Wheat.M4[,sample(1:ncol(Wheat.M4),50)]
+ Ars<-A.mat(M=rssmallM)
+  modelrs<-emmreml(y=Wheat.Y4Train$plant.height,
+                  X=matrix(rep(1,nrow(Ztrain)), ncol=1),
+                  Z=Ztrain, K=Ars+1e-9*diag(nrow(Ars)))
+ rsVus<-c(rsVus,modelrs$Vu)
+ }
> mean(rsVus)
\end{Sinput}
\begin{Soutput}
[1] 32.18926
\end{Soutput}
\begin{Sinput}
> modeloptA<-emmreml(y=Wheat.Y4Train$plant.height,
+                    X=matrix(rep(1,nrow(Ztrain)), ncol=1),
+                    Z=Ztrain, K=optA+1e-9*diag(nrow(optA)))
> modeloptA$Vu
\end{Sinput}
\begin{Soutput}
[1] 30.73143
\end{Soutput}
\begin{Sinput}
> predictmatrix<-as.matrix(Ztest%*%cbind(modelfull$uhat,
+                 modelrs$uhat,modeloptA$uhat))
> colnames(predictmatrix)<-c("All", "rs 50", "opt 50")
\end{Sinput}
\end{Schunk}
\end{GrayBox}

\begin{GrayBox}
 \scriptsize
\textbf{Box 18: Comparing accuracy of predictions for optimally selected anchor markers, randomly selected markers and all markers}
\begin{Schunk}
\begin{Sinput}
> pairs(predictmatrix)
> cor(predictmatrix)
\end{Sinput}
\begin{Soutput}
             All     rs 50    opt 50
All    1.0000000 0.7793880 0.9094115
rs 50  0.7793880 1.0000000 0.8656134
opt 50 0.9094115 0.8656134 1.0000000
\end{Soutput}
\end{Schunk}
\includegraphics{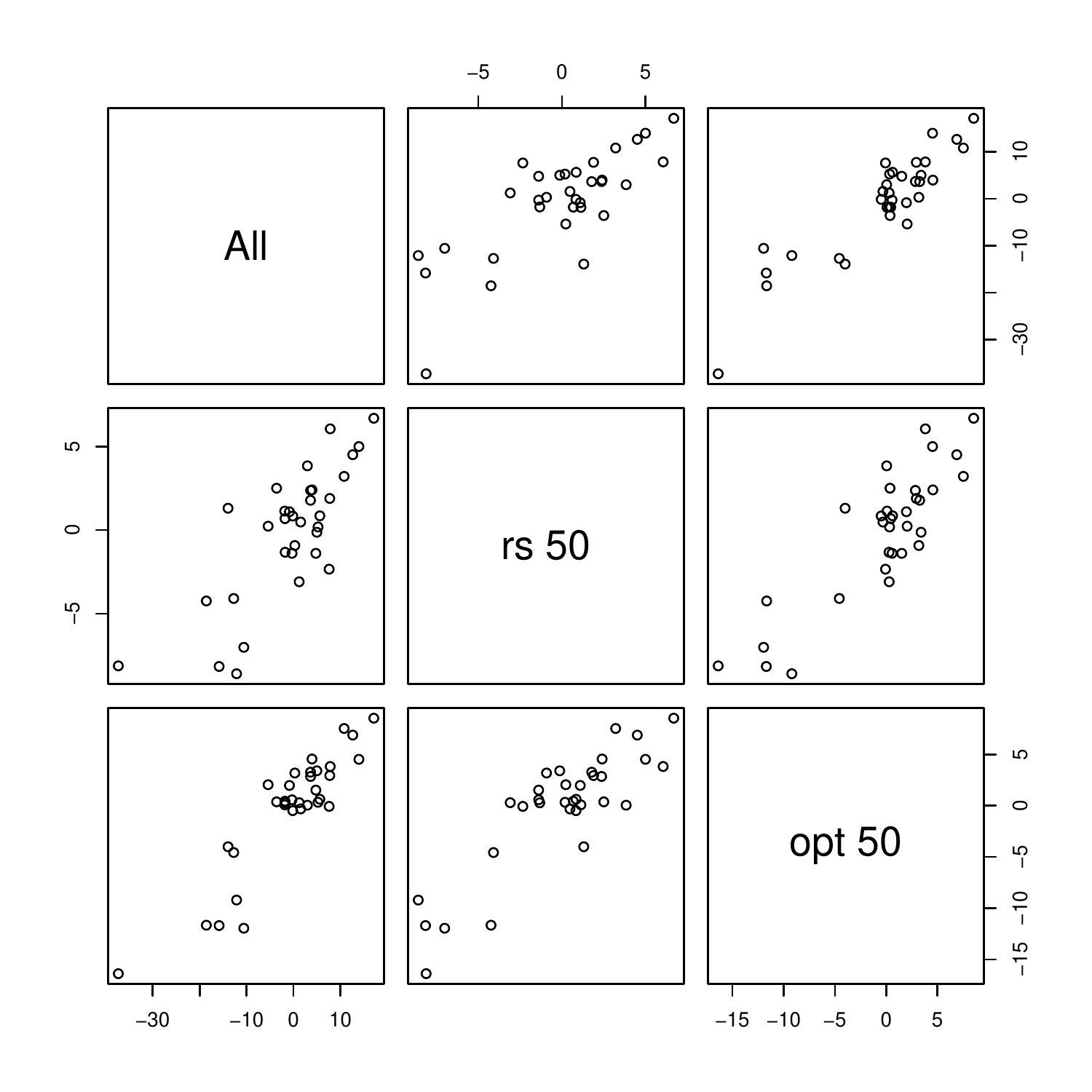}
\end{GrayBox}

\textbf{Minimize inbreeding while maximizing gain}

Many authors \citep{goddard2009genomic, jannink2010dynamics,sun2013mating, akdemir2016efficient} have expressed the importance of reducing inbreeding in PS and GS for long-term success of breeding programs. They argued that GS is likely to lead to a more rapid decline in the selection response unless new alleles are continuously added to the calculation of GEBVs, stressing the importance of balancing short and long term gains by controlling inbreeding in selection. 

A quadratic programming (QP) problem has the generic form $$Q(\bx) = \frac{1}{2} \bx' D \bx -\bd'\bx+ c.$$ Here, $\bx$ is a vector in $\mathbb{R}^n$, $D$ is an $n \times n$ symmetric positive definite matrix, $\bd$ is a constant vector in $\mathbb{R}^n$ and $c$ is a scalar constant. QPs usually come with  a system of linear constraints on the vector $\bx \in \mathbb{R}^n$ which can be written as $$ A\bx = \bff \qquad B\bx \geq g.$$  Here $A$ is an $m_1 \times n$ matrix with $m_1 \leq n$ and $B$ is a $m_2 \times n$ matrix.  The vectors $\bff$ and $\bg$ have lengths $m_1$ and $m_2$ respectively.  QP can be more compactly stated as compactly as: $$\left\{ \begin{array}{l} \mathrm{minimize}_{\bx \in \mathbb{R}^n}: \qquad Q(\bx) = \frac{1}{2} \bx' D \bx - \bd' \bx + c \\ \mathrm{subject\; to:} \qquad A\bx = \bff \qquad B\bx \geq \bg \end{array} \right.$$ There are many efficient algorithms that solves QP's so there is in practice little difficulty in calculating the optimal solution for any particular data set. In this example, I will be using the package \pkg{quadprog} \citep{turlach2007quadprog}. 

Now, let $A$ be a matrix of pedigree based numerator relationships or the additive genetic relationships between the individuals in a breeding population (this matrix can be obtained from a pedigree of genome-wide markers for the individuals) and let $\bc$ be the vector of proportional contributions of individuals to the next generation under a random mating scheme. The average inbreeding and co-ancestry for a given choice of  $\bc$ can be defined as $r=\frac{1}{2}\bc'A\bc.$ If $\bb$ is the vector of GEBV's, i.e., the vector of BLUP (best linear unbiased predictor) estimated BV's of the candidates for selection. The expected gain is defined as $g=\bc'\bb.$ Without loss of generality, we will assume that the breeder's long term goal is to increase the value of $g.$ 

In \cite{wray1994moet,brisbane1995balancing,meuwissen1997maximizing}, an approach that seeks minimizing  the average inbreeding and co-ancestry while restricting the genetic gain is proposed. The optimization problem can be stated as
\begin{equation}\begin{array}{lc}
\underset{\bc}{\text{minimize}}& r=\bc' \frac{A}{2} \bc \\ [10pt]
\mbox{subject to} & \begin{array}[t]{rcl}
\bc' \bb=\rho  &  \\ [10pt]
\bc'\bone=1 & \\ [10pt]
\bc\geq 0, &
\end{array}\end{array}\label{eq:1}\end{equation}
where $\rho$ is the desired level of gain. 
 
This problem is easily recognized as a QP.  Recently, several parental percentage allocation strategies were tested using QP's in \citep{goddard2009genomic, pryce2012novel, schierenbeck2011controlling}.

\begin{GrayBox}
 \scriptsize
\textbf{Box 19: Preparing the wheat data for analysis, setting up the optimization function}
\begin{Schunk}
\begin{Sinput}
> Wheat.Ysc<-Wheat.Y
> Wheat.Ysc[,2]<-(Wheat.Ysc[,2]-mean(Wheat.Ysc[,2]))/sd(Wheat.Ysc[,2])
> P<-cbind(Wheat.Ysc, Wheat.M)
> rownames(P)<-rownames(Wheat.M)
> impinbreed=.1
> STPGAUSERDEFFUNC<-function(Train,Test=NULL, P, 
+                            lambda=NULL, C=NULL){
+   trsmallM<-P[rownames(P)%in%Train,-c(1:2)]
+   g<-matrix(P[rownames(P)%in%Train,2], ncol=1)
+   A<-A.mat(M=trsmallM+1)
+   return(-(1-impinbreed)*mean(g)+
+            impinbreed*mean(c(A[lower.tri(A, diag=TRUE)])))
+ }
\end{Sinput}
\end{Schunk}
\end{GrayBox}

By solving the QP in Equation (\ref{eq:1}) for varying values of $\rho,$ or using the similar criteria in the mate selection approaches, we can trace out an efficient frontier curve, a smooth non-decreasing curve that gives the best possible trade-off of genetic variance against gain. This curve represents the set of optimal allocations and it is called the efficiency frontier (EF) curve in finance \citep{markowitz1952portfolio} and breeding literature. 

Many practical applications require additional constraints, ans it is possible to extend these formulations to introduce additional constraints as positiveness, minimum-maximum for proportions, minimum-maximum for number of lines (cardinality constraints). It is not too difficult to use \pkg{STPGA} to solve a version of this problem.

Suppose the breeder wants to select a subset of the individuals to become parents of the next generation increasing gain while controlling for coancestory and inbreeding. Note that the breeder only wants to select a subset and therefore we can assume that the parental contributions will be the same for each of the selected individuals. The following code illustrates how to select 10 individuals from the wheat data set for changing values of $\lambda.$

\begin{GrayBox}
 \scriptsize
\textbf{Box 20: Selected individuals for changing values of $\lambda$}
\begin{Schunk}
\begin{Sinput}
> GAOUTLIST<-vector(mode="list", length=5)
> i=1
> for (impinbreed in c(0.01,.95,.99)){
+   STPGAUSERDEFFUNC<-function(Train,Test=NULL, P, lambda=NULL, C=NULL){
+     trsmallM<-P[rownames(P)%in%Train,-c(1:2)]
+     g<-P[rownames(P)%in%Train,2]
+     A<-A.mat(M=trsmallM+1)
+     return(-(1-impinbreed)*mean(g)+
+              impinbreed*mean(c(A[lower.tri(A, diag=TRUE)])))
+   }
+   GAOUT<-GenAlgForSubsetSelectionNoTest(P=P, ntoselect=10,npop=100, 
+             nelite=10, mutprob=.5, mutintensity = 1,niterations=100,
+             minitbefstop=50, tabumemsize = 1, plotiters=FALSE,tabu=FALSE,
+            lambda=1e-9,errorstat="STPGAUSERDEFFUNC", mc.cores=4)
+   GAOUTLIST[[i]]<-GAOUT
+   i=i+1
+ }
> GAMINvec<-c()
> GAMINvecgain<-c()
> GAMINvecinbreeding<-c()
> for (i in 1:3){
+ Train<-GAOUTLIST[[i]]$`Solution with rank 1`
+ trsmallM<-P[rownames(P)%in%Train,-c(1:2)]
+ g<-P[rownames(P)%in%Train,2]
+ A<-A.mat(M=trsmallM+1)
+ GAMINvec<-c(GAMINvec,
+             min(GAOUTLIST[[i]]$`Best criterion values over iterarions`))
+ GAMINvecgain<-c(GAMINvecgain, mean(g))
+ GAMINvecinbreeding<-c(GAMINvecinbreeding,mean(c(A[lower.tri(A, diag=TRUE)])))
+ }
\end{Sinput}
\end{Schunk}
\end{GrayBox}

After that I extract the average gain and inbreeding values for each value of $\lambda$ and trace the frontier curve. 
\begin{GrayBox}
 \scriptsize
\textbf{Box 21: Points on the frontier surface}
\begin{Schunk}
\begin{Sinput}
> plot(GAMINvecgain,GAMINvecinbreeding, type="b",
+      xlab="gain", ylab="inbreeding")
\end{Sinput}
\end{Schunk}
\includegraphics{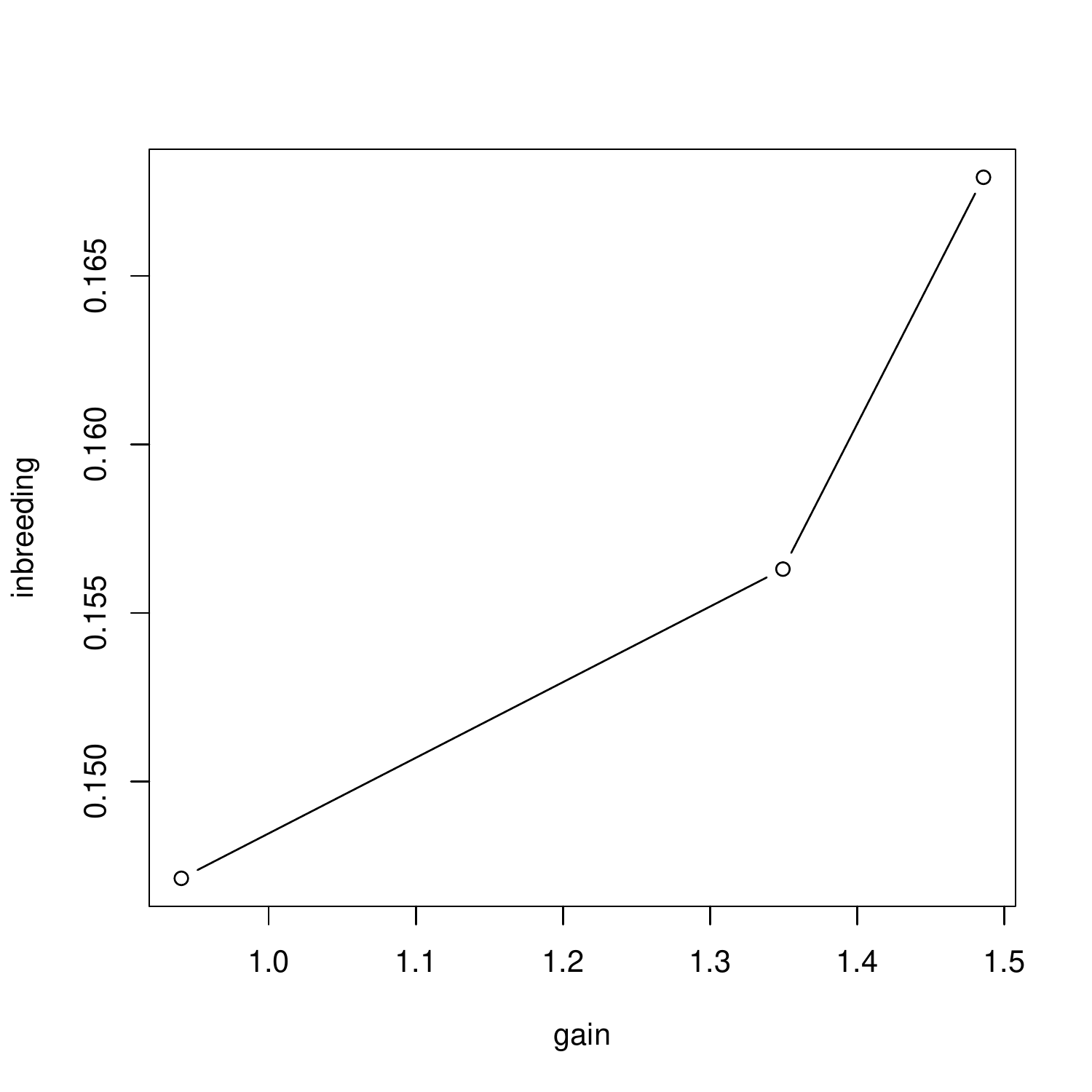}
\end{GrayBox}

Suppose now the breeder wants us to pick 10 individuals, but also asks for the best parental contribution proportions. This is a mixed integer quadratic programming problem.  Let $\epsilon_i$ be the minimum proportion that must be allocated to line $i,$ $\delta_i$ be the maximum proportion that must be allocated to line $i$ if any of line $i$ will be conserved, where we must have $0\leq\epsilon_i\leq\delta_i\leq 1.$ Introduce the binary variables: 
$$z_i = \left\{
     \begin{array}{ll}
       1 & \text{if any of line i is included,} \\
       0 & \text{otherwise.}
     \end{array}
   \right.$$
The cardinality constrained optimization problem is given by 
\[ \begin{array}{ll}
\mbox{minimize} & r=c' \frac{A}{2} c \\ [10pt]
\mbox{subject to} & \begin{array}[t]{rcl}
c' b=\rho,  &  \\ [10pt]
c'\bone=1, & \\ [10pt]
\bc\geq 0, & \\[10pt]
\sum_{i=1}^{m}z_i=K,& \\[10pt]
\epsilon_i z_i\leq c_i\leq\delta_i z_i,& i=1,2,\ldots,m,\\[10pt]
z_i \in \{0,1\}, & i=1,2,\ldots,m. 
\end{array}\end{array}\]

Here is the code for doing the same thing we have done with the previous example, except we provide parental contributions.
\begin{GrayBox}
 \scriptsize
\textbf{Box 22: Estimating the parental contributions for the cardinality constrained problem for changing values of $\lambda$}
\begin{Schunk}
\begin{Sinput}
> require(quadprog)
> GAOUTLIST<-vector(mode="list", length=5)
> i=1
> for (impinbreed in c(0.01,.95,.99)){
+   STPGAUSERDEFFUNC<-function(Train,Test=NULL, P, lambda=1e-5, C=NULL){
+     trsmallM<-P[rownames(P)%in%Train,-c(1:2)]
+     g<-c(P[rownames(P)%in%Train,2])
+     A<-A.mat(M=trsmallM+1)
+     n=length(g)
+     sol  <- solve.QP(Dmat=(n^2/2)*impinbreed*(A+lambda*diag(n)),
+             dvec=(1/n)*(1-impinbreed)*g,Amat=cbind(rep(1,n),diag(n),-diag(n)),
+             bvec=rbind(1,matrix(0,ncol=1,nrow=n),
+                        matrix(-1,ncol=1,nrow=n)), meq=1)
+     names(sol$solution)<-rownames(trsmallM)
+     return(sol$value)
+   }
+   GAOUT<-GenAlgForSubsetSelectionNoTest(P=P, ntoselect=10,npop=50, 
+             nelite=5, mutprob=.5, mutintensity = 1,niterations=100,
+             minitbefstop=50, tabumemsize = 1,plotiters=FALSE,tabu=FALSE,
+             lambda=1e-5,errorstat="STPGAUSERDEFFUNC", mc.cores=4)
+   GAOUTLIST[[i]]<-GAOUT
+   i=i+1
+ }
\end{Sinput}
\end{Schunk}
\end{GrayBox}

\begin{GrayBox}
 \scriptsize
\textbf{Box 23: Estimating the parental contributions for the cardinality constrained problem for changing values of $\lambda$}
\begin{Schunk}
\begin{Sinput}
> GAMINvec<-c()
> GAsols<-vector(mode="list")
> for (i in 1:3){
+   trsmallM<-P[rownames(P)%in%GAOUTLIST[[i]]$`Solution with rank 1`,-c(1:2)]
+   g<-c(P[rownames(P)%in%GAOUTLIST[[i]]$`Solution with rank 1`,2])
+   A<-A.mat(M=trsmallM+1)
+   n=length(g)
+   impinbreed<-c(0.01,.95,.99)[i]
+   
+    sols<-solve.QP(Dmat=(n^2/2)*impinbreed*(A+1e-9*diag(n)),
+           dvec=(1/n)*(1-impinbreed)*g,Amat=cbind(rep(1,n),diag(n),-diag(n)),
+           bvec=rbind(1,matrix(0,ncol=1,nrow=n),
+                      matrix(-1,ncol=1,nrow=n)), meq=1)$solution
+   names(sols)<-GAOUTLIST[[i]]$`Solution with rank 1`
+   GAsols[[i]] <-sols
+   GAMINvec<-c(GAMINvec, 
+               min(GAOUTLIST[[i]]$`Best criterion values over iterarions`))
+ }
> print(round(GAsols[[2]], digits=3))
\end{Sinput}
\begin{Soutput}
IWA8606856      PPG-1     AGRISS IWA8606779       466A    H86-708    PI94479 
       0.1        0.1        0.1        0.1        0.1        0.1        0.1 
 ZG4163/73    NORIN10   PI254037 
       0.1        0.1        0.1 
\end{Soutput}
\end{Schunk}
\end{GrayBox}

\textbf{Variable selection in regression using model selection criteria}

One of the standard use of GA is in variable selection and STPGA can be used in variable selection. 

Another very popular method for variable selection and penalized (shrunk) parameter estimation in high dimensional regression is the lasso approach which can be summarized as finding the best regression coefficients that minimize the regression loss function while minimizing a multiple of the $\ell_1$ norm of the coefficients. The relative stress on the importance of either of these complementary functions is expressed as a linear combination of these two. For example with the squared error loss the lasso optimization criterion is \[(\by-X\bbeta)'(\by-X\bbeta)+\lambda |\bbeta|'\bone.\] The idea, its implications for many different fields of science can not be overlooked, the same goes for the enormous number of methods developed to solve this problem and its variations. However, there is a major statistical problem: there isn't enough principle behind its objective criterion so it is no big magic. It does not have certain properties we would like from a good regression. For example, the residuals of a model using the coefficients obtained by lasso are not orthogonal to the design of the variables that are selected by lasso. This last issue can be circumvented by using lasso only as a selection operator and then fitting the coefficients with least squares without answering the main question: Why this criterion, not another one? Parsimony, yes, why this one? Is this the one that gives the best generalization error (minimize the prediction error)?  The only good argument for the defense of this criterion I can think of is a Bayesian one and the so called ''oracle'' property which is only true under restrictive assumptions such as no dependency among the explanatory variables. we should be careful not to confuse the actual problem with the method, such as the carpenter who sees everything as nails since he/she is good at using hammers.

In my opinion, the whole area of norm penalized estimation neglects the many model selection criteria (AIC \citep{akaike1974new}, BIC \citep{schwarz1978estimating}, ICOMP(IFIM) \citep{bozdogan1987icomp}, etc,...) introduced during the 1970's and onward by many prominent statisticians whose derivations depends on the solid theory of likelihoods, divergence measures, etc.,... I am not sure why we should develop elegant theories and then ignore them for not so elegant ones. So while lasso \cite{tibshirani1996regression} is great and can be solved at great speed in serial (after investment a huge amounts of energy and resources in the last 10 or more years) and the methods to find its solutions can only be partially parallelized. I am not sure the shrinkage approach can take on the next challenge of solving extremely large and complex problems. In addition to not being able to adopt well to parallel computer architectures, their application areas are limited by the problems these methods can address and addressing new problems with the same techniques involves inventing newly crafted methodologies which consumes the most amount of time and resources.

Here is an example using STPGA. I am going to demonstrate this with a classic body fat data set. The data is available in the package \pkg{UsingR} \citep{usingr2015}. I will use the AIC criterion, find the best subsets of size two through 12 (there are 13 explanatory variables) and obtain their AIC values.   

\begin{GrayBox}
 \scriptsize
\textbf{Box 24: Variable selection for regression}
\begin{Schunk}
\begin{Sinput}
> data("fat", package = "UsingR")
> mod <- lm(body.fat.siri ~ age + weight + height + neck + chest + abdomen +
+ hip + thigh + knee + ankle + bicep + forearm + wrist, data = fat)
> x <- model.matrix(mod)
> y <- model.response(model.frame(mod))
> fitnessfuncforSTPGA <- function(Train,Test=NULL, P, lambda=1e-6, C=NULL) {
+     X <- t(P[rownames(P)%in%Train,])
+     mod <- lm.fit(X, y)
+     class(mod) <- "lm"
+ return(AIC(mod))
+ }
> stpgaoutlist<-vector(mode="list")
> ii=1
> for (i in 2:(ncol(x)-2)){
+ stpgaoutlist[[ii]]<-GenAlgForSubsetSelectionNoTest(P=t(x[,-1]),
+         ntoselect=i,npop=200, 
+         nelite=10, mutprob=.5, mutintensity = 1,
+         niterations=200, minitbefstop=50,tabu=FALSE,
+         tabumemsize = 1,plotiters=FALSE,lambda=1e-9,
+         errorstat="fitnessfuncforSTPGA",mc.cores=4)
+ ii=ii+1
+ }
\end{Sinput}
\end{Schunk}
\end{GrayBox}

\begin{GrayBox}
 \scriptsize
\textbf{Box 25: Extracting results}
\begin{Schunk}
\begin{Sinput}
> GAMINs<-c()
> ii=1
> for (i in 2:(ncol(x)-2)){
+   GAMINs<-c(GAMINs,
+             min(stpgaoutlist[[ii]]$`Best criterion values over iterarions`))
+   ii=ii+1
+ }
> selectedvars6<-stpgaoutlist[[6]]$`Solution with rank 1`
> min(stpgaoutlist[[6]]$`Best criterion values over iterarions`)
\end{Sinput}
\begin{Soutput}
[1] 1460.21
\end{Soutput}
\begin{Sinput}
> selectedvars6
\end{Sinput}
\begin{Soutput}
[1] "abdomen" "forearm" "hip"     "age"     "wrist"   "thigh"   "neck"   
\end{Soutput}
\begin{Sinput}
> selectedvars7<-stpgaoutlist[[7]]$`Solution with rank 1`
> min(stpgaoutlist[[7]]$`Best criterion values over iterarions`)
\end{Sinput}
\begin{Soutput}
[1] 1459.716
\end{Soutput}
\begin{Sinput}
> selectedvars7
\end{Sinput}
\begin{Soutput}
[1] "forearm" "thigh"   "age"     "abdomen" "hip"     "wrist"   "height" 
[8] "neck"   
\end{Soutput}
\begin{Sinput}
> selectedvars8<-stpgaoutlist[[8]]$`Solution with rank 1`
> min(stpgaoutlist[[8]]$`Best criterion values over iterarions`)
\end{Sinput}
\begin{Soutput}
[1] 1459.474
\end{Soutput}
\begin{Sinput}
> selectedvars8
\end{Sinput}
\begin{Soutput}
[1] "thigh"   "abdomen" "weight"  "neck"    "hip"     "forearm" "age"    
[8] "wrist"   "height" 
\end{Soutput}
\end{Schunk}
\end{GrayBox}

Lets plot the results. It looks like we should pick seven or eight variables. 

\begin{GrayBox}
 \scriptsize
\textbf{Box 26: Plotting AIC results for different subset sizes}
\begin{Schunk}
\begin{Sinput}
> plot(2:(ncol(x)-2),GAMINs, type="b", xlab="NVARS", ylab="AIC")
\end{Sinput}
\end{Schunk}
\includegraphics{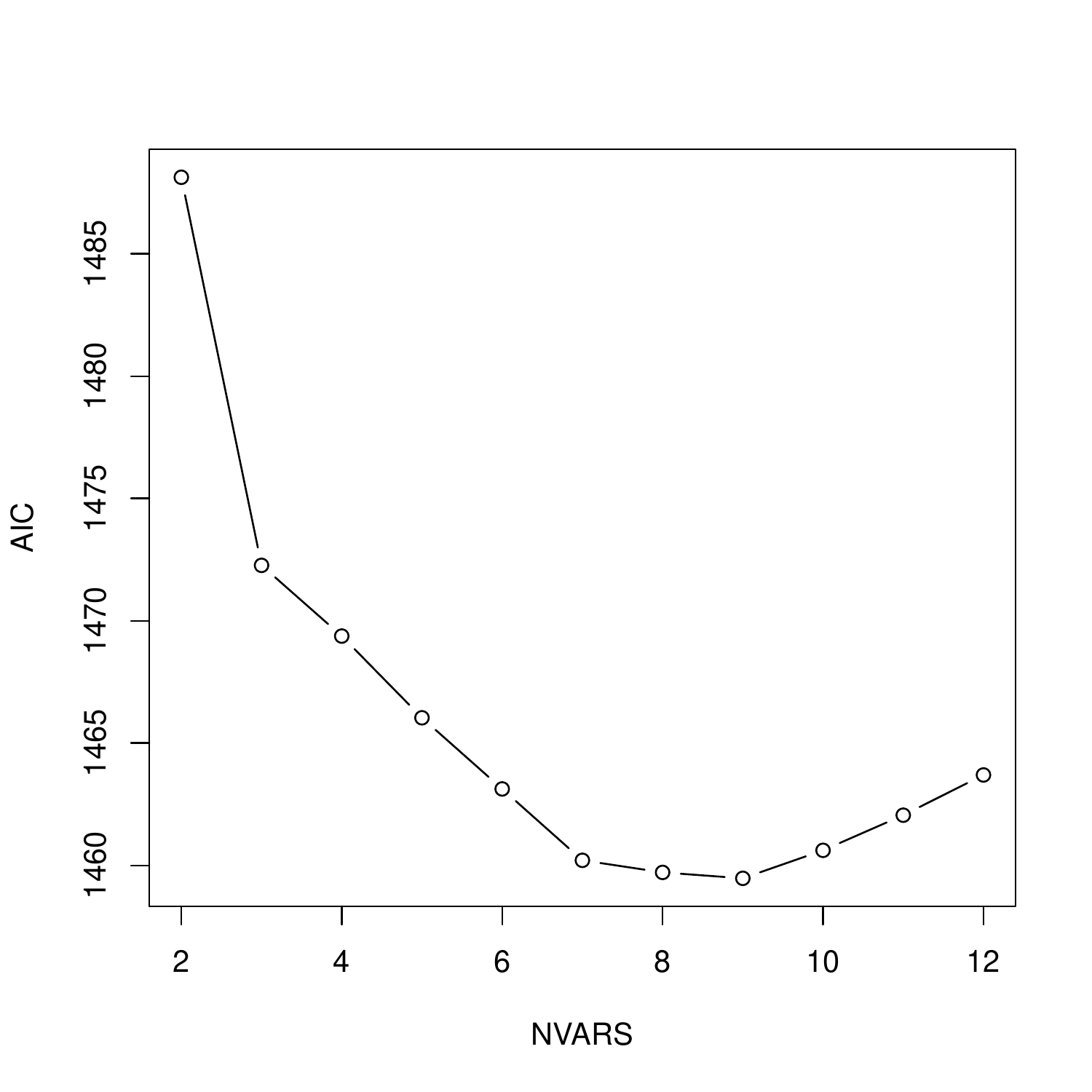}
\end{GrayBox}

Now, I will try to using an unsupervised approach to variable selection, the methodology is new.  Why not pick the subset of variables that result in a kernel matrix that is most aligned with the the kernel that is calculated based on all 13 explanatory variables. Interestingly, most of the variables that are selected by supervised selection are also identified here. 
\begin{GrayBox}
 \scriptsize
\textbf{Box 27: Variable selection- Unsupervised using kernel alignment}
\begin{Schunk}
\begin{Sinput}
> Afull<-tcrossprod(scale(x[,-1], center=TRUE, scale=TRUE))/(ncol(x)-1)
> STPGAUSERDEFFUNC<-function(Train,Test=NULL, P, lambda=1e-6, C=NULL){
+  trsmallx<-t(scale(P[rownames(P)%in%Train,],center=TRUE, scale=TRUE))
+  Atr<-tcrossprod(trsmallx)/(ncol(trsmallx))
+  return(mean((c(Afull[lower.tri(Afull, diag=TRUE)])
+               -c(Atr[lower.tri(Atr, diag=TRUE)]))^2))
+ }
> GAOUT<-GenAlgForSubsetSelectionNoTest(P=t(x[,-1]),
+         ntoselect=7,npop=200, 
+         nelite=5, mutprob=.5, mutintensity = 1,
+         niterations=200, minitbefstop=50,tabu=FALSE,
+         tabumemsize = 1,plotiters=FALSE,lambda=1e-9,
+         errorstat="STPGAUSERDEFFUNC",mc.cores=4)
> selectedunsupervised<-GAOUT$`Solution with rank 1`
> selectedunsupervised
\end{Sinput}
\begin{Soutput}
[1] "knee"    "neck"    "ankle"   "wrist"   "age"     "forearm" "bicep"  
\end{Soutput}
\begin{Sinput}
> selectedvars6
\end{Sinput}
\begin{Soutput}
[1] "abdomen" "forearm" "hip"     "age"     "wrist"   "thigh"   "neck"   
\end{Soutput}
\begin{Sinput}
> intersect(selectedvars6,selectedunsupervised)
\end{Sinput}
\begin{Soutput}
[1] "forearm" "age"     "wrist"   "neck"   
\end{Soutput}
\end{Schunk}
\end{GrayBox}

Here is a similar approach, we only add the response variable to the calculation of full data kernel.

\begin{GrayBox}
 \scriptsize
\textbf{Box 28: Variable selection - supervised using kernel alignment}
\begin{Schunk}
\begin{Sinput}
> newx<-cbind(x[,-1],y)
> colnames(newx)
\end{Sinput}
\begin{Soutput}
 [1] "age"     "weight"  "height"  "neck"    "chest"   "abdomen" "hip"    
 [8] "thigh"   "knee"    "ankle"   "bicep"   "forearm" "wrist"   "y"      
\end{Soutput}
\begin{Sinput}
> Afull2<-tcrossprod(scale(newx, center=TRUE, scale=TRUE))/(ncol(x))
> STPGAUSERDEFFUNC2<-function(Train,Test=NULL, P, lambda=1e-6, C=NULL){
+   trsmallx<-t(scale(P[rownames(P)%in%Train,],center=TRUE, scale=TRUE))
+  Atr<-tcrossprod(trsmallx)/(ncol(trsmallx))
+ 
+  return(mean((c(Afull2[lower.tri(Afull2, diag=TRUE)])
+               -c(Atr[lower.tri(Atr, diag=TRUE)]))^2))
+ }
> GAOUT2<-GenAlgForSubsetSelectionNoTest(P=t(x[,-1]),
+         ntoselect=7,npop=200, 
+         nelite=5, mutprob=.5, mutintensity = 1,
+         niterations=200, minitbefstop=50,tabu=FALSE,
+         tabumemsize = 1,plotiters=FALSE,lambda=1e-9,
+         errorstat="STPGAUSERDEFFUNC2",mc.cores=4)
> selectedsupervised<-GAOUT2$`Solution with rank 1`
> selectedsupervised
\end{Sinput}
\begin{Soutput}
[1] "forearm" "knee"    "ankle"   "wrist"   "age"     "neck"    "bicep"  
\end{Soutput}
\begin{Sinput}
> selectedvars6
\end{Sinput}
\begin{Soutput}
[1] "abdomen" "forearm" "hip"     "age"     "wrist"   "thigh"   "neck"   
\end{Soutput}
\begin{Sinput}
> intersect(selectedvars6,selectedunsupervised)
\end{Sinput}
\begin{Soutput}
[1] "forearm" "age"     "wrist"   "neck"   
\end{Soutput}
\begin{Sinput}
> intersect(selectedvars6,selectedsupervised)
\end{Sinput}
\begin{Soutput}
[1] "forearm" "age"     "wrist"   "neck"   
\end{Soutput}
\begin{Sinput}
> intersect(selectedunsupervised,selectedsupervised)
\end{Sinput}
\begin{Soutput}
[1] "knee"    "neck"    "ankle"   "wrist"   "age"     "forearm" "bicep"  
\end{Soutput}
\end{Schunk}
\end{GrayBox}

We can also align to the kernel matrix obtained from only the response variable(s): 

\begin{GrayBox}
 \scriptsize
\textbf{Box 29: Variable selection- supervised (only by response) using kernel alignment}
\begin{Schunk}
\begin{Sinput}
> Afull3<-tcrossprod(scale(y, center=TRUE, scale=TRUE))
> STPGAUSERDEFFUNC3<-function(Train,Test=NULL, P, lambda=1e-6, C=NULL){
+  trsmallx<-t(scale(P[rownames(P)%in%Train,],center=TRUE, scale=TRUE))
+  Atr<-tcrossprod(trsmallx)/(ncol(trsmallx))
+ 
+  return(mean((c(Afull3[lower.tri(Afull3, diag=TRUE)])
+               -c(Atr[lower.tri(Atr, diag=TRUE)]))^2))
+ }
> GAOUT3<-GenAlgForSubsetSelectionNoTest(P=t(x[,-1]),
+         ntoselect=7,npop=200, 
+         nelite=5, mutprob=.5, mutintensity = 1,
+         niterations=200, minitbefstop=50,tabu=FALSE,
+         tabumemsize = 1,plotiters=FALSE,lambda=1e-9,
+         errorstat="STPGAUSERDEFFUNC3",mc.cores=4)
> selectedsupervised2<-GAOUT3$`Solution with rank 1`
> selectedsupervised2
\end{Sinput}
\begin{Soutput}
[1] "wrist"   "bicep"   "neck"    "forearm" "ankle"   "knee"    "age"    
\end{Soutput}
\begin{Sinput}
> selectedvars6
\end{Sinput}
\begin{Soutput}
[1] "abdomen" "forearm" "hip"     "age"     "wrist"   "thigh"   "neck"   
\end{Soutput}
\begin{Sinput}
> intersect(selectedvars6,selectedunsupervised)
\end{Sinput}
\begin{Soutput}
[1] "forearm" "age"     "wrist"   "neck"   
\end{Soutput}
\begin{Sinput}
> intersect(selectedvars6,selectedsupervised2)
\end{Sinput}
\begin{Soutput}
[1] "forearm" "age"     "wrist"   "neck"   
\end{Soutput}
\begin{Sinput}
> intersect(selectedunsupervised,selectedsupervised2)
\end{Sinput}
\begin{Soutput}
[1] "knee"    "neck"    "ankle"   "wrist"   "age"     "forearm" "bicep"  
\end{Soutput}
\begin{Sinput}
> intersect(selectedsupervised,selectedsupervised2)
\end{Sinput}
\begin{Soutput}
[1] "forearm" "knee"    "ankle"   "wrist"   "age"     "neck"    "bicep"  
\end{Soutput}
\end{Schunk}
\end{GrayBox}

Let's see which variables are picked by lasso and best subsets regressions:
\begin{GrayBox}
 \scriptsize
\textbf{Box 30: Variable selection- supervised using lasso and best subsets regression}
\begin{Schunk}
\begin{Sinput}
> # lasso
> library(glmnet)
> fit.glmnet.lasso.cv <- cv.glmnet(as.matrix(scale(x[,-1], center=T, scale=T)),
+                               as.matrix(y, ncol=1),
+                                  nfold = 5,
+                                  alpha = 1)
> coef(fit.glmnet.lasso.cv, s = fit.glmnet.lasso.cv$lambda.1se)
\end{Sinput}
\begin{Soutput}
14 x 1 sparse Matrix of class "dgCMatrix"
                     1
(Intercept) 19.1507937
age          0.2887603
weight       .        
height      -0.5184681
neck         .        
chest        .        
abdomen      6.2447837
hip          .        
thigh        .        
knee         .        
ankle        .        
bicep        .        
forearm      .        
wrist       -0.1999128
\end{Soutput}
\begin{Sinput}
> GAOUT4<-GenAlgForSubsetSelectionNoTest(P=t(x[,-1]),
+         ntoselect=4,npop=200, 
+         nelite=5, mutprob=.5, mutintensity = 1,
+         niterations=200, minitbefstop=50,tabu=FALSE,
+         tabumemsize = 1,plotiters=FALSE,lambda=1e-9,
+         errorstat="STPGAUSERDEFFUNC3",mc.cores=4)
> selectedsupervised4<-GAOUT4$`Solution with rank 1`
> selectedsupervised4
\end{Sinput}
\begin{Soutput}
[1] "knee"    "neck"    "forearm" "age"    
\end{Soutput}
\end{Schunk}
\end{GrayBox}

\begin{GrayBox}
 \scriptsize
\textbf{Box 31: Variable selection- supervised using lasso and best subsets regression}
\begin{Schunk}
\begin{Sinput}
> # lasso
> library(leaps)
> regsubsets.out <-
+     regsubsets(body.fat.siri ~ age + weight + height + neck + chest + abdomen +
+        hip + thigh + knee + ankle + bicep + forearm + wrist,
+        data = fat,
+        nbest = 1, nvmax = NULL,   
+         force.in = NULL, force.out = NULL,
+        method = "exhaustive")
> sevenvarsselectedbyleaps<-colnames(as.data.frame(summary(regsubsets.out)$outmat))[
+   which(as.data.frame(summary(regsubsets.out)$outmat)[7,]=="*")]
> intersect(sevenvarsselectedbyleaps,selectedunsupervised)
\end{Sinput}
\begin{Soutput}
[1] "age"     "neck"    "forearm" "wrist"  
\end{Soutput}
\begin{Sinput}
> intersect(sevenvarsselectedbyleaps,selectedvars7)
\end{Sinput}
\begin{Soutput}
[1] "age"     "neck"    "abdomen" "thigh"   "forearm" "wrist"  
\end{Soutput}
\end{Schunk}
\end{GrayBox}

Now, again, lets to something that hasn't been done before: I am going to ''make up'' a variable selection criterion for regression that has some nice properties. Lets assume as before $\by$ is the length $n$ response vector, $X$ is the $n\times p $ design matrix; lets denote the generic design of $k\leq p$ variables as $X^{(k)}$ and the set of such k variable designs  as $\mathbf{X}^{(k)}.$ For this generic design matrix we want to find $\bbeta^{(k)}\in\mathbf{R}^k$ such that error sum of squares are minimized, given $X^{(k)}$ this is the least squares solution $\widehat{\bbeta^{(k)}}=({X^{(k)}}'X^{(k)})^{-1}{X^{(k)}}'\by$ with covariance matrix proportional to $({X^{(k)}}'X^{(k)})^{-1}.$ The optimization problem is stated as $$(\widehat{X^{(k)}}, \widehat{\bbeta^{(k)}})=\underset{X^{(k)} \in \mathbf{X}^{(k)}, \bbeta^{(k)} \in \mathbf{R}^k}{\mathrm{argmin}}
(\by-X^{(k)}\bbeta^{(k)})'(\by-X^{(k)}\bbeta^{(k)})-\lambda |{X^{(k)}}'X^{(k)}|$$ and to simplify it further, noting that the penalty term does not depend on $\bbeta^{(k)}$, we can rewrite the above as $$\widehat{X^{(k)}}=\underset{X^{(k)} \in \mathbf{X}^{(k)}}{\mathrm{argmin}}(\by(I-X^{(k)}({X^{(k)}}'X^{(k)})^{-}{X^{(k)}}')\by-\lambda \log |{X^{(k)}}'X^{(k)}|$$ and  $\widehat{\bbeta^{(k)}}=(\widehat{X^{(k)}}'\widehat{X^{(k)}})^{-1}\widehat{X^{(k)}}'\by.$ The interpretation of this optimization criterion is as follows, we want the set of k variables that minimize the squared error loss while also keeping the variances and correlations among the estimated coefficients as low as possible, by increasing $\lambda$ we would be increasing the the relative importance given to the variance and covariances of the coefficients. Note that, when the information matrix is  singular, we can use a generalized inverse obtained by regularization (i.e. add a small constant to the diagonal elements) as described in the previous sections.

\begin{GrayBox}
 \scriptsize
\textbf{Box 32: Variable selection}
\begin{Schunk}
\begin{Sinput}
> mod <- lm(body.fat.siri ~ age + weight + height + neck + chest + abdomen +
+ hip + thigh + knee + ankle + bicep + forearm + wrist, data = fat)
> x <- scale(as.matrix(model.matrix(mod)), center=F, scale=T)
> y <- model.response(model.frame(mod))
> y<-y/sd(y)
> fitnessfuncforSTPGA <- function(Train,Test=NULL, P, lambda=.5, C=NULL) {
+     X <- t(P[rownames(P)%in%Train,])
+     n<-nrow(X)
+ 		p<-ncol(X)
+ 		mindim<-min(p,n)
+ 		rownames(X)<-NULL
+ 		svdX<-svd(X, nu=mindim,nv=mindim)
+ 		insvdnonzero<-1:mindim
+ 		diagvecforinv<-(svdX$d[insvdnonzero])/((svdX$d[insvdnonzero])^2+1e-7)
+ 		coef<-tcrossprod(svdX$v%*%diag(diagvecforinv),svdX$v)%*%t(X)%*%(y-mean(y))
+ 		resids<-y-X%*%coef-mean(y)
+     out<-(1-lambda)*mean(resids^2)-
+     lambda*determinant(crossprod(X), logarithm = T )$modulus
+ return(out)
+ }
> GAMINSmat<-c()
> lambda=1e-6
> stpgaoutlist<-vector(mode="list")
> ii=1
> for (i in 2:(ncol(x)-1)){
+ stpgaoutlist[[ii]]<-GenAlgForSubsetSelectionNoTest(P=t(x[,-1]),
+         ntoselect=i,npop=100, 
+         nelite=10, mutprob=.5, mutintensity = 1,
+         niterations=100, minitbefstop=50,tabu=FALSE,
+         tabumemsize = 1,plotiters=F,lambda=lambda,
+         errorstat="fitnessfuncforSTPGA",mc.cores=4)
+ ii=ii+1
+ }
> GAMINs<-c()
> ii=1
> for (i in 2:(ncol(x)-1)){
+   GAMINs<-c(GAMINs,
+             min(stpgaoutlist[[ii]]$`Best criterion values over iterarions`))
+   ii=ii+1
+ }
> GAMINSmat<-cbind(GAMINSmat,GAMINs)
> 
\end{Sinput}
\end{Schunk}
\end{GrayBox}

\begin{GrayBox}
 \scriptsize
\textbf{Box 33: Variable selection}
\begin{Schunk}
\begin{Sinput}
> plot(2:(ncol(x)-1),GAMINs, type="b", xlab="NVARS", ylab="")
> for (lambda in c(seq(1e-5,.2, length=3),seq(0.21,.90, length=3),seq(0.91,1, length=10))){
+   par(new=T)
+   stpgaoutlist<-vector(mode="list")
+   ii=1
+ for (i in 2:(ncol(x)-1)){
+ stpgaoutlist[[ii]]<-GenAlgForSubsetSelectionNoTest(P=t(x[,-1]),
+         ntoselect=i,npop=100, 
+         nelite=10, mutprob=.5, mutintensity = 1,
+         niterations=100, minitbefstop=50,tabu=FALSE,
+         tabumemsize = 1,plotiters=FALSE,lambda=lambda,
+         errorstat="fitnessfuncforSTPGA",mc.cores=4)
+ ii=ii+1
+ }
+ GAMINs<-c()
+ ii=1
+ for (i in 2:(ncol(x)-1)){
+   GAMINs<-c(GAMINs,min(stpgaoutlist[[ii]]$`Best criterion values over iterarions`))
+   ii=ii+1
+ }
+ GAMINSmat<-cbind(GAMINSmat,GAMINs)
+ plot(2:(ncol(x)-1),GAMINs, type="b", xlab="NVARS", ylab="", axes=F)
+ }
\end{Sinput}
\end{Schunk}
\includegraphics{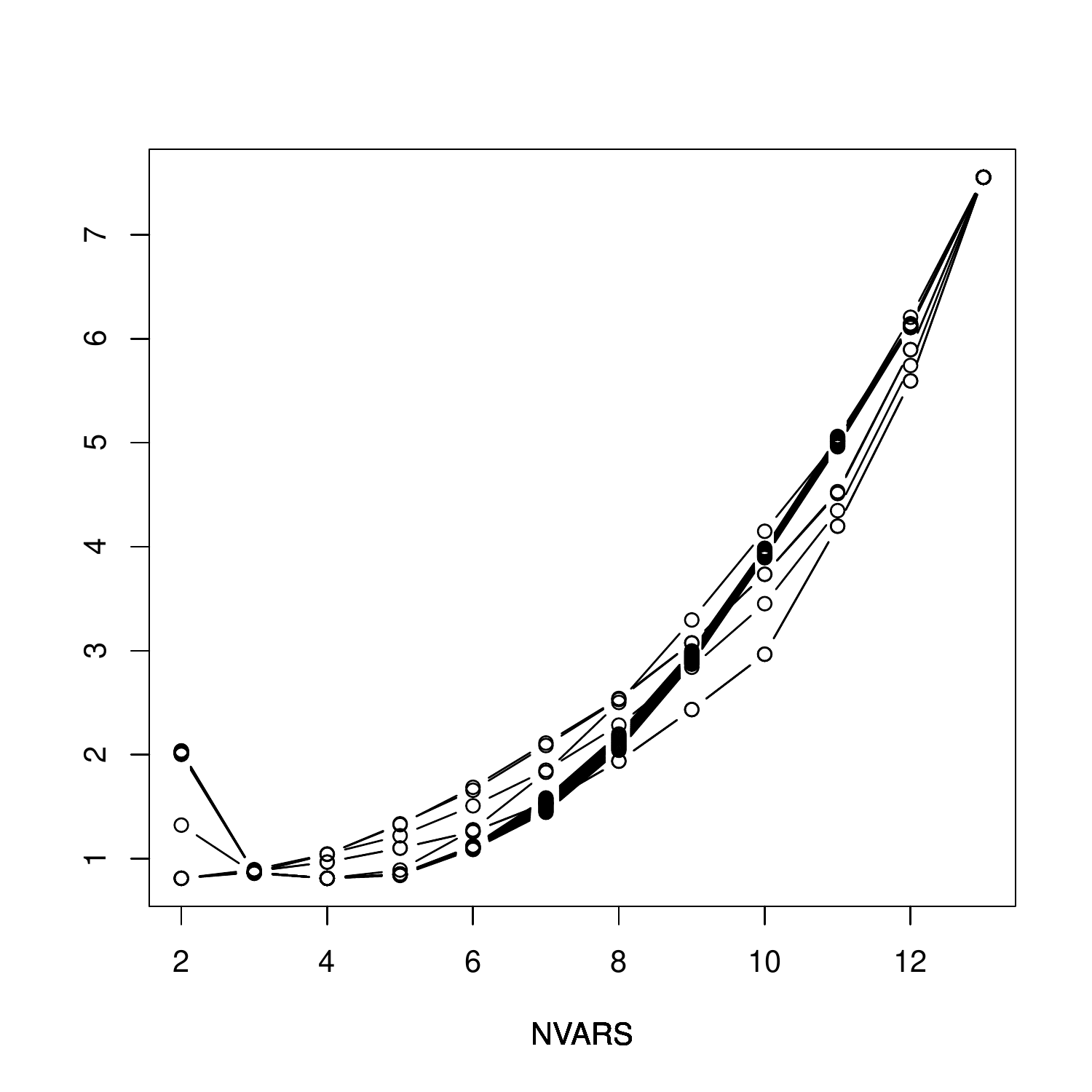}
\end{GrayBox}

\begin{GrayBox}
 \scriptsize
\textbf{Box 34: Variable selection- new criterion}
\begin{Schunk}
\begin{Sinput}
> apply(GAMINSmat, 2, which.min)
\end{Sinput}
\begin{Soutput}
GAMINs GAMINs GAMINs GAMINs GAMINs GAMINs GAMINs GAMINs GAMINs GAMINs GAMINs 
     1      1      1      1      1      3      3      3      3      3      3 
GAMINs GAMINs GAMINs GAMINs GAMINs GAMINs 
     3      3      3      3      3      3 
\end{Soutput}
\begin{Sinput}
> apply(GAMINSmat, 2, min)
\end{Sinput}
\begin{Soutput}
     GAMINs      GAMINs      GAMINs      GAMINs      GAMINs      GAMINs 
  0.8111831   0.8111216   0.1281011  -0.5696401  -0.6409756  -4.4104119 
     GAMINs      GAMINs      GAMINs      GAMINs      GAMINs      GAMINs 
 -9.0394933  -9.1736695  -9.3078458  -9.4420221  -9.5761984  -9.7103746 
     GAMINs      GAMINs      GAMINs      GAMINs      GAMINs 
 -9.8445509  -9.9787272 -10.1129034 -10.2470797 -10.3812560 
\end{Soutput}
\begin{Sinput}
> newselcritout<-GenAlgForSubsetSelectionNoTest(P=t(x[,-1]),
+         ntoselect=5,npop=200, 
+         nelite=10, mutprob=.5, mutintensity = 1,
+         niterations=200, minitbefstop=50,tabu=FALSE,
+         tabumemsize = 1,plotiters=F,lambda=.1,
+         errorstat="fitnessfuncforSTPGA",mc.cores=4)
> newselcritout$`Solution with rank 1`
\end{Sinput}
\begin{Soutput}
[1] "forearm" "ankle"   "height"  "chest"   "wrist"  
\end{Soutput}
\end{Schunk}
\end{GrayBox}

Note that the design criterion above defines a class of optimal solutions whose criterion values that can be plotted as a 3-dimensional surface for changing $\lambda$ and $n;$ this is similar to the frontier curve discussed above for balancing gains and inbreeding problem. In general, any multiobjective optimization problem defines a surface of Pareto optimal points that is called the frontier surface which may or may not be continuous. Frontier surfaces are inspected for identifying a good solution among and in the case of variable selection should be useful for identifying a single model from the class of optimal solutions.

It is easy to make criteria like the one I have made up. Here is another one $$\widehat{X^{(k)}}=\underset{X^{(k)} \in \mathbf{X}^{(k)}}{\mathrm{argmin}}(\by(I-X^{(k)}({X^{(k)}}'X^{(k)})^{-}{X^{(k)}}')\by+\lambda trace ({X^{(k)}}'X^{(k)})^{-1}$$ and  $\widehat{\bbeta^{(k)}}=(\widehat{X^{(k)}}'\widehat{X^{(k)}})^{-1}\widehat{X^{(k)}}'\by;$ and another one: $$\widehat{X^{(k)}}=\underset{X^{(k)} \in \mathbf{X}^{(k)}}{\mathrm{argmin}}(\by(I-X^{(k)}({X^{(k)}}'X^{(k)})^{-}{X^{(k)}}')\by+\lambda trace X^{(k)}({X^{(k)}}'X^{(k)})^{-1}{X^{(k)}}'$$ and  $\widehat{\bbeta^{(k)}}=(\widehat{X^{(k)}}'\widehat{X^{(k)}})^{-1}\widehat{X^{(k)}}'\by.$ Both of these have nice interpretation as well. In fact, any optimal design criteria of Section \ref{sec:crit} could be relevant in this context, for example, try $$\widehat{X^{(k)}}=\underset{X^{(k)} \in \mathbf{X}^{(k)}}{\mathrm{argmin}}(\by(I-X^{(k)}({X^{(k)}}'X^{(k)})^{-}{X^{(k)}}')\by+\lambda trace X^{*(k)}({X^{(k)}}'X^{(k)})^{-1}{X^{*(k)}}',$$ and  $\widehat{\bbeta^{(k)}}=(\widehat{X^{(k)}}'\widehat{X^{(k)}})^{-1}\widehat{X^{(k)}}'\by.$ I encourage the user to program these criterion or their own, perhaps the one you invent will make more sense for the problem that you are trying to solve than this or that other criterion.

The main point of these examples is that variable selection problem can be viewed as a multiobjective optimization problem, where a loss function and a penalty function are being simultaneously optimized. The loss function measures the fit of data to the training data and its optimal value for the optimal fixed sized set of variables improves as more variables are added to that model. The penalty function measures the overall quality of the model. Usually deteriorates as more variables are added to a model. In general, we can assume that the loss function and the penalty function are conflicting objectives, i.e.,  there does not exist a single solution that simultaneously optimizes both functions. This leads to a possibly infinite number of Pareto optimal solutions. The set of non-dominated solutions (none of the objective functions can be improved in value without degrading some other) are called Pareto optimal, and define a surface called the Pareto frontier.

\textbf{Identifying influential observations in regression (and keeping the most consistent regression data)}

When a regression model is fitted to data, if a few of the observations are different in some way from the bulk of the data then using all observations into the model might not be appropriate. The inclusion or exclusion of a few of these observations make a significant change in the parameter estimates or predictions. These are referred to as influential observations. One statistic that is used to identify the influence of an observation is the so called DFBETAS which measures the effect of deletion of single observations on the estimated model coefficients. This measure can be generalized to deletion of a group of individuals as  \[DFBETAS(X_{(-)})=\frac{1}{\widehat{{\sigma_{(-)}^2}}}(\widehat{\bbeta}-\widehat{\bbeta_{(-)}})'({X_{(-)}}'{X_{(-)}}')(\widehat{\bbeta}-\widehat{\bbeta_{(-)}}),\] where $\widehat{\bbeta}$ is the estimated regression coefficients from full data, $\widehat{\bbeta_{(-)}}$ is the same estimated using a part of the data (leaving out its complement from the analysis),  $\widehat{{\sigma_{(-)}}^2}$ is the residual variance estimate obtained using only partial data and ${X_{(-)}}$ is the design matrix for partial data. Here is the application of this to the ''iris'' data set (available with base \proglang{R}) to locate the influential observations if there are any:

\begin{GrayBox}
 \scriptsize
\textbf{Box 35: Deleting influential obs in regression}
\begin{Schunk}
\begin{Sinput}
> library(STPGA)
> STPGAUSERFUNCI<-function(Train,Test=NULL, P, lambda=1e-6, C=NULL){
+   PTrain<-P[rownames(P)%in%Train,]
+   PtP<-crossprod(cbind(1,P[,-1]))
+   PTtPT<-crossprod(cbind(1,PTrain[,-1]))
+   B<-solve(PtP+lambda*diag(nrow(PtP)),crossprod(cbind(1,P[,-1]),P[,1]))
+   BT<-solve(PTtPT+lambda*diag(nrow(PtP)),
+             crossprod(cbind(1,PTrain[,-1]),PTrain[,1]))
+   resT<-PTrain[,1]-cbind(1,PTrain[,-1])%*%BT
+   meanD<-mean((1/sd(resT))*diag(crossprod((B-BT),
+             (PTtPT+lambda*diag(nrow(PtP)))%*%(B-BT))))
+   return(-meanD)
+ }
> data(iris)
> P<-as.matrix(scale(iris[,1:4], center=TRUE, scale=TRUE))
> rownames(P)<-rownames(iris)
> STPGAoutlist<-vector(mode="list", length=20)
> ii=1
> for (i in 135:149){
+ STPGAoutlist[[ii]]<-GenAlgForSubsetSelectionNoTest(P=P,ntoselect=i, 
+                 mutprob = .8, npop = 100, nelite = 5, keepbest = TRUE,
+                 tabu = F, tabumemsize = 0, mutintensity = 1,plotiters=FALSE,
+                 niterations=200, minitbefstop = 50,
+                 errorstat = "STPGAUSERFUNCI", mc.cores=4)
+ ii=ii+1
+ }
> minsvec<-c()
> ii=1
> for (i in 135:149){
+   minsvec<-c(minsvec,
+              min(STPGAoutlist[[ii]]$`Best criterion values over iterarions`))
+ ii=ii+1
+ }
\end{Sinput}
\end{Schunk}
\end{GrayBox}

\begin{GrayBox}
 \scriptsize
\textbf{Box 36: Deleting influential obs in regression}
\begin{Schunk}
\begin{Sinput}
> plot(135:150,c(minsvec,0), type="b")
\end{Sinput}
\end{Schunk}
\includegraphics{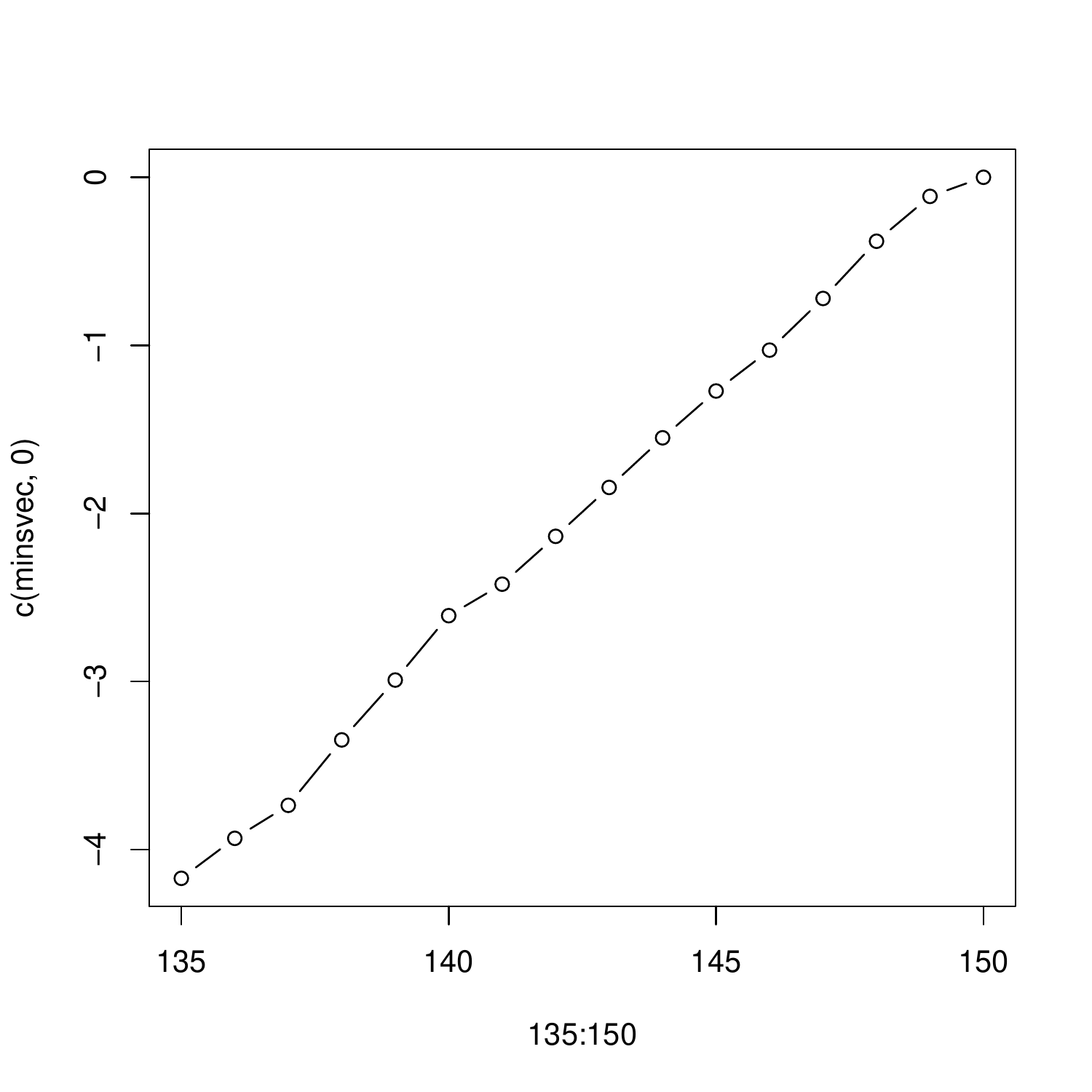}
\end{GrayBox}

What we are looking in the above graph is a point of sharp increase. I don't see such clear change point,  so let's replace replace six observations' values so that they are different than the rest:

\begin{GrayBox}
 \scriptsize
\textbf{Box 37: Deleting influential observations in regression}
\begin{Schunk}
\begin{Sinput}
> data(iris)
> P<-as.matrix(scale(iris[,1:4], center=TRUE, scale=TRUE))
> rownames(P)<-rownames(iris)
> P[c(5,30,55,80,105,130),c(1:4)]<-
+   1.5*P[c(5,30,55,80,105,130),c(1:4)]%*%matrix(rnorm(16),nrow=4, ncol=4)
> STPGAoutlist<-vector(mode="list", length=20)
> ii=1
> for (i in 135:149){
+ STPGAoutlist[[ii]]<-GenAlgForSubsetSelectionNoTest(P=P,ntoselect=i, 
+                  mutprob = .8, npop = 100, nelite = 5, keepbest = TRUE,
+                 tabu = F, tabumemsize = 0, mutintensity = 1,plotiters=FALSE,
+                 niterations=200, minitbefstop = 50,
+                 errorstat = "STPGAUSERFUNCI", mc.cores=4)
+ ii=ii+1
+ }
> minsvec<-c()
> ii=1
> for (i in 135:149){
+   minsvec<-c(minsvec,
+              min(STPGAoutlist[[ii]]$`Best criterion values over iterarions`))
+ ii=ii+1
+ }
\end{Sinput}
\end{Schunk}
\end{GrayBox}

The following plot shows that the criterion value shows a sharp increase when we go from 144 to 145 observations and there are total of 150 observations in the data set, six observations that aren't included in the set of 144 observations cause large change in the value of the estimated coefficients. We observe that these are the observations we have manipulated. 
\begin{GrayBox}
 \scriptsize
\textbf{Box 38: Deleting influential observations in regression}
\begin{Schunk}
\begin{Sinput}
> par(mfrow=c(2,1))
> twopcsiris<-(P%*%svd(P, nu=0,nv=2)$v)
> plot(twopcsiris[,1],twopcsiris[,2], col=(1:150)%in%c(5,30,55,80,105,130)+1)
> plot(135:150,c(minsvec,0), type="b")
> par(mfrow=c(1,1))
> STPGAout<-GenAlgForSubsetSelectionNoTest(P=P,ntoselect=144, 
+                  mutprob = .8, npop = 50, nelite = 5, keepbest = TRUE,
+                 tabu = F, tabumemsize = 0, mutintensity = 1,plotiters=FALSE,
+                 niterations=200, minitbefstop=50,
+                 errorstat = "STPGAUSERFUNCI", mc.cores=4)
> intersect(STPGAout$`Solution with rank 1`, as.character(c(5,30,55,80,105,130)))
\end{Sinput}
\begin{Soutput}
[1] "55" "5" 
\end{Soutput}
\end{Schunk}
\includegraphics{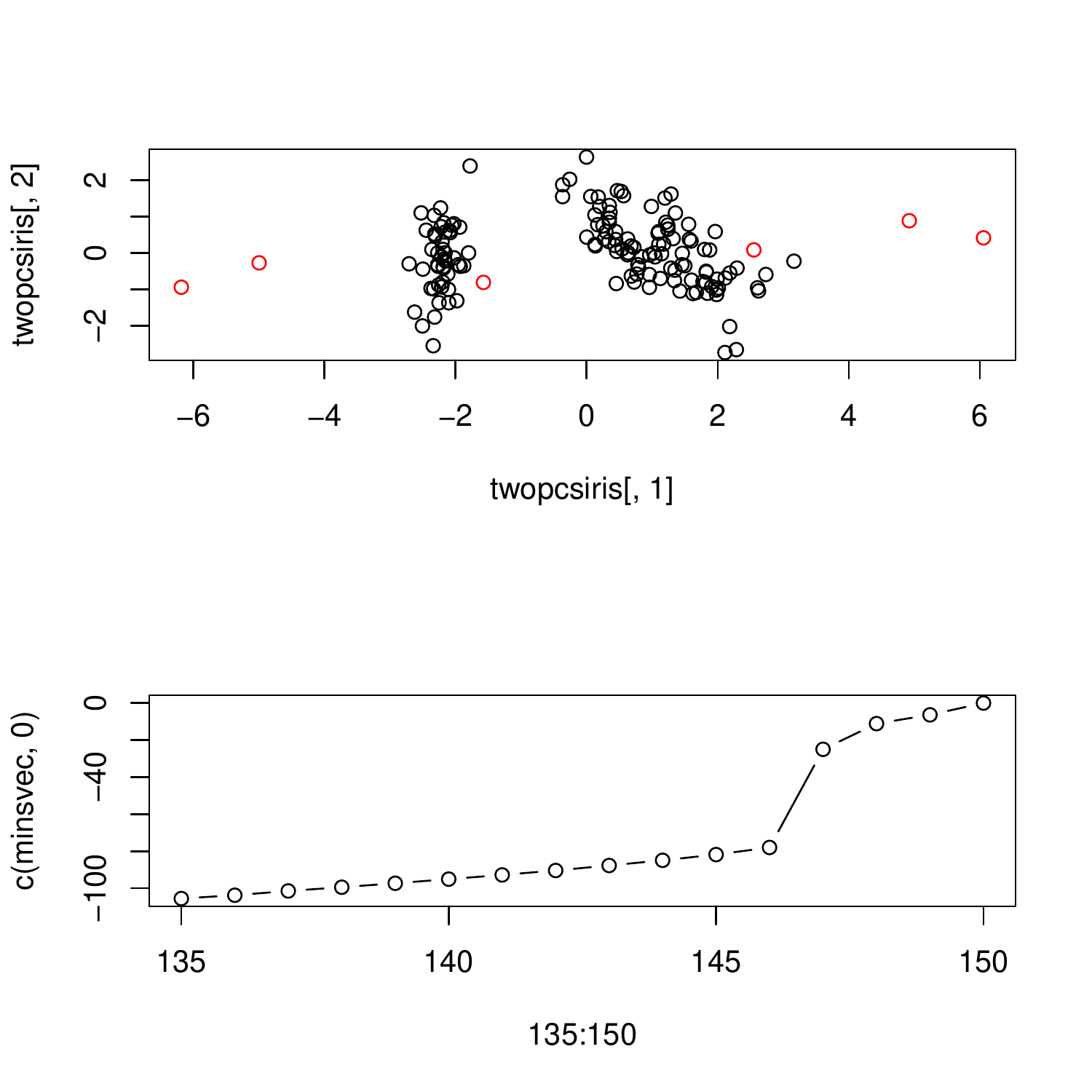}
\end{GrayBox}

\section{Concluding Remarks}
\label{sec:concluding}

I hope that I have provided enough examples to show that generic optimal subset selection problem is a useful tool. The list of examples and applications can easily be extended. However, I think this should take some time to digest. 

In STPGA package I have implemented a GA algorithm for subset selection which was inspired by the recent developments in the genomic breeding of plants, animals and other organisms. The main advantage of this algorithm is that it can be run in parallel to solve the subset selection problem for any given objective function. 

I also did not compare the relative speed of this particular implementation of LA-GA-T algorithm to any other software.  I admit that this software could be written so that it functioned faster using one processor and, no matter what I do, it will never work as fast as some other algorithms that are specialized in their tasks. However, I note that LA-GA-T algorithm can be run in parallel on many computers to solve problems as fast as some problem specific algorithms that have to be performed in serial. 

Prediction accuracy from genomic models can be improved by targeting more informative individuals in the data set used to generate the predictors, this result has been exemplified by several papers in the area (\cite{akdemir2015optimization, yu2016genomic}. Nevertheless, the subject  deserves more attention. 
 
Some of the criteria I have mentioned in the selection of training populations section of this paper are not among the default criteria listed in Table \ref{tab:STPGAOPTFUNCS}, however, I have included them as user defined criteria in the help files that is provided with the package.

I have plans to migrate all of this program to a more efficient programming language for performance improvements, and there is room for improvement redesigning parts of the algorithm for   However, there are some advantages of using \proglang{R}. Some of these include like the accessibility, availability, being able to define functions on the fly. Therefore, I will still be supporting this pure \proglang{R} version. As a statistician, I have the feeling that writing efficient software is partially out of my expertise, immediate interests and surely can be done much easier with collaboration with people with the right skills.  As of now, I can use STPGA with moderately complex problems, either by allowing the algorithm to take its time or by using a highly parallelizable machine. I am open to any suggestions, collaborations in this respect.

\section*{Acknowledgements} 
I am grateful to family, Mehmet Ali and G{\"u}ler and {\"U}mit {\"O}zg{\"u}r, Pelin May , Kristy Akdemir.

My work was also supported by tax payers through the grants: R01GM099992, R01GM101219 and USDA-NIFA-AFRI Triticeae Coordinated Agricultural Project- award number 2011-68002-30029.


\section*{Author contributions} 
Every idea, code in this manuscript was conceived by Deniz Akdemir unless they were acknowledged by the references within the article. If a person or a group makes any claims that they have contributed to this work in any way it should not be taken seriously. 

\bibliography{references.bib}

\begin{thebibliography}{82}
\newcommand{\enquote}[1]{``#1''}
\providecommand{\natexlab}[1]{#1}
\providecommand{\url}[1]{\texttt{#1}}
\providecommand{\urlprefix}{URL }
\expandafter\ifx\csname urlstyle\endcsname\relax
  \providecommand{\doi}[1]{doi:\discretionary{}{}{}#1}\else
  \providecommand{\doi}{doi:\discretionary{}{}{}\begingroup
  \urlstyle{rm}\Url}\fi
\providecommand{\eprint}[2][]{\url{#2}}

\bibitem[{Akaike(1974)}]{akaike1974new}
Akaike H (1974).
\newblock \enquote{A new look at the statistical model identification.}
\newblock \emph{IEEE transactions on automatic control}, \textbf{19}(6),
  716--723.

\bibitem[{Akdemir and Godfrey(2015)}]{akdemir2015EMMREML}
Akdemir D, Godfrey OU (2015).
\newblock \emph{EMMREML: Fitting Mixed Models with Known Covariance
  Structures}.
\newblock R package version 3.1,
  \urlprefix\url{https://CRAN.R-project.org/package=EMMREML}.

\bibitem[{Akdemir and Jannink(2015)}]{akdemir2015locally}
Akdemir D, Jannink JL (2015).
\newblock \enquote{Locally epistatic genomic relationship matrices for genomic
  association and prediction.}
\newblock \emph{Genetics}, \textbf{199}(3), 857--871.

\bibitem[{Akdemir and S{\'a}nchez(2016)}]{akdemir2016efficient}
Akdemir D, S{\'a}nchez JI (2016).
\newblock \enquote{Efficient Breeding by Genomic Mating.}
\newblock \emph{Frontiers in Genetics}, \textbf{7}.

\bibitem[{Akdemir \emph{et~al.}(2015)Akdemir, Sanchez, and
  Jannink}]{akdemir2015optimization}
Akdemir D, Sanchez JI, Jannink JL (2015).
\newblock \enquote{Optimization of genomic selection training populations with
  a genetic algorithm.}
\newblock \emph{Genet Sel Evol}, \textbf{47}, 38.

\bibitem[{Amdahl(1967)}]{amdahl1967validity}
Amdahl GM (1967).
\newblock \enquote{Validity of the single processor approach to achieving large
  scale computing capabilities.}
\newblock In \emph{Proceedings of the April 18-20, 1967, spring joint computer
  conference}, pp. 483--485. ACM.

\bibitem[{Ardia \emph{et~al.}(2016)Ardia, Mullen, Peterson, Ulrich, Boudt, and
  Mullen}]{ardia2016package}
Ardia D, Mullen K, Peterson B, Ulrich J, Boudt K, Mullen MK (2016).
\newblock \enquote{Package ‘DEoptim’.}

\bibitem[{Atkinson and Donev(1992)}]{atkinson1992optimum}
Atkinson A, Donev A (1992).
\newblock \enquote{Optimum experimental designs. Clarendon.}

\bibitem[{Box \emph{et~al.}(1978)Box, Hunter, and Hunter}]{box1978statistics}
Box GE, Hunter WG, Hunter JS (1978).
\newblock \emph{Statistics for experimenters: an introduction to design, data
  analysis, and model building}, volume~1.
\newblock JSTOR.

\bibitem[{Bozdogan(1987)}]{bozdogan1987icomp}
Bozdogan H (1987).
\newblock \enquote{ICOMP: A new model-selection criterion.}
\newblock In \emph{1. Conference of the International Federation of
  Classification Societies}, pp. 599--608.

\bibitem[{Brisbane and Gibson(1995)}]{brisbane1995balancing}
Brisbane J, Gibson J (1995).
\newblock \enquote{Balancing selection response and rate of inbreeding by
  including genetic relationships in selection decisions.}
\newblock \emph{Theoretical and Applied Genetics}, \textbf{91}(3), 421--431.

\bibitem[{Burton \emph{et~al.}(2007)Burton, Clayton, Cardon, Craddock,
  Deloukas, Duncanson, Kwiatkowski, McCarthy, Ouwehand, Samani
  \emph{et~al.}}]{burton2007genome}
Burton PR, Clayton DG, Cardon LR, Craddock N, Deloukas P, Duncanson A,
  Kwiatkowski DP, McCarthy MI, Ouwehand WH, Samani NJ, \emph{et~al.} (2007).
\newblock \enquote{Genome-wide association study of 14,000 cases of seven
  common diseases and 3,000 shared controls.}
\newblock \emph{Nature}, \textbf{447}(7145), 661--678.

\bibitem[{Chun and Kele{\c{s}}(2010)}]{chun2010sparse}
Chun H, Kele{\c{s}} S (2010).
\newblock \enquote{Sparse partial least squares regression for simultaneous
  dimension reduction and variable selection.}
\newblock \emph{Journal of the Royal Statistical Society: Series B (Statistical
  Methodology)}, \textbf{72}(1), 3--25.

\bibitem[{Cressie(1988)}]{Cressie:1988}
Cressie N (1988).
\newblock \enquote{Spatial prediction and ordinary kriging.}
\newblock \emph{Mathematical Geology}, \textbf{20}(4), 405--421.
\newblock ISSN 0882-8121.
\newblock \doi{10.1007/BF00892986}.
\newblock \urlprefix\url{http://dx.doi.org/10.1007/BF00892986}.

\bibitem[{Crossa \emph{et~al.}(2010)Crossa, de~los Campos, P\'erez, Gianola,
  Burgueno, Araus, Makumbi, Singh, Dreisigacker, Yan, Arief, Banziger, and
  Braun}]{Crossa:2010}
Crossa J, de~los Campos G, P\'erez P, Gianola D, Burgueno J, Araus JL, Makumbi
  D, Singh RP, Dreisigacker S, Yan JB, Arief V, Banziger M, Braun HJ (2010).
\newblock \enquote{Prediction of Genetic Values of Quantitative Traits in Plant
  Breeding Using Pedigree and Molecular Markers.}
\newblock \emph{Genetics}, \textbf{186}(2), 713--U406.

\bibitem[{Crossa \emph{et~al.}(2016)Crossa, Jarqu{\'\i}n, Franco,
  P{\'e}rez-Rodr{\'\i}guez, Burgue{\~n}o, Saint-Pierre, Vikram, Sansaloni,
  Petroli, Akdemir \emph{et~al.}}]{crossa2016genomic}
Crossa J, Jarqu{\'\i}n D, Franco J, P{\'e}rez-Rodr{\'\i}guez P, Burgue{\~n}o J,
  Saint-Pierre C, Vikram P, Sansaloni C, Petroli C, Akdemir D, \emph{et~al.}
  (2016).
\newblock \enquote{Genomic prediction of gene bank wheat landraces.}
\newblock \emph{G3: Genes| Genomes| Genetics}, \textbf{6}(7), 1819--1834.

\bibitem[{Daetwyler \emph{et~al.}(2013)Daetwyler, Calus, Pong-Wong, de~los
  Campos, and Hickey}]{daetwyler2013genomic}
Daetwyler HD, Calus MP, Pong-Wong R, de~los Campos G, Hickey JM (2013).
\newblock \enquote{Genomic prediction in animals and plants: simulation of
  data, validation, reporting, and benchmarking.}
\newblock \emph{Genetics}, \textbf{193}(2), 347--365.

\bibitem[{Draper and Pukelsheim(1996)}]{draper1996overview}
Draper NR, Pukelsheim F (1996).
\newblock \enquote{An overview of design of experiments.}
\newblock \emph{Statistical Papers}, \textbf{37}(1), 1--32.

\bibitem[{Fedorov(1972)}]{fedorov1972theory}
Fedorov VV (1972).
\newblock \emph{Theory of optimal experiments}.
\newblock Elsevier.

\bibitem[{Fisher(1960)}]{fisher1960design}
Fisher RA (1960).
\newblock \enquote{The design of experiments. ed.}
\newblock \emph{New York: Hafner}.

\bibitem[{Fisher(1992)}]{fisher1992arrangement}
Fisher RA (1992).
\newblock \enquote{The arrangement of field experiments.}
\newblock In \emph{Breakthroughs in statistics}, pp. 82--91. Springer.

\bibitem[{Furnival and Wilson(1974)}]{furnival1974regressions}
Furnival GM, Wilson RW (1974).
\newblock \enquote{Regressions by leaps and bounds.}
\newblock \emph{Technometrics}, \textbf{16}(4), 499--511.

\bibitem[{Gianola \emph{et~al.}(2006)Gianola, Fernando, and
  Stella}]{Gianola:2006}
Gianola D, Fernando RL, Stella A (2006).
\newblock \enquote{Genomic-assisted Prediction of Genetic Value with
  Semiparametric Procedures.}
\newblock \emph{Genetics}, \textbf{173}(3), 1761--1776.

\bibitem[{Goddard(2009)}]{goddard2009genomic}
Goddard M (2009).
\newblock \enquote{Genomic selection: prediction of accuracy and maximisation
  of long term response.}
\newblock \emph{Genetics}, \textbf{136}(2), 245--257.

\bibitem[{Goldberg and Holland(1988)}]{goldberg1988genetic}
Goldberg DE, Holland JH (1988).
\newblock \enquote{Genetic algorithms and machine learning.}
\newblock \emph{Machine learning}, \textbf{3}(2), 95--99.

\bibitem[{Gonz\'alez-Recio \emph{et~al.}(2008)Gonz\'alez-Recio, Gianola, Long,
  Weigel, Rosa, and Avendano}]{Gonzalez:2008}
Gonz\'alez-Recio O, Gianola D, Long N, Weigel KA, Rosa GJM, Avendano S (2008).
\newblock \enquote{Nonparametric Methods for Incorporating Genomic Information
  Into Genetic Evaluations: An Application to Mortality in Broilers.}
\newblock \emph{Genetics}, \textbf{178}(4), 2305--2313.
\newblock \doi{10.1534/genetics.107.084293}.
\newblock \eprint{http://www.genetics.org/content/178/4/2305.full.pdf+html},
  \urlprefix\url{http://www.genetics.org/content/178/4/2305.abstract}.

\bibitem[{Gruska(1999)}]{gruska1999quantum}
Gruska J (1999).
\newblock \emph{Quantum computing}, volume 2005.
\newblock McGraw-Hill London.

\bibitem[{Haines(1987)}]{haines1987application}
Haines LM (1987).
\newblock \enquote{The application of the annealing algorithm to the
  construction of exact optimal designs for linear--regression models.}
\newblock \emph{Technometrics}, \textbf{29}(4), 439--447.

\bibitem[{Hayes \emph{et~al.}(2009)Hayes, Bowman, Chamberlain, and
  Goddard}]{Hayes:2009}
Hayes B, Bowman P, Chamberlain A, Goddard M (2009).
\newblock \enquote{Invited review: Genomic selection in dairy cattle: Progress
  and challenges.}
\newblock \emph{Journal of Dairy Science}, \textbf{92}(2), 433 -- 443.
\newblock ISSN 0022-0302.

\bibitem[{Henderson(1953)}]{henderson1953estimation}
Henderson CR (1953).
\newblock \enquote{Estimation of variance and covariance components.}
\newblock \emph{Biometrics}, \textbf{9}(2), 226--252.

\bibitem[{Henderson(1975)}]{Henderson:1975}
Henderson CR (1975).
\newblock \enquote{Best linear Unbiased Estimation and Prediction Under a
  Selection Model.}
\newblock \emph{Biometrics}, \textbf{31}(2), 423--447.

\bibitem[{Henderson \emph{et~al.}(1959)Henderson, Kempthorne, Searle, and
  Von~Krosigk}]{henderson1959estimation}
Henderson CR, Kempthorne O, Searle SR, Von~Krosigk C (1959).
\newblock \enquote{The estimation of environmental and genetic trends from
  records subject to culling.}
\newblock \emph{Biometrics}, \textbf{15}(2), 192--218.

\bibitem[{Holland(1992{\natexlab{a}})}]{holland1992adaptation}
Holland JH (1992{\natexlab{a}}).
\newblock \emph{Adaptation in natural and artificial systems: an introductory
  analysis with applications to biology, control, and artificial intelligence}.
\newblock MIT Press.

\bibitem[{Holland(1992{\natexlab{b}})}]{holland1992genetic}
Holland JH (1992{\natexlab{b}}).
\newblock \enquote{Genetic algorithms.}
\newblock \emph{Scientific american}, \textbf{267}(1), 66--72.

\bibitem[{Isidro \emph{et~al.}(2015)Isidro, Jannink, Akdemir, Poland, Heslot,
  and Sorrells}]{isidro2015training}
Isidro J, Jannink JL, Akdemir D, Poland J, Heslot N, Sorrells ME (2015).
\newblock \enquote{Training set optimization under population structure in
  genomic selection.}
\newblock \emph{Theoretical and Applied Genetics}, \textbf{128}(1), 145--158.

\bibitem[{Jannink(2010)}]{jannink2010dynamics}
Jannink JL (2010).
\newblock \enquote{Dynamics of long-term genomic selection.}
\newblock \emph{Genetics Selection Evolution}, \textbf{42}(1), 35.

\bibitem[{Kempthorne \emph{et~al.}(1957)}]{kempthorne1957introduction}
Kempthorne O, \emph{et~al.} (1957).
\newblock \enquote{An introduction to genetic statistics.}
\newblock \emph{An introduction to genetic statistics.}

\bibitem[{Kiefer(1959)}]{kiefer1959optimum}
Kiefer J (1959).
\newblock \enquote{Optimum experimental designs.}
\newblock \emph{Journal of the Royal Statistical Society. Series B
  (Methodological)}, pp. 272--319.

\bibitem[{Kiefer \emph{et~al.}(1985)Kiefer, Brown, Olkin, and
  Sacks}]{kiefer1985jack}
Kiefer JC, Brown L, Olkin I, Sacks J (1985).
\newblock \emph{Jack Carl Kiefer Collected Papers: Design of Experiments}.
\newblock Springer.

\bibitem[{Lalo{\"e}(1993)}]{laloe1993precision}
Lalo{\"e} D (1993).
\newblock \enquote{Precision and information in linear models of genetic
  evaluation.}
\newblock \emph{Genetics Selection Evolution}, \textbf{25}(6), 557--576.

\bibitem[{Lalo{\"e} and Phocas(2003)}]{Laloe2003241}
Lalo{\"e} D, Phocas F (2003).
\newblock \enquote{A proposal of criteria of robustness analysis in genetic
  evaluation.}
\newblock \emph{Livestock Production Science}, \textbf{80}(3), 241 -- 256.
\newblock ISSN 0301-6226.
\newblock \doi{http://dx.doi.org/10.1016/S0301-6226(02)00092-1}.
\newblock
  \urlprefix\url{//www.sciencedirect.com/science/article/pii/S0301622602000921}.

\bibitem[{Leuenberger and Loss(2001)}]{leuenberger2001quantum}
Leuenberger MN, Loss D (2001).
\newblock \enquote{Quantum computing in molecular magnets.}
\newblock \emph{Nature}, \textbf{410}(6830), 789--793.

\bibitem[{Liu \emph{et~al.}(2000)Liu, Wang, Frutos, Condon, Corn, and
  Smith}]{liu2000dna}
Liu Q, Wang L, Frutos AG, Condon AE, Corn RM, Smith LM (2000).
\newblock \enquote{DNA computing on surfaces.}
\newblock \emph{Nature}, \textbf{403}(6766), 175--179.

\bibitem[{Makowsky \emph{et~al.}(2011)Makowsky, Pajewski, Klimentidis, Vazquez,
  Duarte, Allison, and de~Los~Campos}]{makowsky2011beyond}
Makowsky R, Pajewski NM, Klimentidis YC, Vazquez AI, Duarte CW, Allison DB,
  de~Los~Campos G (2011).
\newblock \enquote{Beyond missing heritability: prediction of complex traits.}
\newblock \emph{PLoS Genet}, \textbf{7}(4), e1002051.

\bibitem[{Markowitz(1952)}]{markowitz1952portfolio}
Markowitz H (1952).
\newblock \enquote{Portfolio selection.}
\newblock \emph{The journal of finance}, \textbf{7}(1), 77--91.

\bibitem[{Mebane~Jr \emph{et~al.}(2015)Mebane~Jr, Sekhon, and
  Sekhon}]{mebane2015package}
Mebane~Jr WR, Sekhon JS, Sekhon MJS (2015).
\newblock \enquote{Package ‘rgenoud’.}

\bibitem[{Meuwissen(1997)}]{meuwissen1997maximizing}
Meuwissen T (1997).
\newblock \enquote{Maximizing the response of selection with a predefined rate
  of inbreeding.}
\newblock \emph{Journal of animal science}, \textbf{75}(4), 934--940.

\bibitem[{Meuwissen \emph{et~al.}(2001)Meuwissen, Hayes, and
  Goddard}]{Meuwissen:2001}
Meuwissen THE, Hayes BJ, Goddard ME (2001).
\newblock \enquote{Prediction of Total Genetic Value Using Genome-Wide Dense
  Marker Maps.}
\newblock \emph{Genetics}, \textbf{157}(4), 1819--1829.

\bibitem[{Mitchell(1974)}]{mitchell1974algorithm}
Mitchell TJ (1974).
\newblock \enquote{An algorithm for the construction of “D-optimal”
  experimental designs.}
\newblock \emph{Technometrics}, \textbf{16}(2), 203--210.

\bibitem[{Nguyen and Miller(1992)}]{nguyen1992review}
Nguyen NK, Miller AJ (1992).
\newblock \enquote{A review of some exchange algorithms for constructing
  discrete D-optimal designs.}
\newblock \emph{Computational Statistics \& Data Analysis}, \textbf{14}(4),
  489--498.

\bibitem[{Paun \emph{et~al.}(2005)Paun, Rozenberg, and Salomaa}]{paun2005dna}
Paun G, Rozenberg G, Salomaa A (2005).
\newblock \emph{DNA computing: new computing paradigms}.
\newblock Springer Science \& Business Media.

\bibitem[{Pritchard and Bacon(1978)}]{pritchard1978prospects}
Pritchard DJ, Bacon DW (1978).
\newblock \enquote{Prospects for reducing correlations among parameter
  estimates in kinetic models.}
\newblock \emph{Chemical Engineering Science}, \textbf{33}(11), 1539--1543.

\bibitem[{Pronzato(2008)}]{pronzato2008optimal}
Pronzato L (2008).
\newblock \enquote{Optimal experimental design and some related control
  problems.}
\newblock \emph{Automatica}, \textbf{44}(2), 303--325.

\bibitem[{Pronzato and M{\"u}ller(2012)}]{pronzato2012design}
Pronzato L, M{\"u}ller WG (2012).
\newblock \enquote{Design of computer experiments: space filling and beyond.}
\newblock \emph{Statistics and Computing}, \textbf{22}(3), 681--701.

\bibitem[{Pryce \emph{et~al.}(2012)Pryce, Hayes, and Goddard}]{pryce2012novel}
Pryce J, Hayes B, Goddard M (2012).
\newblock \enquote{Novel strategies to minimize progeny inbreeding while
  maximizing genetic gain using genomic information.}
\newblock \emph{Journal of dairy science}, \textbf{95}(1), 377--388.

\bibitem[{Pukelsheim(2006)}]{pukelsheim2006optimal}
Pukelsheim F (2006).
\newblock \emph{Optimal design of experiments}.
\newblock SIAM.

\bibitem[{Rietveld \emph{et~al.}(2013)Rietveld, Medland, Derringer, Yang, Esko,
  Martin, Westra, Shakhbazov, Abdellaoui, Agrawal
  \emph{et~al.}}]{rietveld2013gwas}
Rietveld CA, Medland SE, Derringer J, Yang J, Esko T, Martin NW, Westra HJ,
  Shakhbazov K, Abdellaoui A, Agrawal A, \emph{et~al.} (2013).
\newblock \enquote{GWAS of 126,559 individuals identifies genetic variants
  associated with educational attainment.}
\newblock \emph{science}, \textbf{340}(6139), 1467--1471.

\bibitem[{Rincent \emph{et~al.}(2012)Rincent, Lalo{\"e}, Nicolas, Altmann,
  Brunel, Revilla, Rodriguez, Moreno-Gonzalez, Melchinger, Bauer
  \emph{et~al.}}]{rincent2012maximizing}
Rincent R, Lalo{\"e} D, Nicolas S, Altmann T, Brunel D, Revilla P, Rodriguez
  VM, Moreno-Gonzalez J, Melchinger A, Bauer E, \emph{et~al.} (2012).
\newblock \enquote{Maximizing the reliability of genomic selection by
  optimizing the calibration set of reference individuals: comparison of
  methods in two diverse groups of maize inbreds (Zea mays L.).}
\newblock \emph{Genetics}, \textbf{192}(2), 715--728.

\bibitem[{Risch \emph{et~al.}(1996)Risch, Merikangas
  \emph{et~al.}}]{risch1996future}
Risch N, Merikangas K, \emph{et~al.} (1996).
\newblock \enquote{The future of genetic studies of complex human diseases.}
\newblock \emph{Science}, \textbf{273}(5281), 1516--1517.

\bibitem[{Satman(2013)}]{satman2013galts}
Satman M (2013).
\newblock \enquote{galts: Genetic algorithms and C-steps based LTS (Least
  Trimmed Squares) estimation.}
\newblock \emph{R package version}, \textbf{1}.

\bibitem[{Schierenbeck \emph{et~al.}(2011)Schierenbeck, Pimentel, Tietze,
  K{\"o}rte, Reents, Reinhardt, Simianer, and
  K{\"o}nig}]{schierenbeck2011controlling}
Schierenbeck S, Pimentel E, Tietze M, K{\"o}rte J, Reents R, Reinhardt F,
  Simianer H, K{\"o}nig S (2011).
\newblock \enquote{Controlling inbreeding and maximizing genetic gain using
  semi-definite programming with pedigree-based and genomic relationships.}
\newblock \emph{Journal of dairy science}, \textbf{94}(12), 6143--6152.

\bibitem[{Sch{\"o}lkopf and Smola(2002)}]{scholkopflearning}
Sch{\"o}lkopf B, Smola AJ (2002).
\newblock \emph{Learning with kernels: support vector machines, regularization,
  optimization, and beyond}.
\newblock MIT press.

\bibitem[{Schwarz \emph{et~al.}(1978)}]{schwarz1978estimating}
Schwarz G, \emph{et~al.} (1978).
\newblock \enquote{Estimating the dimension of a model.}
\newblock \emph{The annals of statistics}, \textbf{6}(2), 461--464.

\bibitem[{Scrucca \emph{et~al.}(2013)}]{scrucca2013ga}
Scrucca L, \emph{et~al.} (2013).
\newblock \enquote{GA: a package for genetic algorithms in R.}
\newblock \emph{Journal of Statistical Software}, \textbf{53}(4), 1--37.

\bibitem[{Sivanandam and Deepa(2007)}]{sivanandam2007introduction}
Sivanandam S, Deepa S (2007).
\newblock \emph{Introduction to genetic algorithms}.
\newblock Springer Science \& Business Media.

\bibitem[{Sun \emph{et~al.}(2013)Sun, VanRaden, O'Connell, Weigel, and
  Gianola}]{sun2013mating}
Sun C, VanRaden P, O'Connell J, Weigel K, Gianola D (2013).
\newblock \enquote{Mating programs including genomic relationships and
  dominance effects.}
\newblock \emph{Journal of dairy science}, \textbf{96}(12), 8014--8023.

\bibitem[{Tendys(2002)}]{tendys2002gafit}
Tendys T (2002).
\newblock \enquote{gafit: Genetic Algorithm for Curve Fitting.}
\newblock \emph{R package version 0.4, URL http://CRAN. R-project.
  org/src/contrib/Archive/gafit}.

\bibitem[{Tian \emph{et~al.}(2011)Tian, Bradbury, Brown, Hung, Sun,
  Flint-Garcia, Rocheford, McMullen, Holland, and Buckler}]{tian2011genome}
Tian F, Bradbury PJ, Brown PJ, Hung H, Sun Q, Flint-Garcia S, Rocheford TR,
  McMullen MD, Holland JB, Buckler ES (2011).
\newblock \enquote{Genome-wide association study of leaf architecture in the
  maize nested association mapping population.}
\newblock \emph{Nature genetics}, \textbf{43}(2), 159--162.

\bibitem[{Tibshirani(1996)}]{tibshirani1996regression}
Tibshirani R (1996).
\newblock \enquote{Regression shrinkage and selection via the lasso.}
\newblock \emph{Journal of the Royal Statistical Society. Series B
  (Methodological)}, pp. 267--288.

\bibitem[{Turlach and Weingessel(2007)}]{turlach2007quadprog}
Turlach BA, Weingessel A (2007).
\newblock \enquote{quadprog: Functions to solve quadratic programming
  problems.}
\newblock \emph{R package version}, pp. 1--4.

\bibitem[{VanRaden \emph{et~al.}(2009)VanRaden, Tassell, Wiggans, Sonstegard,
  Schnabel, Taylor, and Schenkel}]{VanRaden:2009}
VanRaden P, Tassell CV, Wiggans G, Sonstegard T, Schnabel R, Taylor J, Schenkel
  F (2009).
\newblock \enquote{Invited Review: Reliability of genomic predictions for North
  American Holstein bulls.}
\newblock \emph{Journal of Dairy Science}, \textbf{92}(1), 16 -- 24.
\newblock ISSN 0022-0302.

\bibitem[{VanRaden(2008)}]{VanRaden:2008}
VanRaden PM (2008).
\newblock \enquote{Efficient Methods to Compute Genomic Predictions.}
\newblock \emph{Journal of Dairy Science}, \textbf{91}(11), 4414--23.

\bibitem[{Vapnik(1998)}]{Vapnik:1998}
Vapnik V (1998).
\newblock \emph{Statistical learning theory}.
\newblock 1 edition. Wiley.
\newblock ISBN 0471030031.

\bibitem[{Verzani(2015)}]{usingr2015}
Verzani J (2015).
\newblock \emph{UsingR: Data Sets, Etc. for the Text "Using R for Introductory
  Statistics", Second Edition}.
\newblock R package version 2.0-5,
  \urlprefix\url{https://CRAN.R-project.org/package=UsingR}.

\bibitem[{Wahba(1990)}]{Wahba:1990}
Wahba G (1990).
\newblock \emph{Spline Models for Observational Data}.
\newblock Society for Industrial and Applied Mathematics.
\newblock \doi{10.1137/1.9781611970128}.
\newblock \eprint{http://epubs.siam.org/doi/pdf/10.1137/1.9781611970128},
  \urlprefix\url{http://epubs.siam.org/doi/abs/10.1137/1.9781611970128}.

\bibitem[{Welch(1982)}]{welch1982branch}
Welch WJ (1982).
\newblock \enquote{Branch-and-bound search for experimental designs based on D
  optimality and other criteria.}
\newblock \emph{Technometrics}, \textbf{24}(1), 41--48.

\bibitem[{Willighagen(2005)}]{willighagen2005genalg}
Willighagen E (2005).
\newblock \enquote{Genalg: R based genetic algorithm.}
\newblock \emph{R package version 0.1}, \textbf{1}.

\bibitem[{Witten \emph{et~al.}(2009)Witten, Tibshirani, and
  Hastie}]{witten2009penalized}
Witten DM, Tibshirani R, Hastie T (2009).
\newblock \enquote{A penalized matrix decomposition, with applications to
  sparse principal components and canonical correlation analysis.}
\newblock \emph{Biostatistics}, p. kxp008.

\bibitem[{Wray and Goddard(1994)}]{wray1994moet}
Wray N, Goddard M (1994).
\newblock \enquote{MOET breeding schemes for wool sheep 1. Design
  alternatives.}
\newblock \emph{Animal Production}, \textbf{59}(01), 71--86.

\bibitem[{Yang \emph{et~al.}(2010)Yang, Benyamin, McEvoy, Gordon, Henders,
  Nyholt, Madden, Heath, Martin, Montgomery, Goddard, and Visscher}]{Yang:2010}
Yang J, Benyamin B, McEvoy BP, Gordon S, Henders AK, Nyholt DR, Madden PA,
  Heath AC, Martin NG, Montgomery GW, Goddard ME, Visscher PM (2010).
\newblock \enquote{Common SNPs explain a large proportion of the heritability
  for human height.}
\newblock \emph{Nature Genetics}, \textbf{42}(7), 565--569.
\newblock \doi{10.1038/ng.608}.
\newblock \urlprefix\url{http://dx.doi.org/10.1038/ng.608}.

\bibitem[{Yates(1935)}]{yates1935complex}
Yates F (1935).
\newblock \enquote{Complex experiments.}
\newblock \emph{Supplement to the Journal of the Royal Statistical Society},
  \textbf{2}(2), 181--247.

\bibitem[{Yu \emph{et~al.}(2016)Yu, Li, Guo, Zhu, Wu, Mitchell, Roozeboom,
  Wang, Wang, Pederson \emph{et~al.}}]{yu2016genomic}
Yu X, Li X, Guo T, Zhu C, Wu Y, Mitchell SE, Roozeboom KL, Wang D, Wang ML,
  Pederson GA, \emph{et~al.} (2016).
\newblock \enquote{Genomic prediction contributing to a promising global
  strategy to turbocharge gene banks.}
\newblock \emph{Nature Plants}, \textbf{2}, 16150.

\end{thebibliography}

\appendix
\appendixpage

\textbf{\large{1. Initial solutions and an island model}}
\vspace{0.5cm}
\textbf{A.1}
Default design criteria implemented in STPGA.
\begin{sidewaystable}[!h]
  \centering
  \caption{Default design criteria implemented in STPGA}
    \begin{tabular}{llll}
    \textbf{criterion name} & \textbf{formula} & \textbf{type} & \textbf{equation} \bigstrut[b]\\
    \hline
    \hline
    \textit{AOPT} & $trace[C(X'_{Train}X_{Train}+\lambda*I)^{-1}C']$ & X & Equation \ref{eq:ridge1}\bigstrut\\
    \hline
    \textit{CDMAX} & $max[diag(CX_{Test}(X'_{Train}X_{Train}+\lambda*I)^{-1}X'_{Test}C')/$ & X & Equation \ref{eq:ridge2} \bigstrut[t]\\
          & $diag(CX_{Test}X'_{Test}C')]$ &  & \bigstrut[b]\\
    \hline
    \textit{CDMAX0} & $max[diag(CX_{Train}(X'_{Train}X_{Train}+\lambda*I)^{-1}X'_{Train}C')/$ & X & Equation \ref{eq:ridge2} \bigstrut[t]\\
          & $diag(CX_{Train}X'_{Train}C')]$ &  & \bigstrut[b]\\
    \hline
    \textit{CDMAX2} & $max[diag(CX_{Test}(X'_{Train}X_{Train}+\lambda*I)^{-1}X'_{Train}X_{Train}$ & X & Equation \ref{eq:ridge2} \bigstrut[t]\\
          & $(X'_{Train}X_{Train}+\lambda*I)^{-1}X'_{Test}C')/diag(CX_{Test}X'_{Test}C')]$ &  & \bigstrut[b]\\
    \hline
    \textit{CDMEAN} & $mean[diag(CX_{Test}(X'_{Train}X_{Train}+\lambda*I)^{-1}X'_{Test}C')/$ & X & Equation \ref{eq:ridge2} \bigstrut[t]\\
          & $diag(CX_{Test}X'_{Test}C')]$ &  & \bigstrut[b]\\
    \hline
    \textit{CDMEAN0} & $mean[diag(CX_{Train}(X'_{Train}X_{Train}+\lambda*I)^{-1}X'_{Train}C')/$ & X & Equation \ref{eq:ridge2} \bigstrut[t]\\
          & $diag(CX_{Train}X'_{Train}C')]$ &  & \bigstrut[b]\\
    \hline
    \textit{CDMEAN2} & $mean[diag(CX_{Test}(X'_{Train}X_{Train}+\lambda*I)^{-1}X'_{Train}X_{Train}$ & X & Equation \ref{eq:ridge2} \bigstrut[t]\\
          & $(X'_{Train}X_{Train}+\lambda*I)^{-1}X'_{Test}C')/diag(CX_{Test}X'_{Test}C')]$ &  & \bigstrut[b]\\
    \hline
    \textit{CDMEANMM} & $-mean[diag(CZ_{Test}(K-lambda*(Z_{Train}'MZ_{Train})^{-1}+\lambda*Kinv)Z_{Test}'C')/$  & K & Equation \ref{eq:mixed2}\bigstrut[t]\\
          & $(diag(CZ_{Test}KZ_{Test}'C'))],$ $M=I-W(W'W)^{-}W'$  &  & \bigstrut[b]\\
    \hline
    \textit{DOPT} & $logdet(C(X'_{Train}X_{Train}+\lambda*I)^{-1}C')$ & X & Equation \ref{eq:ridge1}\bigstrut[b]\\
    \hline
    \textit{EOPT} & $max(eigenval(C(X'_{Train}X_{Train}+\lambda*I)^{-1}C'))$  & X & Equation \ref{eq:ridge1} \bigstrut[b]\\
    \hline
    \textit{GAUSSMEANMM} & $-mean(diag(Z_{Test}KZ_{Test}'-$ & K & Equation \ref{eq:rkhs1}\bigstrut[t]\\
          & $Z_{Test}KZ_{Train}'(Z_{Train}KZ_{Train}'+\lambda*I)^{-1}Z_{Train}KZ_{Test}')$  &  & \bigstrut[b]\\
    \hline
    \textit{GOPTPEV} & $max(eigenval(CX_{Test}(X'_{Train}X_{Train}+\lambda*I)^{-1}X'_{Test}C'))$ & X  & Equation \ref{eq:ridge2} \bigstrut[b] \\
     \hline
    \textit{GOPTPEV2 } & $mean(eigenval(CX_{Test}(X'_{Train}X_{Train}+\lambda*I)^{-1}X'_{Test}C'))$ & X  & Equation \ref{eq:ridge2} \bigstrut[b] \\
     \hline
    \textit{PEVMAX} & $max(diag(CX_{Test}(X'_{Train}X_{Train}+\lambda*I)^{-1}X'_{Test}C'))$ & X  & Equation \ref{eq:ridge2} \bigstrut[b] \\
     \hline
    \textit{PEVMAX0} & $max(diag(CX_{Train}(X'_{Train}X_{Train}+\lambda*I)^{-1}X'_{Train}C'))$ & X  & Equation \ref{eq:ridge2} \bigstrut[b] \\
     \hline
    \textit{PEVMAX2} & $max[diag(CX_{Test}(X'_{Train}X_{Train}+\lambda*I)^{-1}$ & X & Equation \ref{eq:ridge2} \bigstrut[t]\\
          & $X'_{Train}X_{Train}(X'_{Train}X_{Train}+\lambda*I)^{-1}X'_{Test}C']$ &  & \bigstrut[b]\\
     \hline
    \textit{PEVMEAN} & $mean(diag(CX_{Test}(X'_{Train}X_{Train}+\lambda*I)^{-1}X'_{Test}C'))$ & X & Equation \ref{eq:ridge2} \bigstrut[b] \\
     \hline
    \textit{PEVMEAN0} &  $mean(diag(CX_{Train}(X'_{Train}X_{Train}+\lambda*I)^{-1}X'_{Train}C'))$ & X & Equation \ref{eq:ridge2} \bigstrut[b]\\
    \hline
    \textit{PEVMEAN2} & $mean[diag(CX_{Test}(X'_{Train}X_{Train}+\lambda*I)^{-1}$ & X & Equation \ref{eq:ridge2} \bigstrut[t]\\
          & $X'_{Train}X_{Train}(X'_{Train}X_{Train}+\lambda*I)^{-1}X'_{Test}C']$ &  & \bigstrut[b]\\
    \hline
    \textit{PEVMEANMM} & $mean(diag(CZ_{test}(Z_{Train}'MZ_{Train}+\lambda*Kinv)^{-1}Z_{Test}'C')))$ & K  & Equation \ref{eq:mixed1} \bigstrut[t] \\
           & $M=I-W(W'W)^{-}W'$  &  & \bigstrut[b]\\
    \hline
    \end{tabular}%
  \label{tab:STPGAOPTFUNCS}%
\end{sidewaystable}%

\textbf{A.2}
Using initial solutions and implementing an island model using GA. The scenario in the island model is not necessarily fixed, ingenious or advanced. It is for demonstration purposes and the users are encouraged to play around with it or change it completely.

\begin{GrayBox}
 \scriptsize
\textbf{Box A1: Loading the wheat data set included in STPGA}
\begin{Schunk}
\begin{Sinput}
> data(WheatData)
> svdWheat<-svd(Wheat.K, nu=50, nv=50)
> PC50WHeat<-Wheat.K%*%svdWheat$v
> rownames(PC50WHeat)<-rownames(Wheat.K)
> DistWheat<-dist(PC50WHeat)
> TreeWheat<-cutree(hclust(DistWheat), k=4)
> Test<-rownames(PC50WHeat)[TreeWheat==2]
> Candidates<-setdiff(rownames(PC50WHeat), Test)
> deptest<-Wheat.Y[Wheat.Y$id%in%Test,]
> Ztest<-model.matrix(~-1+deptest$id)
\end{Sinput}
\end{Schunk}
\end{GrayBox}

\begin{GrayBox}
 \scriptsize
\textbf{Box A2: Function that implements a simple island model}
\begin{Schunk}
\begin{Sinput}
> repeatgenalg<-function(numrepsouter,numrepsinner){
+   StartingPopulation2=NULL 
+   for (i in 1:numrepsouter){
+     StartingPopulation<-lapply(1:numrepsinner, function(x){
+       GenAlgForSubsetSelectionNoTest(P=PC50WHeat,
+       ntoselect=50, InitPop=StartingPopulation2,
+  npop=200, nelite=5, mutprob=.5, mutintensity = rpois(1,1),
+  niterations=200,minitbefstop=5, tabumemsize = 0, tabu=FALSE,
+  plotiters=FALSE, lambda=1e-9,
+  errorstat="CDMEAN", mc.cores=4)})
+     StartingPopulation2<-vector(mode="list", length = numrepsouter*1)
+     ij=1
+     for (i in 1:numrepsinner){
+       for (j in 1:1){
+         StartingPopulation2[[ij]]<-StartingPopulation[[i]][[j]]
+         ij=ij+1
+       }
+     }
+   }
+   ListTrain<-GenAlgForSubsetSelectionNoTest(
+             P=PC50WHeat[rownames(PC50WHeat)%in%Candidates,],
+             ntoselect=50,
+             InitPop=StartingPopulation2,npop=200, 
+             nelite=10, mutprob=.5, mutintensity = 1,niterations=200,
+             minitbefstop=50, tabumemsize = 0,tabu=FALSE, plotiters=FALSE,
+             lambda=1e-9,errorstat="DOPT", mc.cores=4)
+   return(ListTrain)
+ }
\end{Sinput}
\end{Schunk}
\end{GrayBox}

\begin{GrayBox}
 \scriptsize
\textbf{Box A3: Function that implements a simple island model}
\begin{Schunk}
\begin{Sinput}
> ListTrain<-repeatgenalg(10, 4)
> min(ListTrain$`Best criterion values over iterarions`)
\end{Sinput}
\begin{Soutput}
[1] -47.44094
\end{Soutput}
\begin{Sinput}
> min(Train4$`Best criterion values over iterarions`)
\end{Sinput}
\begin{Soutput}
[1] 5.123539
\end{Soutput}
\begin{Sinput}
> deptrainopt<-Wheat.Y[(Wheat.Y$id%in%ListTrain$`Solution with rank 1`),]
> Ztrain<-model.matrix(~-1+deptrainopt$id)
> modelopt<-emmreml(y=deptrainopt$plant.height,
+                    X=matrix(1, nrow=nrow(deptrainopt), ncol=1), 
+                   Z=Ztrain, K=Wheat.K)
> predictopt<-Ztest%*%modelopt$uhat
> cor(predictopt, deptest$plant.height)
\end{Sinput}
\begin{Soutput}
          [,1]
[1,] 0.3155026
\end{Soutput}
\end{Schunk}
\end{GrayBox}

\begin{GrayBox}
 \scriptsize
\textbf{Box A4: Function that implements a simple island model}
\begin{Schunk}
\begin{Sinput}
> TreeWheatTrain<-TreeWheat
> TreeWheatTrain[names(TreeWheatTrain)%in%ListTrain$`Solution with rank 1`]<-5
\end{Sinput}
\end{Schunk}
\end{GrayBox}

\begin{GrayBox}
 \scriptsize
\textbf{Box A5: Function that implements a simple island model}
\begin{Schunk}
\begin{Sinput}
> plot(PC50WHeat[,1],PC50WHeat[,2], col=TreeWheatTrain,
+      pch=as.character(TreeWheatTrain), xlab="pc1", ylab="pc2")
\end{Sinput}
\end{Schunk}
\includegraphics{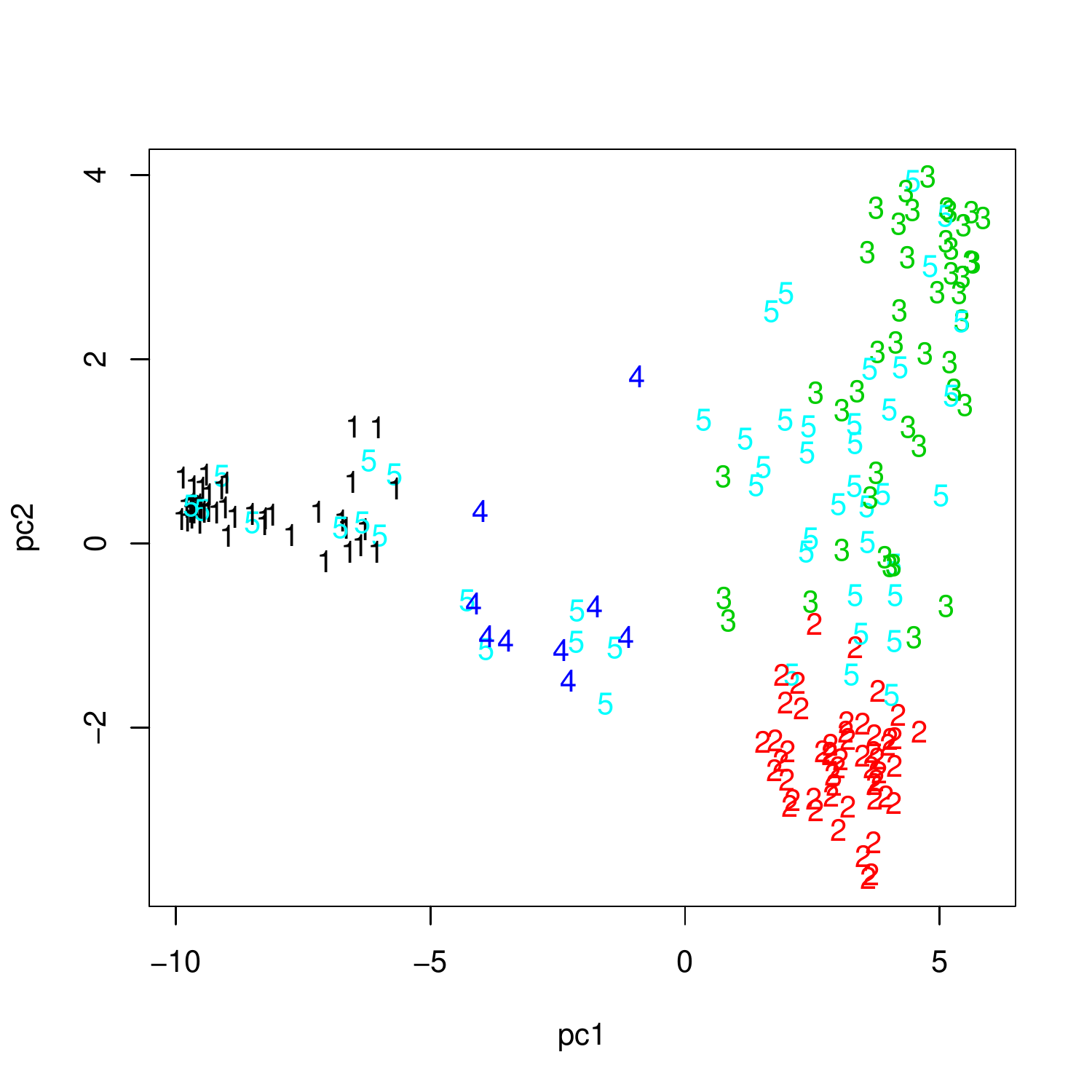}
\end{GrayBox}

\textbf{A.3}
A mixed Integer quadratic programming function for proportions
\begin{GrayBox}
 \scriptsize
\textbf{Box A6: A mixed Integer quadratic programming function for proportions}
\begin{Schunk}
\begin{Sinput}
> require(quadprog)
> MIQP<-function(Dmat, dvec, cardinality,npop=200, 
+           nelite=10, mutprob=.5, mutintensity = 1,niterations=200,
+           minitbefstop=50, tabumemsize = 1,plotiters=FALSE,tabu=FALSE,
+           lambda=1e-5, mc.cores=4){
+ 	P<-cbind(dvec,Dmat)
+ 	rownames(P)<-rownames(Dmat)
+ STPGAUSERDEFFUNC<-function(Train,Test=NULL, P, lambda=1e-5, C=NULL){
+     smallD<-P[rownames(P)%in%Train,-1]
+     smallD=smallD+lambda*diag(nrow(smallD))
+     smalld<-P[rownames(P)%in%Train,1]
+     n=length(smalld)
+     sol  <- solve.QP(Dmat=smallD,
+                      dvec=smalld,Amat=cbind(rep(1,n),diag(n),-diag(n)),
+                      bvec=rbind(1,matrix(0,ncol=1,nrow=n),matrix(-1,ncol=1,nrow=n)),
+                      meq=1)
+     names(sol$solution)<-rownames(smallD)
+     return(sol$value)
+   }
+   GAOUT<-GenAlgForSubsetSelectionNoTest(P=P, ntoselect=cardinality,npop=npop, 
+     nelite=nelite, mutprob=mutprob,
+     mutintensity = mutintensity,niterations=niterations,
+     minitbefstop=minitbefstop, tabumemsize = tabumemsize,
+     plotiters=plotiters,tabu=tabu,
+     lambda=lambda,errorstat="STPGAUSERDEFFUNC",
+     mc.cores=mc.cores)
+     smallD<-P[rownames(P)%in%GAOUT$`Solution with rank 1`,-1]
+     smallD=smallD+lambda*diag(nrow(smallD))
+     smalld<-P[rownames(P)%in%GAOUT$`Solution with rank 1`,1]
+     n=length(smalld)
+     sol  <- solve.QP(Dmat=smallD, dvec=smalld,
+                      Amat=cbind(rep(1,n),diag(n),-diag(n)),
+                      bvec=rbind(1,matrix(0,ncol=1,nrow=n),
+                                 matrix(-1,ncol=1,nrow=n)), meq=1)
+     names(sol$solution)<-rownames(smallD)
+     return(sol$value)
+ }
\end{Sinput}
\end{Schunk}
\end{GrayBox}

\end{document}